\documentclass[11pt,a4paper]{article}
\pdfoutput=1

\usepackage{jheppub}
\usepackage{latexsym}
\usepackage{multirow}
\usepackage{color}
\usepackage[usenames,dvipsnames,table]{xcolor}
\usepackage{graphicx}
\usepackage{subcaption}
\usepackage{epsfig} 
\usepackage{epsf}   
\usepackage{dcolumn}
\usepackage{bm}
\usepackage{dcolumn}
\usepackage{textcomp}
\usepackage{float}
\usepackage{hypcap}
\usepackage[]{hyperref}
\usepackage{makecell}
\usepackage{color}
\usepackage{pifont}
\usepackage{appendix}
\usepackage{amsmath}
\usepackage{mathrsfs}
\usepackage{multirow,bigdelim}  
\usepackage{lineno}
\usepackage[normalem]{ulem}
\graphicspath{{fig/}}
\usepackage{physics}
\usepackage{simpler-wick}
\usepackage{slashed}

\def\be{\begin{equation}}
\def\ee{\end{equation}}
\definecolor{maroon}{rgb}{0.5, 0.0, 0.0}	
\definecolor{arsenic}{rgb}{0.23, 0.27, 0.29}

\newcommand{\comment}[1]{}
\usepackage{amssymb}
\hypersetup{
	bookmarks=true,         
	unicode=false,          
	pdftoolbar=true,        
	pdfmenubar=true,        
	pdffitwindow=true,   
	pdfstartview={FitH},    
	pdfsubject={dark NSI},   
	pdfnewwindow=true,      
	pdfcreator={RevTeX},
	colorlinks=true,     
	linkcolor=Blue,         
	citecolor=Blue,        
	filecolor=Blue,     
	urlcolor=Blue,           
}

\usepackage{tikz,xcolor,hyperref}

\definecolor{lime}{HTML}{A6CE39}
\DeclareRobustCommand{\orcidicon}{\hspace{-4pt}
	\begin{tikzpicture}
		\draw[lime, fill=lime] (0,0) 
		circle [radius=0.16] 
		node[white] {\hspace{0.1mm}{\fontfamily{qag}\selectfont \tiny ID}};
		\draw[white, fill=white] (-0.07,0.1) 
		circle [radius=0.01];
	\end{tikzpicture}
	\hspace{-3.2mm}
}

\foreach \x in {A, ..., Z}{\expandafter\xdef\csname 
	orcid\x\endcsname{\noexpand\href{https://orcid.org/\csname orcidauthor\x\endcsname}
		{\noexpand\orcidicon}}
}

\preprint{}

\title{Constraining and Resolving Lorentz-Violating New Physics at ESSnuSB Using Complementarity with DUNE}

\author[a, b]{Himanshu Bora\orcidA{},} 
\author[a]{Debajyoti Dutta\orcidB{},}
\author[c]{Abinash Medhi\orcidC{}}

\affiliation[a]{Department of Physics, Bhattadev University, Bajali, Pathsala, 781325, India}
\affiliation[b]{Department of Physics, Kamrup College, Chamata, Nalbari, Assam - 78306, India}
\affiliation[c]{Department of Physics, G.L. Choudhury College, Barpeta Road, Assam 781315, India}

\emailAdd{hb@kamrupcollege.ac.in}
\emailAdd{phy.debajyoti@bhattadevuniversity.ac.in}
\emailAdd{abinashmedhi0@gmail.com}

\date{\today}

\abstract{
We examine the sensitivity of the ESSnuSB and DUNE long-baseline neutrino experiments to isotropic, CPT-violating Lorentz Invariance Violation (LIV). Using detailed simulations for the 360 km and 540 km ESSnuSB baselines and the 1300 km DUNE setup, we assess how LIV parameters influence oscillation probabilities, event spectra, and degeneracies among oscillation parameters. We find that LIV-induced modifications can closely mimic variations in $\theta_{23}$ and $\delta_{\rm CP}$, potentially leading to incorrect determination of the atmospheric mixing angle octant and the leptonic CP phase if LIV effects are not accounted for. Although combining the two ESSnuSB baselines improves overall sensitivity, it does not fully remove these degeneracies. In contrast, a joint ESSnuSB+DUNE analysis benefiting from the synergy between second-maximum sensitivity at ESSnuSB and first-maximum, matter-enhanced sensitivity at DUNE can successfully resolve all these degeneracies and can yield significantly stronger constraints on all the LIV parameters. The results presented here highlights the essential role of multi-baseline, multi-energy experimental strategies to probe Planck-suppressed Lorentz-violating new physics.
}

\keywords{Lorentz Invariance Violation, CP-Violation, Neutrino Physics, Beyond Standard Model, Long Baseline Experiments}
\arxivnumber{25xx.xxxxx}

\begin{document}

\maketitle

\section{Introduction}\label{sec:intro}

The Standard Model (SM) of particle physics arguably stands as the most successful theory in fundamental physics, describing the fundamental interactions with remarkable precision. However, the discovery of neutrino oscillations, established through definitive observations of atmospheric and solar neutrinos by the Super-Kamiokande (SK) \cite{Super-Kamiokande:1998kpq} and Sudbury Neutrino Observatory (SNO) \cite{SNO:2002tuh} experiments, respectively, provided the first concrete experimental evidence of its incompleteness. The fact that neutrinos have mass, as confirmed by a wealth of data from subsequent experiments \cite{Super-Kamiokande:2001ljr,SNO:2001kpb, SNO:2002tuh2,Super-Kamiokande:1999vcj,Super-Kamiokande:2000bnn,CHOOZ:1997zlf,DayaBay:2012fng,DoubleChooz:2012eiq,RENO:2012mkc,KamLAND:2004mhv,KamLAND:2002uet,KamLAND:2008dgz,OPERA:2018nar,T2K:2011qtm, T2K:2013ppw, NOvA:2019cyt} spanning over decades, lead to an extension of the SM framework with the three-flavour framework. Current and future experimental efforts, like those at  DUNE\cite{DUNE:2015lol}, T2HK\cite{Abe:2018uyc}, T2HKK\cite{Abe:2016ero}, ESSnuSB\cite{ESSnuSB:2023design} etc. aim at resolving the remaining unknowns within the three-flavor paradigm: the value of the leptonic CP phase ($\delta_{CP}$), the neutrino mass ordering (normal vs. inverted) and the precise value, and particularly, the octant of the atmospheric mixing angle ($\theta_{23}$).

The exploration of Beyond Standard Model (BSM) physics \cite{Ge:2018uhz, Miranda:2015dra, Medhi:2021wxj, Medhi:2023ebi, Medhi:2022qmu,Babu:2019iml,Farzan:2017xzy,Masud:2016nuj,PhysRevD.94.013014,Coloma:2017zpg,PICORETI201670,SNO:2018pvg} encompasses far more than the quest to explain neutrino mass. A crucial aspect that requires close examination is Lorentz invariance, one of the foundational symmetries upon which both the Standard Model (SM) and Einstein’s theory of General Relativity are built \cite{Tasson:2014dfa}. Despite being verified with exceptional precision, several theoretical considerations indicate that Lorentz invariance may not be an exact symmetry of nature. In attempts to unify quantum mechanics and general relativity at the Planck scale ($M_P \sim 10^{19}$ GeV), it has been suggested that the conventional structure of spacetime could emerge only as a low-energy limit of a more fundamental framework. Within this perspective, the SM can be regarded as a low-energy effective field theory (EFT), wherein potential effects from Planck-scale physics—such as possible Lorentz Invariance Violation (LIV) \cite{PhysRevD.69.016005,PhysRevD.98.112013,PhysRevD.99.104062,PhysRevD.99.123018,ARIAS2007401,LSND:2005oop,MINOS:2008fnv,MINOS:2010kat,IceCube:2010fyu,MiniBooNE:2011pix,DoubleChooz:2012eiq,Pan:2023qln}, neutrino decoherence \cite{PhysRevD.56.6648,Benatti:2000ph,PhysRevD.100.055023,PhysRevD.95.113005,Lisi:2000zt} would manifest as tiny, Planck-suppressed corrections to the SM Lagrangian that might be experimentally accessible. The Standard Model Extension (SME) offers a general and model-independent EFT framework to systematically describe and investigate these potential deviations \cite{Colladay:1996iz, Colladay:1998fq, Kostelecky:2003fs, Kostelecky:1988zi, Ferrari:2018tps}.

The CPT theorem is one of the most profound results of local relativistic quantum field theory, which states that any such theory must be invariant under the combined action of Charge conjugation (C), Parity inversion (P) and Time reversal (T). Its proof relies on three fundamental assumptions: Lorentz invariance, hermiticity of the Hamiltonian and locality (or microcausality) \cite{PhysRevLett.89.231602}. A crucial consequence is that any interacting quantum field theory that violates CPT symmetry must necessarily also violate Lorentz invariance \cite{PhysRevLett.89.231602}. The search for LIV is, thus, intimately connected to the CPT theorem. Neutrinos serve as a uniquely powerful probe for these effects. Their remarkably weak interactions allow them to traverse vast cosmological and terrestrial distances, providing an exceptionally long path that can amplify the phase of tiny, Planck-suppressed LIV effects to a level that may be observable in high-precision experiments \cite{Barenboim:2018ctx, Agarwalla2020}. Consequently, long-baseline (LBL) neutrino oscillation experiments are among the most sensitive tools available for testing Lorentz symmetry \cite{Majhi:2019tfi, Fiza:2022xfw}. 

Stringent constraints on various LIV parameters have already been set by experiments such as Super-Kamiokande \cite{Super-Kamiokande:2014exs}, IceCube \cite{IceCube:2017qyp}, T2K \cite{T2K:2017ega}, MINOS \cite{MINOS:2008fnv, MINOS:2010kat} and Double Chooz \cite{DoubleChooz:2012eiq}. A broad range of experimental parameters—such as neutrino-beam flavor composition, baseline length, direction, and energy—as well as diverse detector techniques, provide complementary sensitivities to the many operators characterizing LIV. Beyond experimental searches \cite{RevModPhys.83.11}, a number of recent phenomenological studies have investigated the specific effects of LIV on long-baseline (LBL) experiments. These include investigations in long-baseline accelerator neutrinos \cite{Barenboim:2018ctx, Majhi:2019tfi, Agarwalla2020, Fiza:2022xfw, Raikwal:2023lzk}, short-baseline reactor antineutrinos \cite{Abrahao:2015rba}, atmospheric neutrinos \cite{Super-Kamiokande:1998kpq, IceCube:2010fyu}, solar neutrinos \cite{SNO:2018pvg, PhysRevD.92.073003}, and high-energy astrophysical neutrinos \cite{PhysRevD.99.043015, PhysRevD.99.043013, PhysRevD.99.123018}. In \cite{Majhi:2019tfi}, the authors explored the impact of CPT-violating LIV parameters on the oscillation probabilities at NOvA and T2K, demonstrating that LIV can significantly alter the sensitivity to CP violation and the mass hierarchy, also highlighting that the synergy between NOvA and T2K yields significantly enhanced sensitivities compared to individual setups. In the context of DUNE, the authors \cite{Agarwalla2020} investigated how LIV affects the determination of the $\theta_{23}$ octant and the reconstruction of the CP phase. They demonstrated that non-zero LIV coefficients generally reduce octant sensitivity; however, interestingly, the simultaneous presence of $a_{e\mu}$ and $a_{e\tau}$ can lead to a mutual nullifying effect that largely restores the octant sensitivity. Furthermore, reference \cite{Rahaman:2021leu} demonstrated that the tension between T2K and NOvA data could be reduced in the presence of LIV, although the octant and mass hierarchy sensitivities for both experiments deteriorated. Expanding the experimental landscape, this study \cite{Fiza:2022xfw} investigated the synergy between DUNE and the proposed Protvino-to-ORCA (P2O) experiment, showing that their combination effectively breaks degeneracies and pushes limits to the $\mathcal{O}(10^{-24})$ GeV level. A global perspective was provided by the authors in \cite{Raikwal:2023lzk}, here they combined atmospheric data from ICAL with beam data from DUNE and T2HK, demonstrating that atmospheric neutrinos provide crucial constraints on the muon sector parameters that complement beam-based measurements.

The determination of the standard oscillation parameters is complicated by a significant challenge - the phenomenological signatures of LIV can be entangled with the effects of standard oscillation parameters. The introduction of new energy dependencies and new CP-violating phases from LIV can mimic the effects of the standard CP phase $\delta_{CP}$, the atmospheric mixing angle $\theta_{23}$ and the neutrino mass ordering \cite{Agarwalla2020}. Furthermore, the Hamiltonian for isotropic, CPT-violating LIV is mathematically analogous to that of Non-Standard Interactions (NSI) in matter, creating a potential ``new physics" degeneracy \cite{Barenboim:2018lpo, Sahoo:2022rns}. These degeneracies may lead to an incorrect determination of the standard parameters if LIV effects are present in nature but not accounted for in the analysis, or conversely, may obscure a genuine LIV signal.

In this work, we investigate the potential of two next-generation facilities: the European Spallation Source neutrino Super Beam (ESSnuSB) \cite{ESSnuSB:2023design} and the Deep Underground Neutrino Experiment (DUNE) \cite{DUNE:2015lol}, both individually and combined, to constrain the isotropic, CPT-violating LIV parameters $a_{\alpha\beta}$ and to resolve the degeneracies highlighted previously. While in \cite{Delgadillo:2025sme}, the authors investigated the constraints on the isotropic CPT-odd coefficients and the correlations of the leptonic CP-violating phase, $\delta_{CP}$, and the atmospheric mixing angle, $\sin ^2 \theta_{23}$, with respect to the said LIV parameters by a combined analysis of DUNE (1300 km baseline) and ESSnuSB (360 km baseline), we offer a more comprehensive analysis. Our analysis differs in the sense that we study the synergy of all available baseline combinations and by using full-range marginalization over $\delta_{CP}$, $\theta_{23}$, and both the mass orderings to identify possible global degeneracies. Recently, the JUNO experiment released its first physics result, determining $\sin^2\theta_{12}$, and the solar mass-squared difference, $\Delta m^2_{21} \equiv m^2_2 - m^2_1$, to unprecedented precision of $\sin^2\theta_{12} = 0.3092 \pm 0.0087 \quad \text{and} \quad \Delta m^2_{21} = (7.50 \pm 0.12) \times 10^{-5} \, \text{eV}^2,$ which reduced the relative $1\sigma$ uncertainty in the determination of $\sin^2\theta_{12}$ to 2.81\% and that in the determination of $\Delta m^2_{21}$ to 1.55\% \cite{Abusleme2025JUNO}. We also include the recent JUNO measurement of $\sin^{2}\theta_{12}$ to ensure that our results use the most up-to-date experimental inputs.

The proposed ESSnuSB facility, with its plan to utilize two distinct baselines (360 km to Zinkgruvan and 540 km to Garpenberg), both operating near the second oscillation maximum, presents an interesting case. This configuration inherently provides some internal complementarity, probing slightly different L/E values and matter effects. A key question is whether this internal synergy within ESSnuSB is sufficient to effectively disentangle LIV parameters from standard oscillations, particularly $\delta_{CP}$ and $\theta_{23}$. 
DUNE contrasts sharply with ESSnuSB, featuring a broad energy spectrum and a very long 1300 km baseline covering the first oscillation maximum and experiencing substantial matter effects ($\hat{A} \sim 0.2-0.3$ near the peak). Our study aims to quantify the capability of the combined two-baseline ESSnuSB setup to constrain LIV and resolve degeneracies, and critically assess the additional power gained by including the complementary information from DUNE.

The paper is structured as follows: section \ref{sec:formalism} outlines the theoretical formalism of LIV within the SME. Section \ref{sec:simulation} details the experimental setups considered and the simulation methodology employed. Section \ref{sec:probabilities} analyzes the impact of these LIV parameters on neutrino oscillation probabilities, highlighting the origins of parameter degeneracies. Section \ref{sec:events} examines how these probability modifications translate into observable effects on event rates. The core results are presented in section \ref{sec:results_discussions}, which details the sensitivity analysis and projected constraints on the LIV parameters (sections \ref{sec:constraints}), specifically demonstrating the power of the combined DUNE and ESSnuSB analysis to resolve critical degeneracies (section \ref{sec:degeneracy}). Finally, section \ref{sec:conclusion} offers a summary of our findings and concluding remarks.

\section{Theoretical Formalism}\label{sec:formalism}

In the Standard Model Extension (SME) framework, potential deviations from exact Lorentz and CPT symmetry are parameterized by adding coordinate-independent terms to the standard Lagrangian \cite{Kostelecky:2003cr, Colladay:1998fq, Barenboim:2018ctx, Majhi:2019tfi, Fiza:2022xfw}. The general Lorentz-violating Lagrangian density for a Dirac fermion, restricted to renormalizable operators (mass dimension $\le 4$), can be written as:

$$
\mathcal{L}_{LIV}=-\frac{1}{2}[a_{\alpha\beta}^{\mu}\overline{\psi}_{\alpha}\gamma_{\mu}\psi_{\beta}+b_{\alpha\beta}^{\mu}\overline{\psi}_{\alpha}\gamma_{5}\gamma_{\mu}\psi_{\beta}-ic_{\alpha\beta}^{\mu\nu}\overline{\psi}_{\alpha}\gamma_{\mu}\partial_{\nu}\psi_{\beta}-id_{\alpha\beta}^{\mu\nu}\overline{\psi}_{\alpha}\gamma_{5}\gamma_{\mu}\partial_{\nu}\psi_{\beta}]+h.c.
$$
Here, $\psi_\alpha$ represents the fermion field for flavor $\alpha$, and the coefficients $a_{\alpha\beta}^\mu$, $b_{\alpha\beta}^\mu$, $c_{\alpha\beta}^{\mu\nu}$, and $d_{\alpha\beta}^{\mu\nu}$ are constant background fields that parameterize the violation of Lorentz symmetry. The terms involving $a_{\alpha\beta}^\mu$ and $b_{\alpha\beta}^\mu$ are CPT-violating, whereas the derivative-coupling terms involving $c_{\alpha\beta}^{\mu\nu}$ and $d_{\alpha\beta}^{\mu\nu}$ are CPT-even \cite{Kostelecky:2003cr}. 

In the context of the Standard Model, only left-handed neutrinos participate in weak interactions. The observable effects of LIV on these neutrinos are controlled by specific combinations of the SME coefficients \cite{Kostelecky:2003cr, Majhi:2019tfi, Agarwalla2020, Barenboim:2018ctx, Fiza:2022xfw}:
\begin{align*}
(a_L)_{\alpha\beta}^\mu &= (a+b)_{\alpha\beta}^\mu \quad (\text{CPT-violating}) \\
(c_L)_{\alpha\beta}^{\mu\nu} &= (c+d)_{\alpha\beta}^{\mu\nu} \quad (\text{CPT-even})
\end{align*}

These coefficients form Hermitian matrices in flavor space. Following common practice for phenomenological studies, we adopt the isotropic approximation, assuming rotational symmetry by considering only the time components ($\mu=\nu=0$)  \cite{Kostelecky:2003cr, Majhi:2019tfi}. Following this, a convenient notational change is adopted: $(a_L)_{\alpha\beta}^0 \to a_{\alpha\beta}$ and $(c_L)_{\alpha\beta}^{00} \to c_{\alpha\beta}$ \cite{Majhi:2019tfi, Agarwalla2020}. These coefficients are defined in the standard Sun-centered celestial equatorial frame. This work focuses exclusively on the effects of the isotropic, CPT-violating $a_{\alpha\beta}$ parameters.

The time evolution of neutrino flavor states $(\nu_e, \nu_\mu, \nu_\tau)$ propagating through matter is described by an effective Hamiltonian:
\begin{equation}
H_{eff}=H_{vac}+H_{mat}+H_{LIV}
\label{eq:HeffTotal}
\end{equation}
where:
\begin{itemize}
    \item $H_{vac}$ is the standard vacuum oscillation Hamiltonian:
    \begin{equation}
    H_{vac}=\frac{1}{2E}U \text{diag}(0, \Delta m_{21}^{2}, \Delta m_{31}^{2}) U^{\dagger}
    \label{eq:Hvac}
    \end{equation}
    (Here, U is the PMNS matrix, E is neutrino energy, and $\Delta m_{ij}^2 = m_i^2 - m_j^2$).
    \item $H_{mat}$ represents the standard MSW matter effect:
    \begin{equation}
    H_{mat}=\sqrt{2}G_{F}N_{e} \text{diag}(1, 0, 0)
    \label{eq:Hmat}
    \end{equation}
    ($G_F$ is the Fermi constant, $N_e$ is the electron number density).
    \item The isotropic LIV Hamiltonian (including both CPT-odd and CPT-even terms (for completeness) \cite{Kostelecky:2003cr} is written as:
    $$
    \begin{pmatrix} a_{ee} & a_{e\mu} & a_{e\tau} \\ a_{e\mu}^* & a_{\mu\mu} & a_{\mu\tau} \\ a_{e\tau}^* & a_{\mu\tau}^* & a_{\tau\tau} \end{pmatrix} - \frac{4}{3}E \begin{pmatrix} c_{ee} & c_{e\mu} & c_{e\tau} \\ c_{e\mu}^* & c_{\mu\mu} & c_{\mu\tau} \\ c_{e\tau}^* & c_{\mu\tau}^* & c_{\tau\tau} \end{pmatrix}
    $$

    As mentioned earlier, the focus of this study is on the first term, which is parameterized by the CPT-violating $a_{\alpha\beta}$ coefficients. This term is notably energy-independent, in contrast to the CPT-even term parameterized by $c_{\alpha\beta}$, which is directly proportional to the neutrino energy $E$ \cite{Kostelecky:2003cr, Barenboim:2018ctx}. This distinction in energy dependence is fundamental; it provides a physical basis to model LIV-effects as energy-independent corrections to the matter potential.
    
    $H_{LIV}$, thus, incorporates the CPT-violating LIV effects focused on in this study:
    \begin{equation}
    H_{LIV} = \begin{pmatrix} a_{ee} & a_{e\mu} & a_{e\tau} \\ a_{e\mu}^{*} & a_{\mu\mu} & a_{\mu\tau} \\ a_{e\tau}^{*} & a_{\mu\tau}^{*} & a_{\tau\tau} \end{pmatrix}
    \label{eq:HLIVa}
    \end{equation}
    The diagonal elements ($a_{ee}, a_{\mu\mu}, a_{\tau\tau}$) are real, while the off-diagonal elements are complex, satisfying $a_{\alpha\beta} = a_{\beta\alpha}^* = |a_{\alpha\beta}| e^{i\phi_{\alpha\beta}}$ for $\alpha \neq \beta$. The phases $\phi_{e\mu}, \phi_{e\tau}, \phi_{\mu\tau}$ represent new potential sources of CP violation beyond the standard phase $\delta_{CP}$ in the PMNS matrix.
\end{itemize} 

Crucially, due to the CPT-violating nature of the $a_{\alpha\beta}$ terms, the effective Hamiltonian for antineutrinos transforms as :
\begin{equation}
H_{eff}(\overline{\nu}) = H_{vac} - H_{mat}^{*} - H_{LIV}^{*}
\label{eq:HeffAntiNu}
\end{equation}
Specifically, the LIV term flips its sign and gets conjugated:
\begin{equation}
H_{LIV}(\overline{\nu}) = -H_{LIV}^{*} = 
-\begin{pmatrix} a_{ee} & a_{e\mu}^{*} & a_{e\tau}^{*} \\ a_{e\mu} & a_{\mu\mu} & a_{\mu\tau}^{*} \\ a_{e\tau} & a_{\mu\tau} & a_{\tau\tau} \end{pmatrix}
\label{eq:HLIVaAntiNu}
\end{equation}

This difference between the neutrino and antineutrino Hamiltonians (in both $H_{mat}$ and $H_{LIV}$) is fundamental to probing CPT violation and requires analyzing both $\nu$ and $\bar{\nu}$ oscillation channels.

\section{Experiment and Simulation Details}\label{sec:simulation}

As mentioned previously, LBL experiments offer an ideal framework for examining LIV effects because the long travel distances can significantly amplify any LIV-induced phase shifts \cite{Barenboim:2018lpo, Agarwalla2020}. In this work, we are exploring the potential constraints on isotropic, CPT-violating LIV parameters ($a_{\alpha\beta}$) using the ESSnuSB and DUNE oscillation experiments and attempting to resolve any degeneracies that may arise due to LIV effects. We are using the General Long Baseline Experiment Simulator (GLoBES) software package \cite{Huber:2004ka, HUBER2007439} to simulate the performance of the two baselines (360 km and 540 km) in the proposed ESSnuSB facility, both individually and in combination, as well as to explore the synergy with a longer 1300 km baseline of the DUNE experiment. 

\subsection{Experiment Details}
In this subsection, we discuss the details of the neutrino experiments used in our study.
\paragraph{ESSnuSB:} The European Spallation Source neutrino Super Beam (ESSnuSB) is a proposed facility leveraging the high-power proton linear accelerator (linac) of the European Spallation Source in Lund, Sweden \cite{ESSnuSB:2023design}. Two potential sites for large Water Cherenkov far detectors (MEMPHYS-like, $\sim 500$ kton fiducial volume) are considered - Zinkgruvan (located 360 km from Lund and Garpenberg (located 540 km from Lund). By upgrading the linac power (potentially to 10 MW) and adding an accumulator ring and target station, ESSnuSB aims to produce a very intense neutrino beam. In this analysis, we have considered a total run time of 10 years equally divided between $\nu$ and $\bar{\nu}$ modes. ESSnuSB is uniquely designed to operate near the second oscillation maximum ($E_\nu \sim 0.35$ GeV for the proposed baselines), a region known to enhance sensitivity to $\delta_{CP}$ compared to the first maximum \cite{ESSnuSB:2023design}. A primary physics goal is the precise measurement of the leptonic CP phase $\delta_{CP}$.

\paragraph{DUNE:} The Deep Underground Neutrino Experiment (DUNE) is an LBL experiment under construction in the US \cite{DUNE:2015lol, DUNE:2016hlj, DUNE:2020ypp, DUNE:2020txw}. It will utilize a high-intensity, broadband neutrino beam generated by the Long-Baseline Neutrino Facility (LBNF) at Fermilab, initially operating at 1.2 MW power (with future upgrades planned) \cite{DUNE:2015lol}. The simulation details of DUNE used in this work can be found in \cite{DUNE:2021cuw}. The beam, peaking around $E_\nu \sim 2.5$ GeV, covers the first oscillation maximum. The far detector consists of four 10-kton (fiducial) Liquid Argon Time Projection Chambers (LArTPCs) located 1300 km away at the Sanford Underground Research Facility (SURF) \cite{DUNE:2015lol, DUNE:2020txw}. DUNE's primary goals include the definitive determination of the neutrino mass ordering, the discovery of leptonic CP violation, and precise measurements of oscillation parameters \cite{DUNE:2015lol, DUNE:2020ypp}. Its very long baseline (1300 km) leads to significant matter effects, which enhance sensitivity to the mass ordering but can also introduce correlations and degeneracies with other parameters, including potential LIV effects \cite{Agarwalla2020}. The DUNE experiment is designed to investigate the phenomenon of proton decay, while simultaneously enhancing its sensitivity to neutrinos originating from core-collapse supernovae. Furthermore, it seeks to probe physics that extends beyond the Standard Model \cite{DUNE:2020fgq}.

\subsection{Simulation Details }

The CPT-violating $a_{\alpha\beta}$ parameters enter the antineutrino Hamiltonian with an opposite sign and complex conjugation compared to the neutrino Hamiltonian (Eq. \ref{eq:HeffAntiNu}). This leads to distinct differences in the oscillation probabilities $P(\nu_\alpha \to \nu_\beta)$ and $P(\bar{\nu}_\alpha \to \bar{\nu}_\beta)$ beyond the standard matter effect differences. Therefore, analyzing dedicated runs in both neutrino and antineutrino modes, as planned for both DUNE and ESSnuSB and simulated in this work, is essential for isolating and constraining these CPT-violating LIV signals. In line with this strategy, all experimental configurations simulated in this work include dedicated runs in both neutrino and antineutrino modes. The simulation details for neutrino and antineutrino runtimes, along with detector uncertainties, are listed in Table \ref{tab:uncertainity}. While some illustrative plots in this paper may show only the neutrino channel for clarity, the full $\Delta\chi^2$ analyses and the final constraints presented in Sections \ref{sec:constraints} and \ref{sec:degeneracy} are derived from the combined statistical power of both neutrino and antineutrino channels.

\begin{table}[H]
    \centering
    \renewcommand{\arraystretch}{1.0}
    \begin{tabular}{|c|c|c|c|}

        \hline
        \multirow{2}{*}{Experiment details} &\multirow{2}{*}{Channels} & \multicolumn{2}{|c|}{Normalization uncertainty} \\ \cline{3-4}
        & & Signal & Background \\
        \hline \hline
        
             \textbf{ESSnuSB}, Baseline: 360 \& 540 km    &  & & \\
         L/E = 1200 \& 1534 km/GeV &  $\nu_e (\bar \nu_e)$ appearance  & 3.2\% (3.9\%) & 5\% (5\%) \\ 
          Fiducial mass = 500 kt (WC)& $\nu_\mu (\bar \nu_\mu)$ disappearance  & 3.6\% (3.6\%) & 5\% (5\%)  \\ 
        \hline
         \textbf{DUNE}, Baseline: 1300 km  &  & &\\
         L/E = 1543 km/GeV &  $\nu_e (\bar \nu_e)$ appearance & 2\% (2\%) & 5\% (5\%) \\
          Fiducial mass = 40 kt (LArTPC) & $\nu_\mu (\bar \nu_\mu)$ disappearance & 5\% (5\%) & 5\% (5\%) \\ 
         \hline
    \end{tabular}
    \caption{Experimental configurations and detector uncertainties for ESSnuSB and DUNE.}
    \label{tab:uncertainity}
\end{table}

\textbf{Parameter Values and Marginalization:} The true values and marginalization ranges for the standard oscillation parameters used in this analysis are listed in Table \ref{tab:marginalization_val}. Values are based on recent global fits, NuFit-6.0 \cite{Esteban:2024eli} and the recently reported JUNO results for $\theta_{12}$ \cite{Abusleme2025JUNO}. The true ordering is assumed to be normal ordering (NO). All LIV parameters ($a_{\alpha\beta}, \phi_{\alpha\beta}$) are set to zero in the true hypothesis and marginalized over appropriate ranges in the test hypothesis. Any other specific marginalization details, if used for analysis, are discussed in the respective sections.
\begin{table*}[!h]
    \centering 
    \renewcommand{\arraystretch}{1.3}
    \begin{tabular}{|c|c|c|}
    \hline 
    \rule{0pt}{12pt} Oscillation Parameter & Best Fit Value & Details of Marginalization \tabularnewline
    \hline \hline
    \rule{0pt}{12pt} $\theta_{12}$       & $33.68^{\circ}$               &  Prior from JUNO \cite{Abusleme2025JUNO} \tabularnewline \hline 
    \rule{0pt}{12pt} $\theta_{13}$       & $8.52^{\circ}$                & within 1$\sigma$ allowed range \tabularnewline \hline 
    \rule{0pt}{12pt} $\theta_{23}$       & $48.5^{\circ}$                & within 3$\sigma$ allowed range\tabularnewline \hline 
    \rule{0pt}{12pt} $\delta_{CP}$       & $-90^{\circ}$                 & $[-180^{\circ},180^{\circ}]$\tabularnewline \hline 
    \rule{0pt}{12pt} $\Delta m_{21}^{2}$ & $7.49\times10^{-5}\rm~eV^{2}$ & within 1$\sigma$ allowed range \tabularnewline \hline
    \renewcommand{\arraystretch}{1.0}
    \rule{0pt}{12pt} $\Delta m_{31}^{2}$ &  $2.53\times10^{-3}\rm~eV^{2}$& $\left( [-2.58,-2.44] \cup [2.46,2.60] \right)\times10^{-3}\rm~eV^{2}$\tabularnewline  
    
    \hline 
    \end{tabular}\\
    \vspace{0.2cm}
    \caption{The oscillation parameter values used in our analysis, along with the corresponding marginalization details (NuFit 6.0 \cite{Esteban:2024eli}). }
    \label{tab:marginalization_val}
\end{table*} 
\section{Results and Discussions}\label{sec:results_discussions}
In this section, we discuss the result of our analysis. In subsection \ref{sec:probabilities}, we show the effect of LIV parameters on the oscillation probabilities. Then in subsection \ref{sec:events}, we discuss the effects of LIV parameters on the events of both detectors. Finally, we show how the sensitivity is affected in the presence of LIV in section \ref{sec:sensitivity_analysis}.

\subsection{LIV and Neutrino Oscillation Probabilities}
\label{sec:probabilities}
The impact of the LIV Hamiltonian on neutrino flavor transitions can be calculated using matter perturbation theory \cite{Kikuchi:2008vq, Arafune:1997hd}. The evolution of the neutrino state $|\nu(x)\rangle$ over a distance $x$ is described by the Schrödinger-like equation $i\frac{d}{dx}|\nu\rangle = H_{eff}|\nu\rangle$. The probability of a transition from flavor $\alpha$ to $\beta$ after traversing a baseline $L$ is then $P_{\alpha\beta} = |\langle\nu_\beta | e^{-iH_{eff}L} | \nu_\alpha \rangle|^2$ \cite{Majhi:2019tfi}. While analytical approximations, such as those derived up to first or second order in perturbation theory, are valuable for understanding the leading-order effects and parameter dependencies \cite{Sarker2023LIV}, the full probability analysis relies on exact numerical calculations performed using the GLoBES software package. To implement the LIV Hamiltonian (Eq. \ref{eq:HLIVa}), we developed an independent probability engine to numerically solve the neutrino evolution equation. The plots presented here illustrate how various LIV parameters affect the oscillation probability differently for a benchmark LIV parameter magnitude $|a_{\alpha\beta}| = 2 \times 10^{-23}$ GeV in a 1300 km baseline for $\nu_\mu \to \nu_e$ appearance and $\nu_\mu \to \nu_\mu$ disappearance channels.

\subsubsection{$\nu_\mu \to \nu_e$ Appearance Probability}

For long-baseline experiments, the $\nu_\mu \to \nu_e$ appearance channel is paramount for measuring CP violation and the mass ordering. In the presence of isotropic, CPT-violating LIV, the probability can be approximated as a sum of the standard interaction (SI) term and new interference terms from the LIV parameters $a_{e\mu}$ and $a_{e\tau}$ \cite{Majhi:2019tfi, Agarwalla2020, Fiza:2022xfw}:
$$P_{\mu e}(SI+LIV) \simeq P_{\mu e}(SI) + P_{\mu e}(a_{e\mu}) + P_{\mu e}(a_{e\tau})$$
The SI term is given by $P_{\mu e}(SI) \simeq X + Y \cos(\delta_{CP} + \Delta)$, where $X$ and $Y$ are functions of the standard oscillation parameters and matter effect, $\Delta = \frac{\Delta m_{31}^2 L}{4E}$, and $\delta_{CP}$ is the standard CP-violating phase \cite{Arafune:1997hd, Majhi:2019tfi}. The LIV interference terms introduce new dependencies on the LIV parameters and their associated phases, $\phi_{e\mu}$ and $\phi_{e\tau}$. The diagonal parameter $a_{ee}$ enters as a sub-leading effect, modifying the effective matter potential term $\hat{A} = \frac{2\sqrt{2}G_F N_e E}{\Delta m_{31}^2}$ \cite{Majhi:2019tfi}.

\begin{figure}[H]
    \centering
    \includegraphics[width=0.48\linewidth]{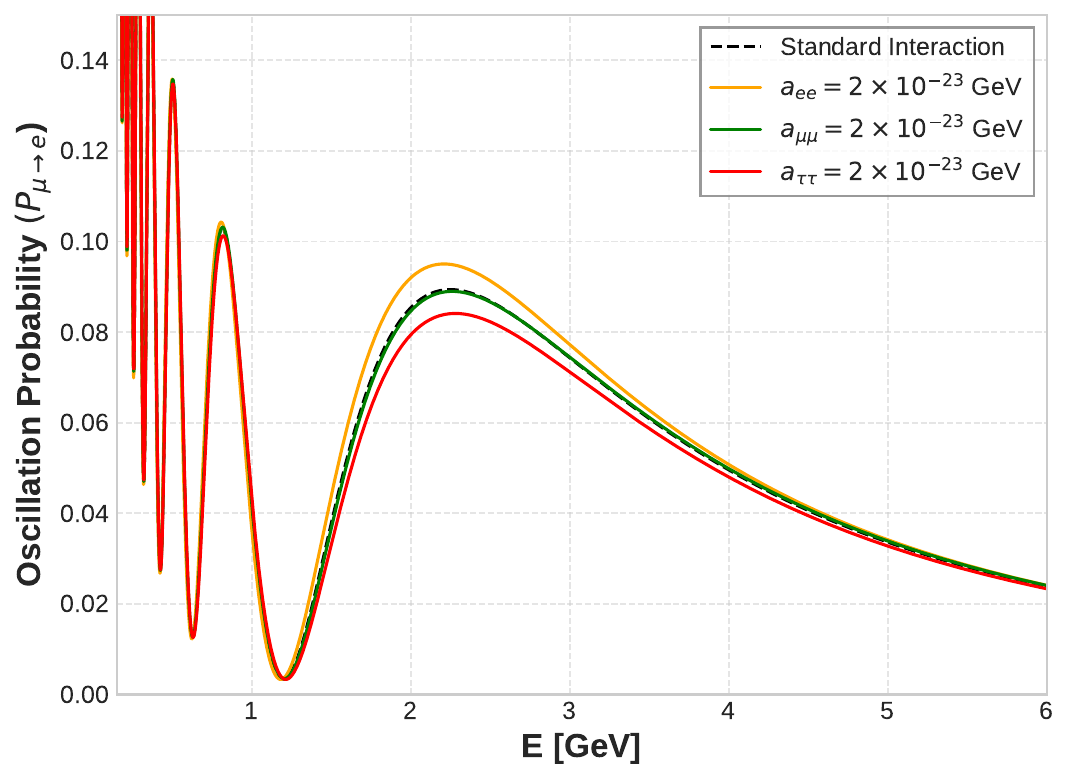}
    \includegraphics[width=0.48\linewidth]{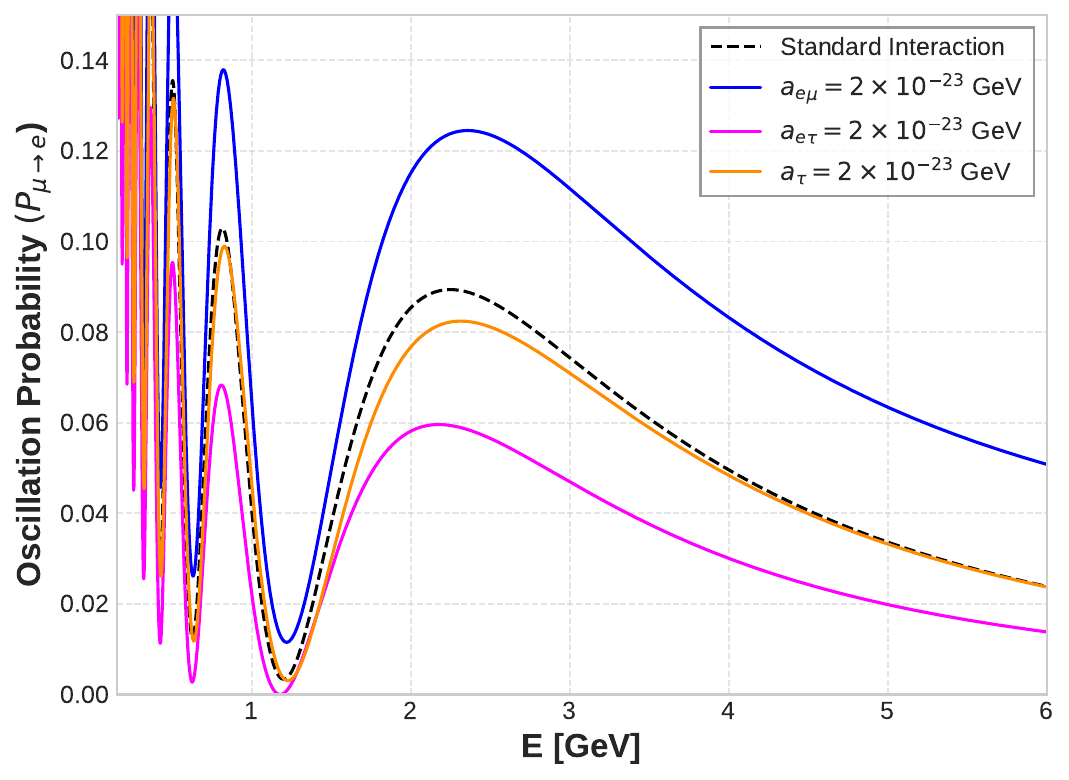}
    \includegraphics[width=0.48\linewidth]{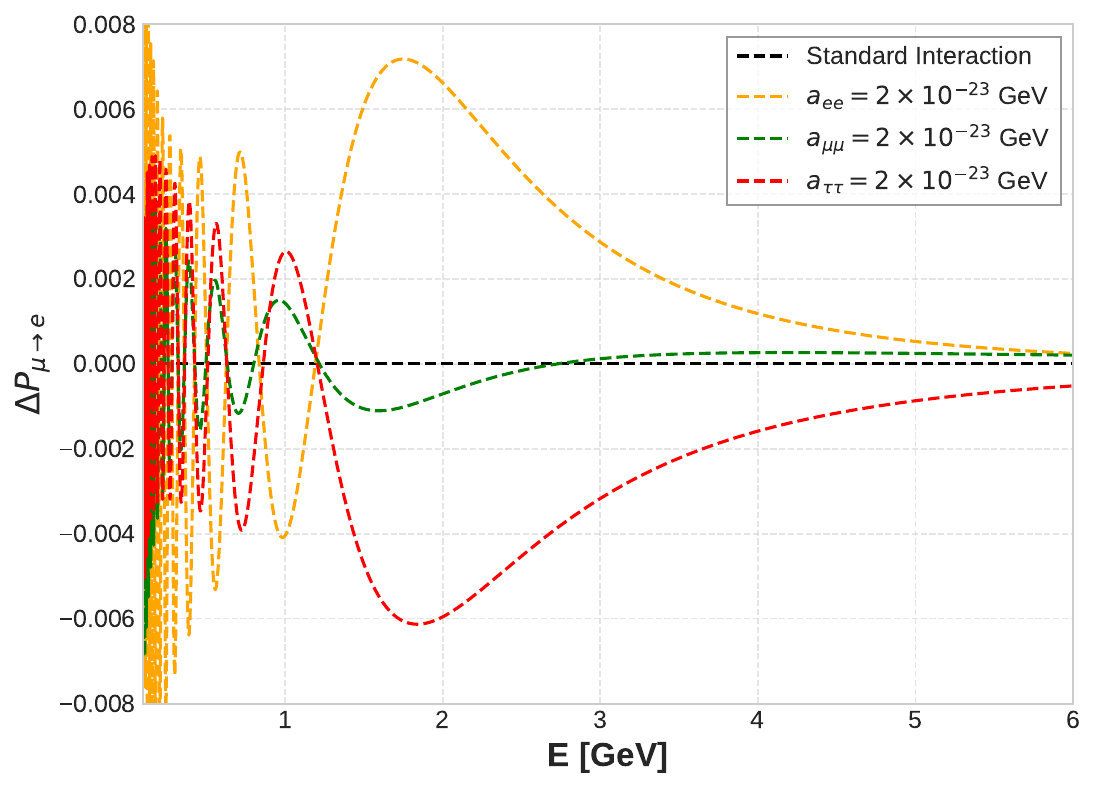}
    \includegraphics[width=0.48\linewidth]{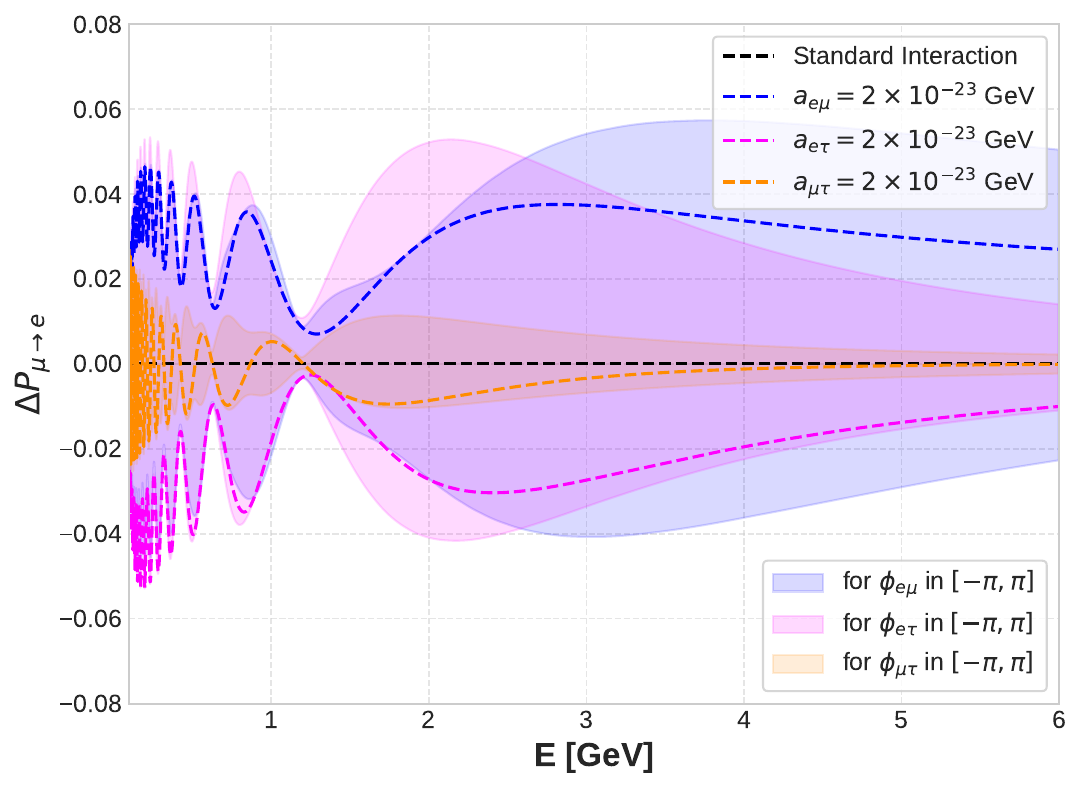}
    \caption{Appearance Probability in presence of LIV parameters, $P_{\mu \rightarrow e}$ and its deviation from Standard Interaction, $\Delta P_{\mu \rightarrow e} = P_{\mu \rightarrow e}({SI+LIV}) - P_{\mu \rightarrow e}(SI)$ as functions of L/E. Effects of individual diagonal LIV parameters ($a_{ee}, a_{\mu\mu}, a_{\tau\tau}$) and effects of off-diagonal LIV parameters ($a_{e\mu}, a_{e\tau}, a_{\mu\tau}$) on $P_{\mu \rightarrow e}$ are shown in the top-left panel and the top-right panel respectively. Bottom-left panel shows $\Delta P_{\mu \rightarrow e}$ in presence of diagonal LIV parameters ($a_{ee}, a_{\mu\mu}, a_{\tau\tau}$) while the bottom-right panel shows that for off-diagonal LIV parameters ($a_{e\mu}, a_{e\tau}, a_{\mu\tau}$), with shaded bands showing the impact of varying the LIV phase $\phi_{\alpha\beta} \in [-\pi, \pi]$. Benchmark LIV parameter magnitude $|a_{\alpha\beta}| = 2 \times 10^{-23}$ GeV for a 1300 km baseline is used for illustration. }
    \label{fig:Prob_App}
\end{figure}
Figure \ref{fig:Prob_App} illustrates these effects. The diagonal parameters $a_{ee}$, $a_{\mu\mu}$, and $a_{\tau\tau}$ show relatively small impacts on $P_{\mu e}$ compared to the off-diagonal ones, consistent with their sub-leading role in the approximate probability expression. The upper-left panel shows slight modifications near the oscillation peaks, while the lower-left panel confirms that the absolute difference $\Delta P_{\mu e}$ induced is typically an order of magnitude smaller than that from off-diagonal parameters. $a_{ee}$ and $a_{\tau\tau}$ produce noticeable deviations, whereas the effect of $a_{\mu\mu}$ is almost negligible in this channel for the benchmark LIV parameter magnitude $|a_{\alpha\beta}| = 2 \times 10^{-23}$ GeV used for illustration.

The parameters $a_{e\mu}$ and $a_{e\tau}$ significantly alter $P_{\mu e}$, as expected from their leading-order contributions. $a_{e\mu}$ tends to enhance the probability (for $\phi_{e\mu}=0$), while $a_{e\tau}$ tends to suppress it (for $\phi_{e\tau}=0$) around the first oscillation maximum (lower $E/L$), as seen in the upper-right panel. The parameter $a_{\mu\tau}$ has a considerably smaller effect on appearance probability. The lower-right panel demonstrates the crucial role of the LIV phases $\phi_{e\mu}$ and $\phi_{e\tau}$. Varying these phases (shaded bands) can drastically change the probability, leading to large enhancements or suppressions relative to the standard case. The impact of the phase $\phi_{\mu\tau}$ is much less pronounced.

\subsubsection{The $\nu_\mu \to \nu_\mu$ Disappearance Probability}

The $\nu_\mu \to \nu_\mu$ survival channel provides a complementary probe of the LIV parameter space. To leading order, this channel is not sensitive to the parameters that dominate the appearance channel ($a_{e\mu}, a_{e\tau}$). Instead, it is primarily affected by a different set of LIV parameters. The approximate probability is given by \cite{Majhi:2019tfi}:
$$
P_{\mu\mu}^{LIV} \simeq 1 - \sin^2(2\theta_{23})\sin^2\Delta - |a_{\mu\tau}|\cos\phi_{\mu\tau} \cdot f(\theta_{23}, \Delta) + (a_{\mu\mu}-a_{\tau\tau}) \cdot g(\theta_{23}, \Delta)
$$
where $f$ and $g$ are functions of the standard oscillation parameters. This expression shows that the survival probability constrains the parameters $|a_{\mu\tau}|$, its phase $\phi_{\mu\tau}$, and the difference between the diagonal elements $(a_{\mu\mu}-a_{\tau\tau})$ \cite{Majhi:2019tfi}.

\begin{figure}[H]
    \centering
    \includegraphics[width=0.48\linewidth]{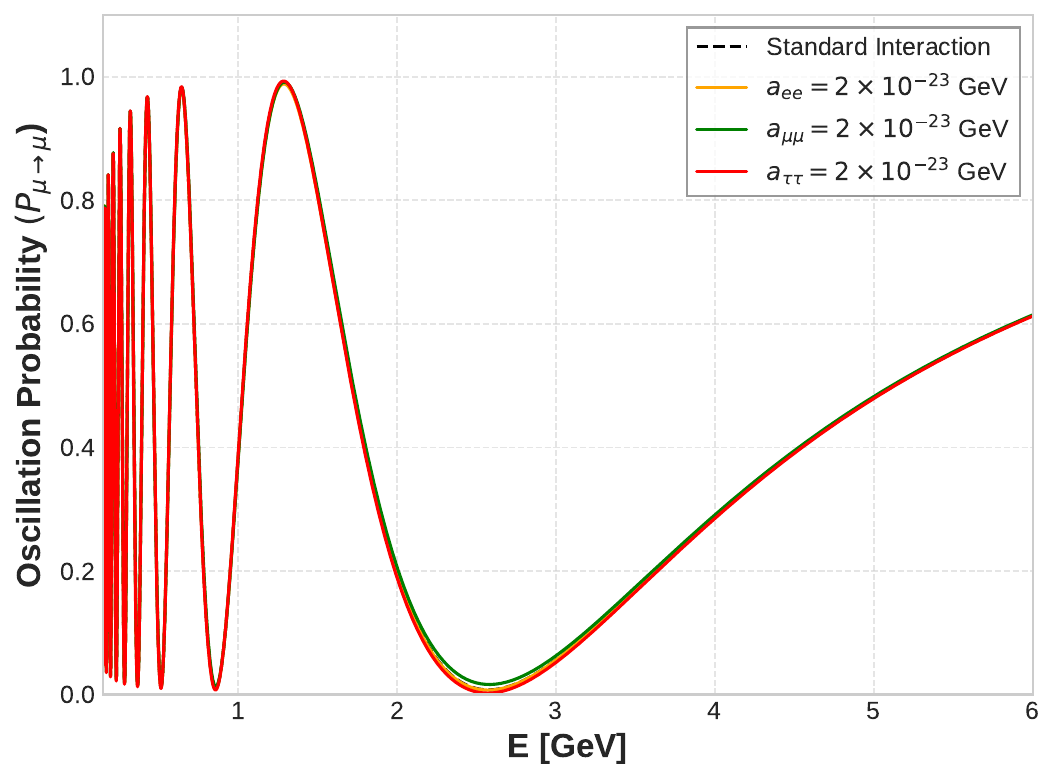}
    \includegraphics[width=0.48\linewidth]{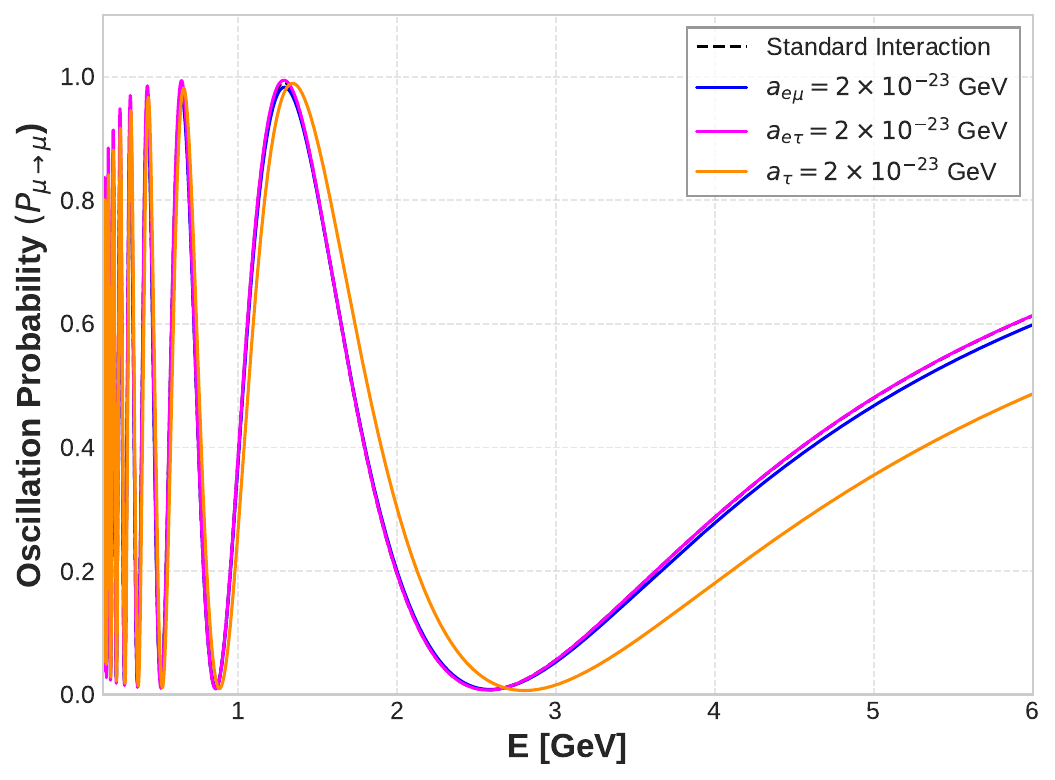}
    \includegraphics[width=0.48\linewidth]{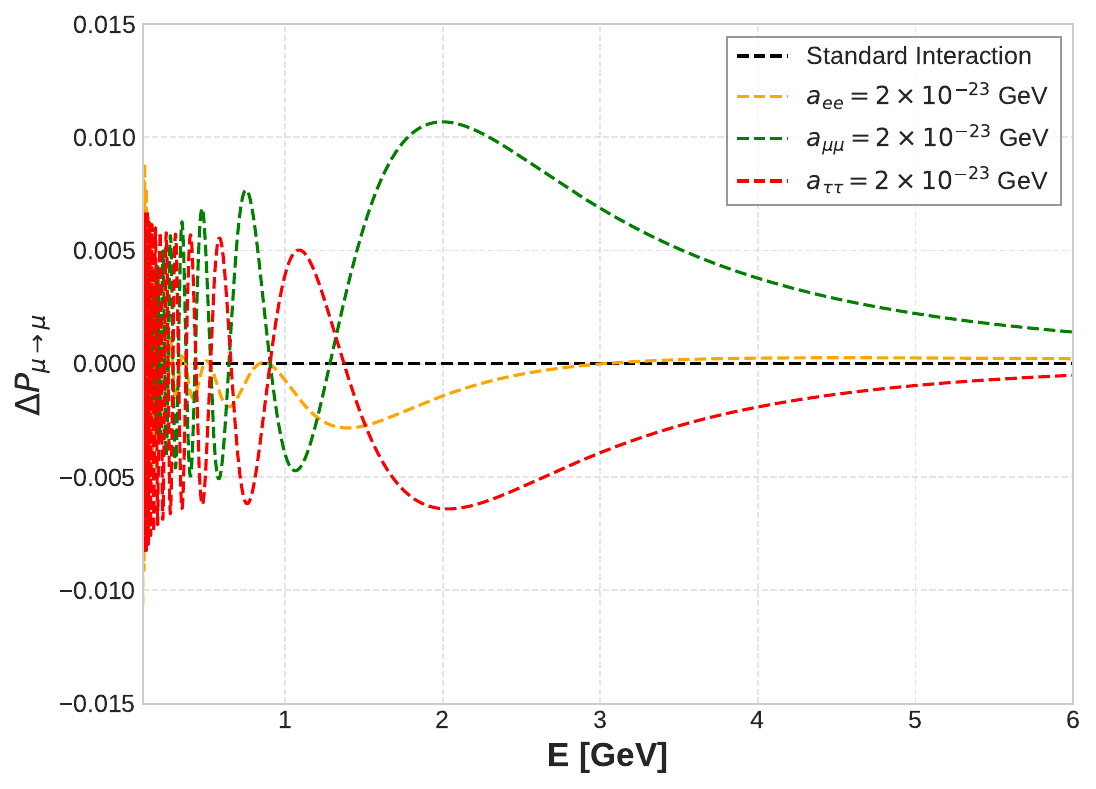}
    \includegraphics[width=0.48\linewidth]{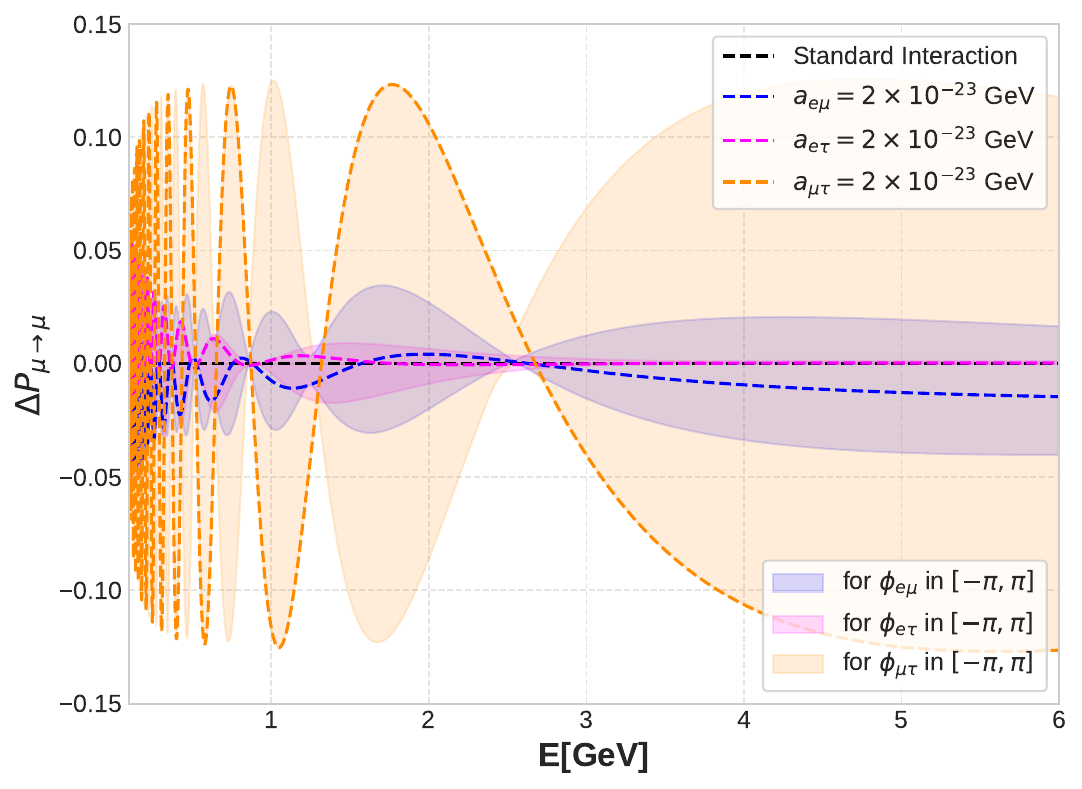}
    \caption{Disappearance Probability in presence of LIV parameters ($P_{\mu \rightarrow \mu}$) and its deviation from Standard Interaction, ($\Delta P_{\mu \rightarrow \mu} = P_{\mu \rightarrow \mu}({SI+LIV}) - P_{\mu \rightarrow \mu}(SI)$) as functions of L/E. Effects of individual diagonal LIV parameters ($a_{ee}, a_{\mu\mu}, a_{\tau\tau}$) and effects of off-diagonal LIV parameters ($a_{e\mu}, a_{e\tau}, a_{\mu\tau}$) on $P_{\mu \rightarrow \mu}$ are shown in the top-left panel and the top-right panel respectively. Bottom-left panel shows $\Delta P_{\mu \rightarrow \mu}$ in presence of diagonal LIV parameters ($a_{ee}, a_{\mu\mu}, a_{\tau\tau}$) while the bottom-right panel shows that for off-diagonal LIV parameters ($a_{e\mu}, a_{e\tau}, a_{\mu\tau}$), with shaded bands showing the impact of varying the LIV phase $\phi_{\alpha\beta} \in [-\pi, \pi]$. Benchmark LIV parameter magnitude $|a_{\alpha\beta}| = 2 \times 10^{-23}$ GeV for a 1300 km baseline is used for illustration. }
    \label{fig:Prob_Disapp}
\end{figure}

Figure \ref{fig:Prob_Disapp} illustrates these dependencies. Compared to the appearance channel, the diagonal parameters $a_{\mu\mu}$ and $a_{\tau\tau}$ show a more visible impact on $P_{\mu\mu}$, particularly away from the oscillation minima. Their effects are opposite in sign, consistent with the dependence on their difference $(a_{\mu\mu}-a_{\tau\tau})$. The effect of $a_{ee}$ remains negligible in this channel, as seen in the lower-left panel showing $\Delta P_{\mu\mu}$.

The parameter $a_{\mu\tau}$ induces the most significant modification to $P_{\mu\mu}$, especially at higher $E/L$ values (beyond the first minimum). The phase $\phi_{\mu\tau}$ (shaded band in lower-right panel) substantially modulates this effect. In contrast, $a_{e\mu}$ and $a_{e\tau}$ have very minor impacts on the survival probability, primarily visible as small deviations in the lower-right panel.

\subsubsection{Phenomenological Implications and Parameter Degeneracies}

The probability analysis highlights distinct phenomenological signatures and potential challenges. The parameters $a_{e\mu}$ and $a_{e\tau}$, along with their phases, are the dominant LIV contributions to the crucial $\nu_e$ appearance channel, while $a_{\mu\mu}, a_{\tau\tau}$ and $a_{\mu\tau}$ (with its phase) primarily affect the $\nu_\mu$ survival channel.

A key feature is the potential for cancellation effects in the appearance probability. The analytical expressions show that $P_{\mu e}(a_{e\mu})$ and $P_{\mu e}(a_{e\tau})$ contribute with opposite signs relative to certain standard terms and possess different dependencies on $\theta_{23}$. As observed in Figure \ref{fig:Prob_App}, $a_{e\mu}$ (for $\phi=0$) enhances $P_{\mu e}$ while $a_{e\tau}$ (for $\phi=0$) suppresses it near the first maximum. This structure allows scenarios where non-zero values of both parameters could yield a combined effect that closely mimics the Standard Interaction (SI) probability, especially when considering the freedom introduced by their phases $\phi_{e\mu}$ and $\phi_{e\tau}$. 

This leads directly to significant parameter degeneracies. The interference between the standard CP phase $\delta_{CP}$ and the LIV phases $\phi_{e\mu}, \phi_{e\tau}$ in the appearance channel means that different combinations of $(\delta_{CP}, \phi_{e\mu}, \phi_{e\tau}, |a_{e\mu}|, |a_{e\tau}|, \theta_{23})$ can produce very similar oscillation probabilities. Consequently, at the event level, a specific non-zero LIV scenario might perfectly mimic the event spectrum of the Standard Interaction (SI) but with different values of $\delta_{CP}$ or $\theta_{23}$. If such LIV effects exist but are not included in the analysis, this could lead to a biased, incorrect determination of $\delta_{CP}$ and $\theta_{23}$. To resolve these degeneracies it is necessary to combine information from multiple channels (appearance and disappearance) and, crucially, from experiments with different characteristics (like baseline and energy spectrum), as we shall explore later in this work.

\subsection{Effects of LIV parameters on Neutrino Events}\label{sec:events}
We first simulated the effects of LIV parameters on the appearance and disappearance channel events for the 360 km and 540 km baselines of ESSnuSB, and the 1300 km baseline of DUNE at the event level, using the runtimes specified in Table \ref{tab:uncertainity}. This simulation aimed to assess the resolution capabilities of each experimental setup at its respective operating energies. Next, we investigated whether specific non-zero LIV scenarios could mimic the Standard Interaction (SI) event spectrum at the event level, but with different values of $\delta_{CP}$ or $\theta_{23}$, thereby identifying potential degeneracies. 

\subsubsection{Events in the $\nu_\mu \to \nu_e$ Appearance Channel}

\begin{figure}[H]
    \centering
    \includegraphics[width=0.45\linewidth]{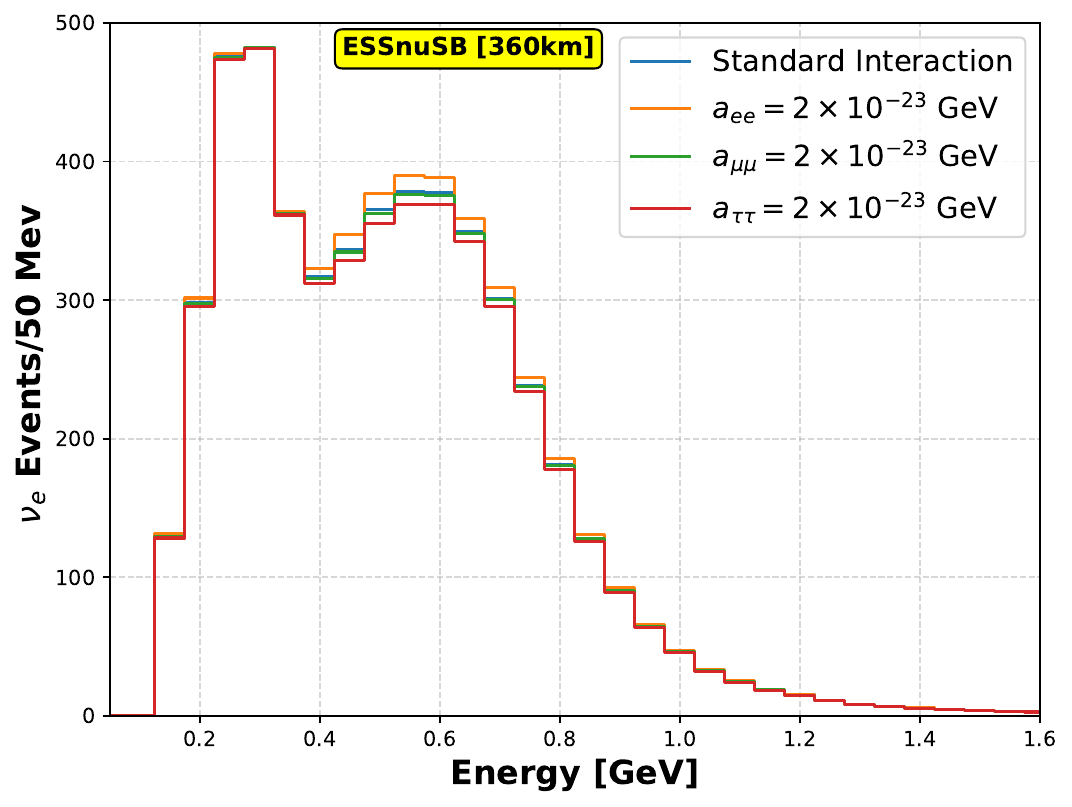}
    \includegraphics[width=0.45\linewidth]{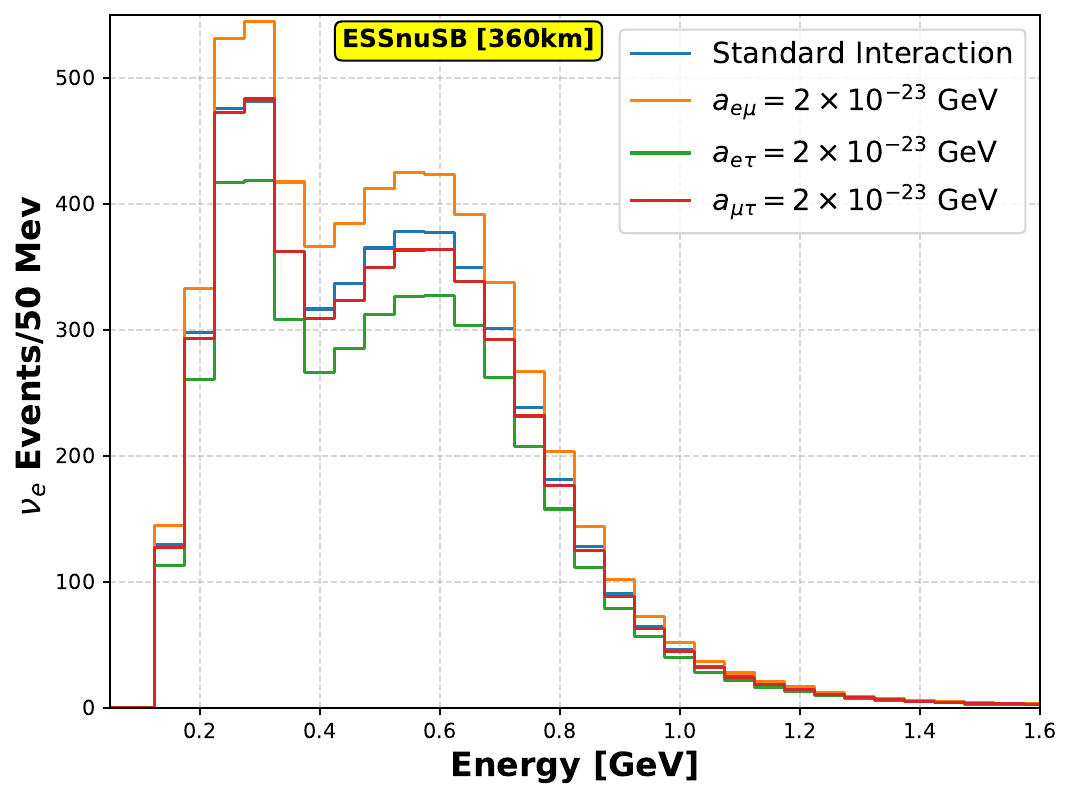}
    \includegraphics[width=0.45\linewidth]{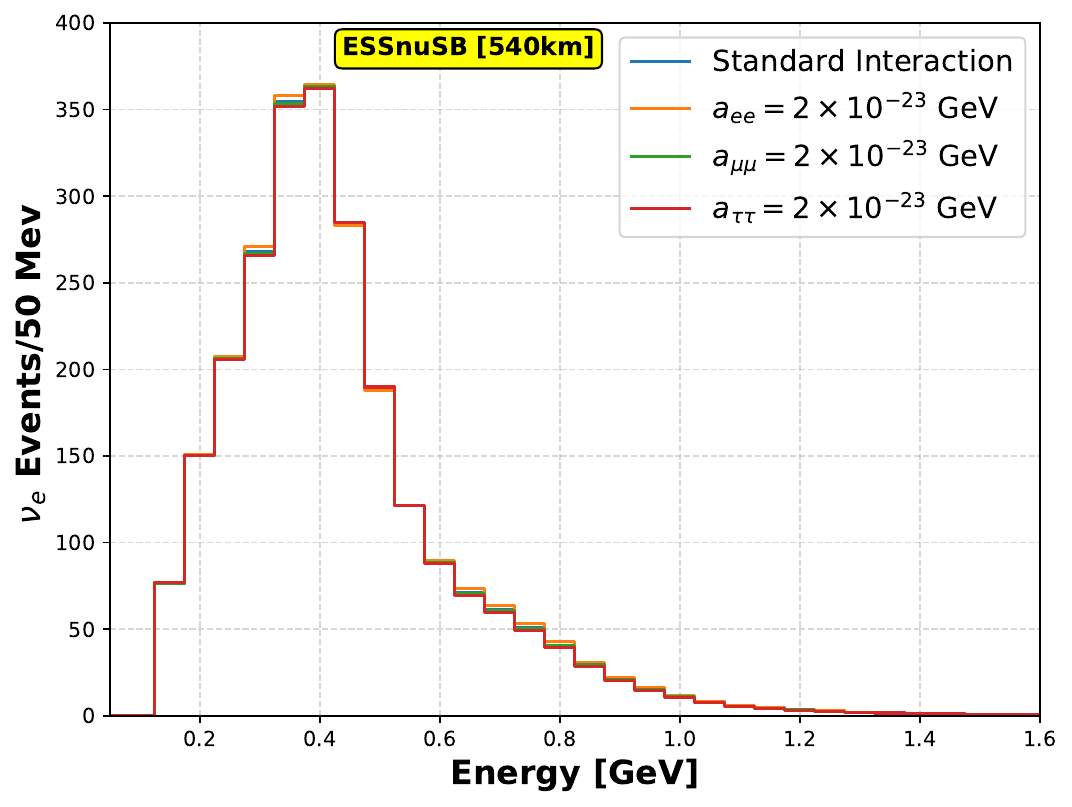}
    \includegraphics[width=0.45\linewidth]{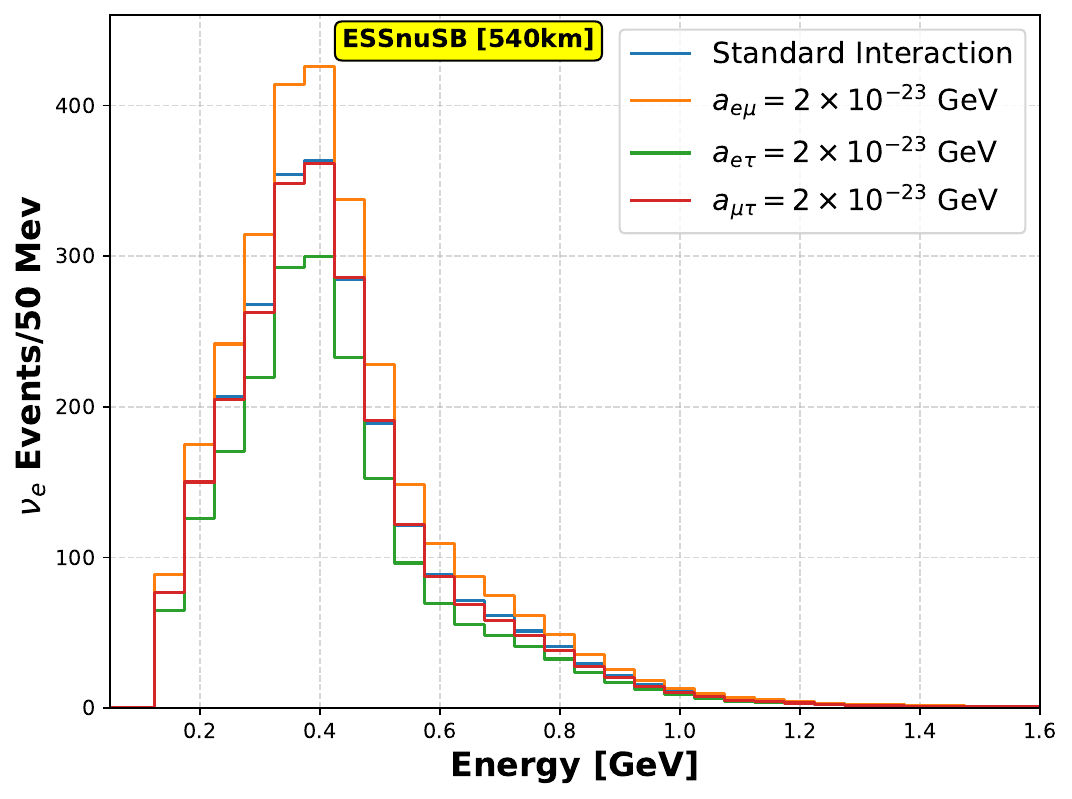}
    \includegraphics[width=0.45\linewidth]{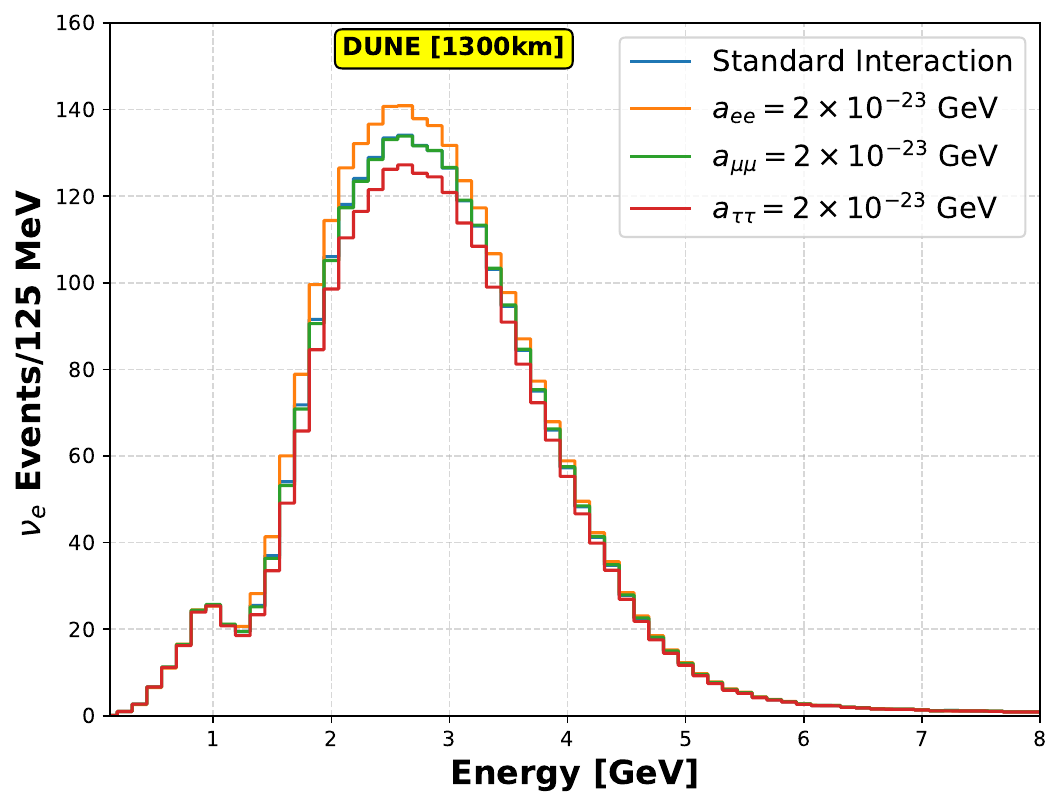}
    \includegraphics[width=0.45\linewidth]{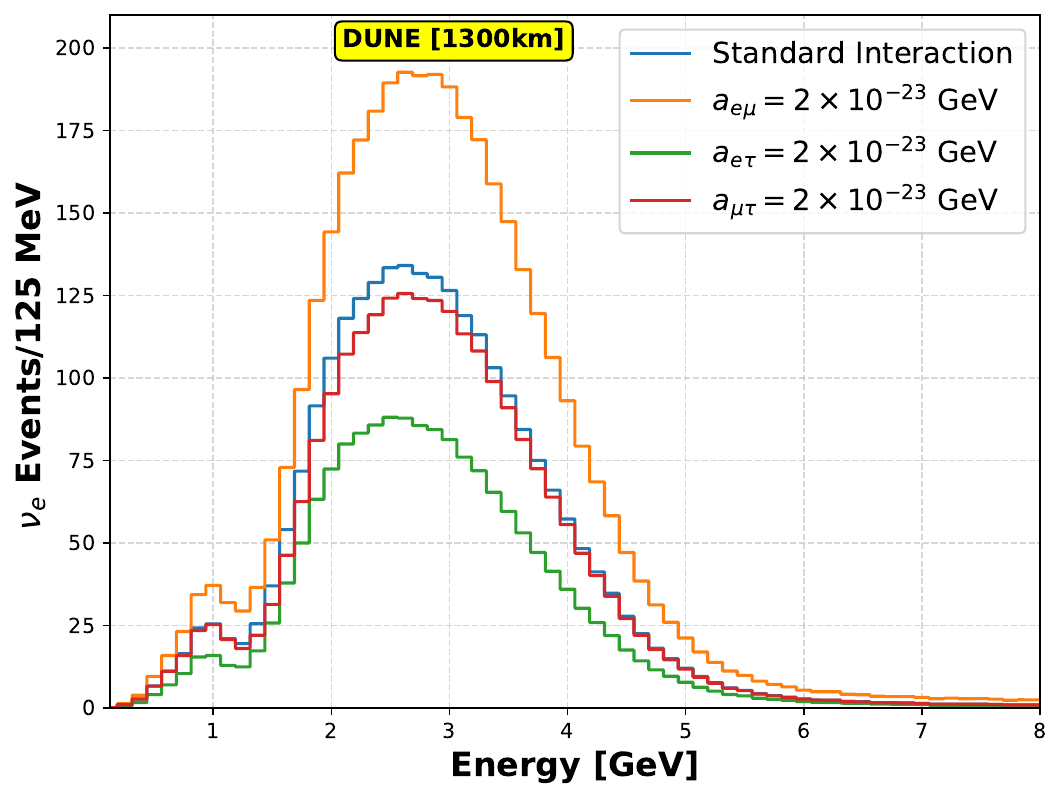}
    \caption{Neutrino appearance ($\nu_e$) events in ESSnuSB (top [L=360km] and middle [L=540km] panel) and DUNE (Bottom panel) in presence of diagonal (left panel) and off-diagonal (right panel) LIV parameters. }
    \label{fig:event_nuapp}
\end{figure}

Figure \ref{fig:event_nuapp} reveals that the event spectrum for the ESSnuSB 360 km baseline (top panels) clearly exhibits a two-peak structure. This is a key feature of operating at the second oscillation maximum, where the first maximum appears as a smaller shoulder (around 0.3 GeV) and the second maximum forms the dominant, higher-energy peak (around 0.6 GeV). In contrast, the ESSnuSB 540 km baseline (middle panels) displays a single, broader peak. At this longer baseline, the first maximum is suppressed, and the second maximum dominates the event spectrum. The DUNE 1300 km baseline (bottom panels) shows an entirely different profile, with its broadband spectrum centered on the first oscillation maximum at a much higher energy ($\sim 2.5$ GeV). Crucially, these event-level plots confirm that the sensitivities observed at the probability level (Section \ref{sec:probabilities}) are directly mirrored in the observable event rates. 

The left panels of Figure \ref{fig:event_nuapp} show that the diagonal LIV parameters ($a_{ee}$, $a_{\mu\mu}$, $a_{\tau\tau}$) cause only minimal deviations from the Standard Interaction (SI) event spectrum (black line). This reflects their sub-leading role in the appearance probability. The right panels demonstrate that the off-diagonal parameters $a_{e\mu}$ and $a_{e\tau}$ induce significant energy-dependent changes to the event spectra, just as they dominated the probability modifications. Specifically, $a_{e\mu}$ (orange line) clearly enhances the event count across the peaks, while $a_{e\tau}$ (green line) causes a notable suppression in all three baseline configurations. The effect of $a_{\mu\tau}$ (red line) is much less pronounced, again matching the probability-level findings. This confirms that the phenomenological impact of each LIV parameter observed in the probability analysis is directly reflected in the predicted event-level sensitivities. The interference between the standard parameters ($\delta_{CP}$, $\theta_{23}$) and the LIV parameters ($a_{e\mu}$, $a_{e\tau}$ and their phases $\phi_{e\mu}$, $\phi_{e\tau}$) at the probability level will therefore manifest as a potential degeneracy at the event level, where a specific non-zero LIV scenario might mimic the event spectrum of the Standard Interaction (SI) but with different values of $\delta_{CP}$ or $\theta_{23}$.

\subsubsection{Events in the $\nu_\mu \to \nu_\mu$ Disappearance Channel}

\begin{figure}[H]
    \centering
    \includegraphics[width=0.45\linewidth]{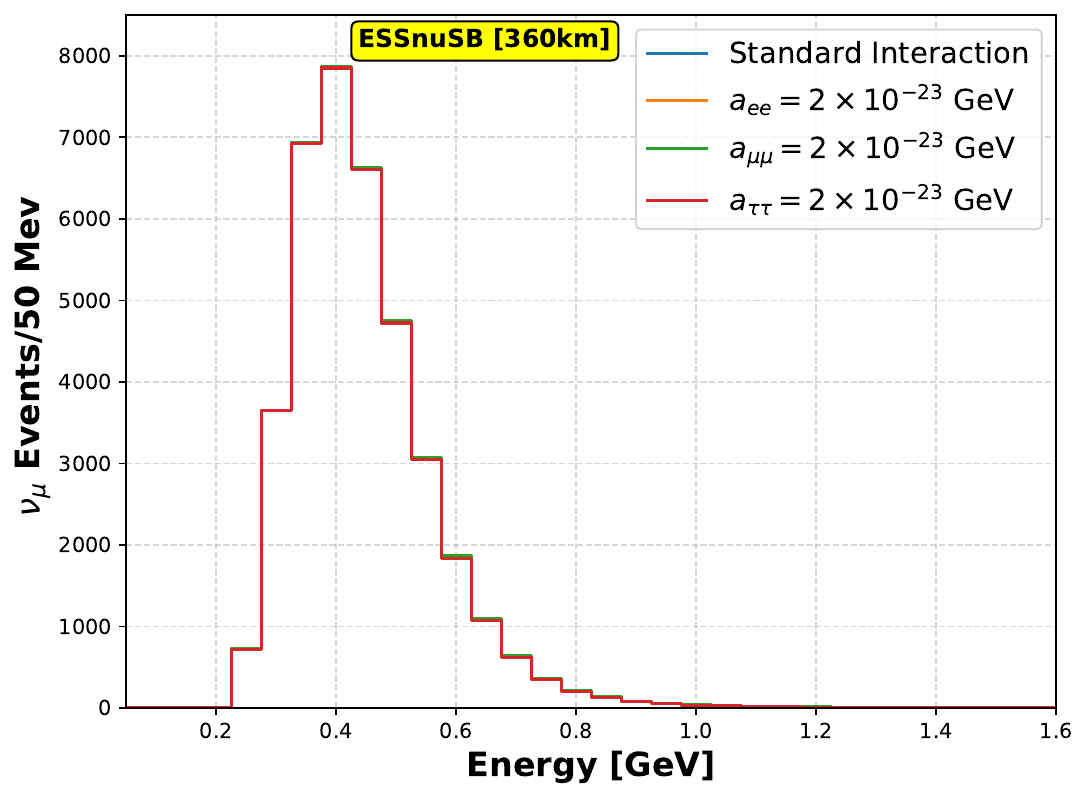}
    \includegraphics[width=0.45\linewidth]{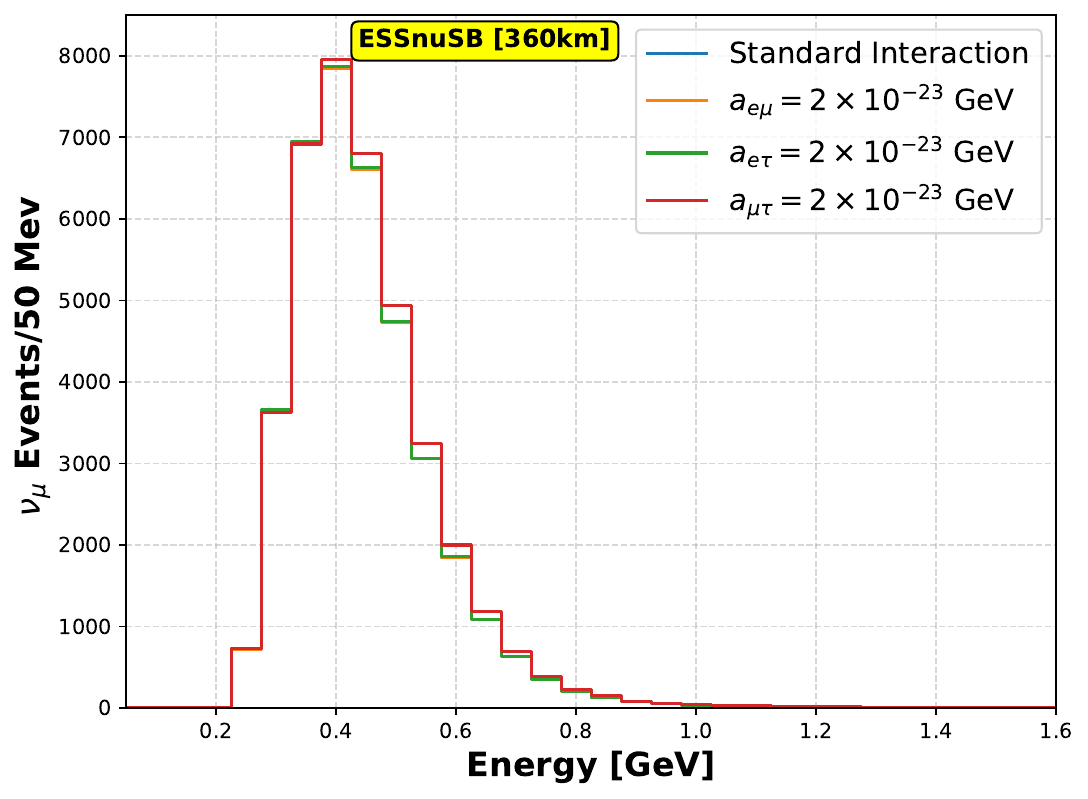}
    \includegraphics[width=0.45\linewidth]{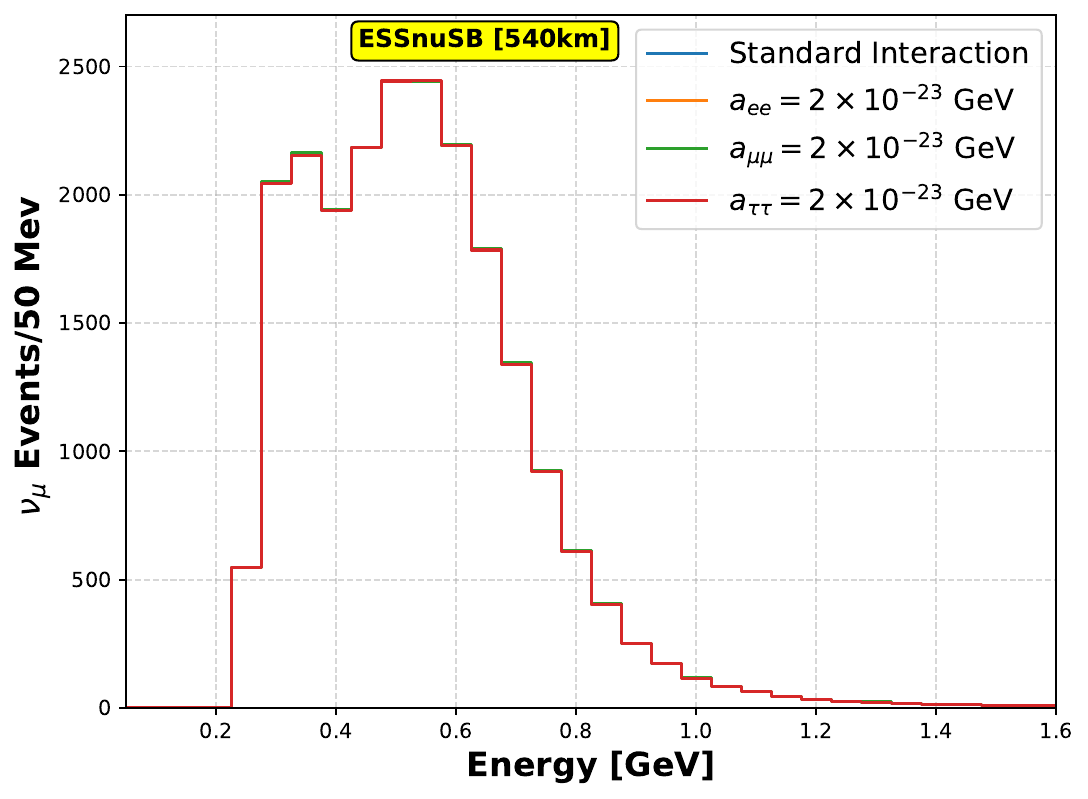}
    \includegraphics[width=0.45\linewidth]{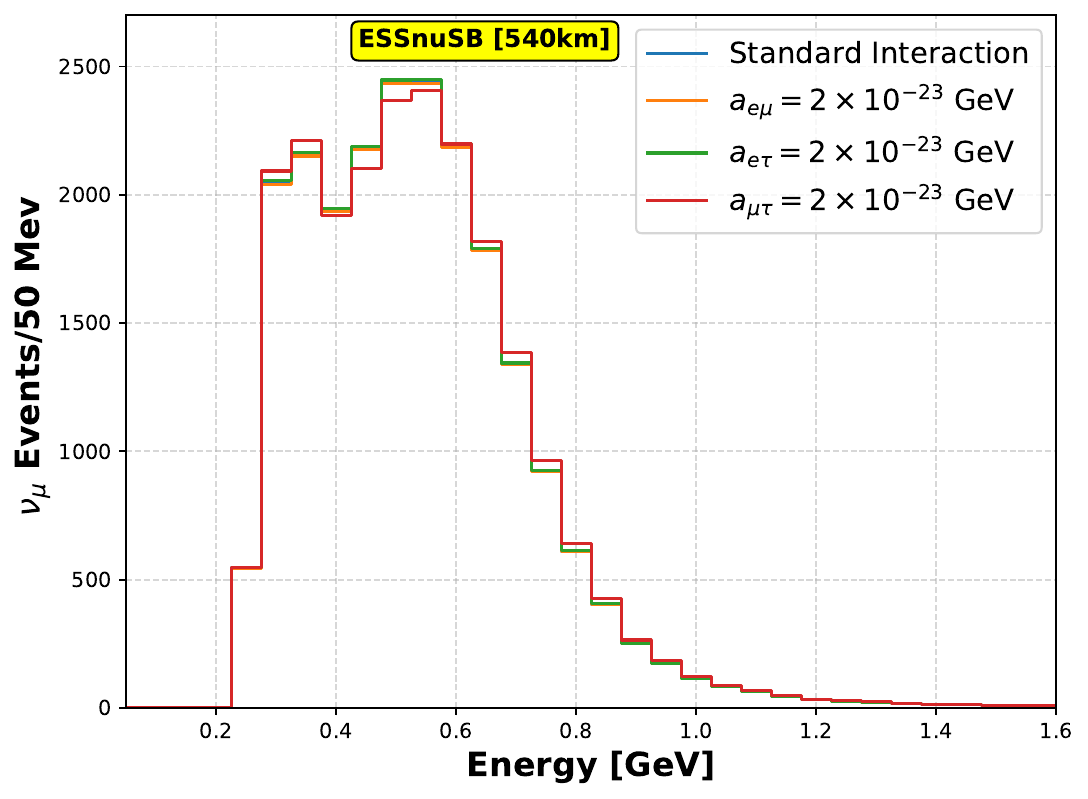}
    \includegraphics[width=0.45\linewidth]{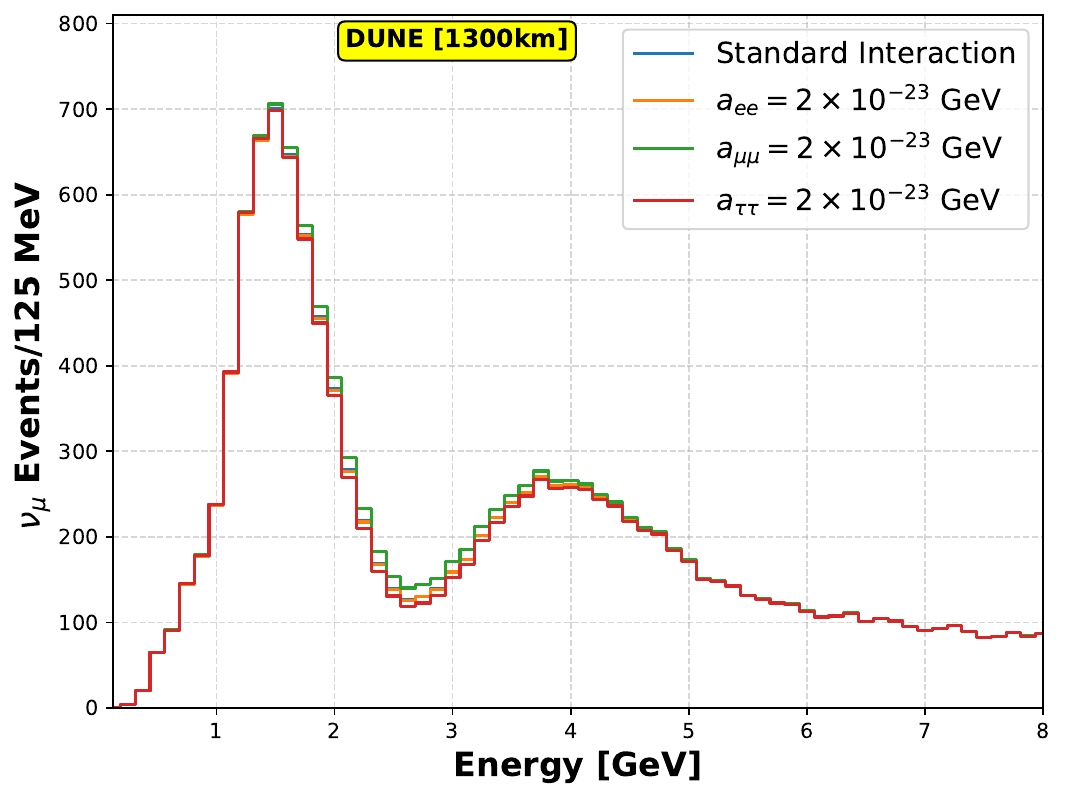}
    \includegraphics[width=0.45\linewidth]{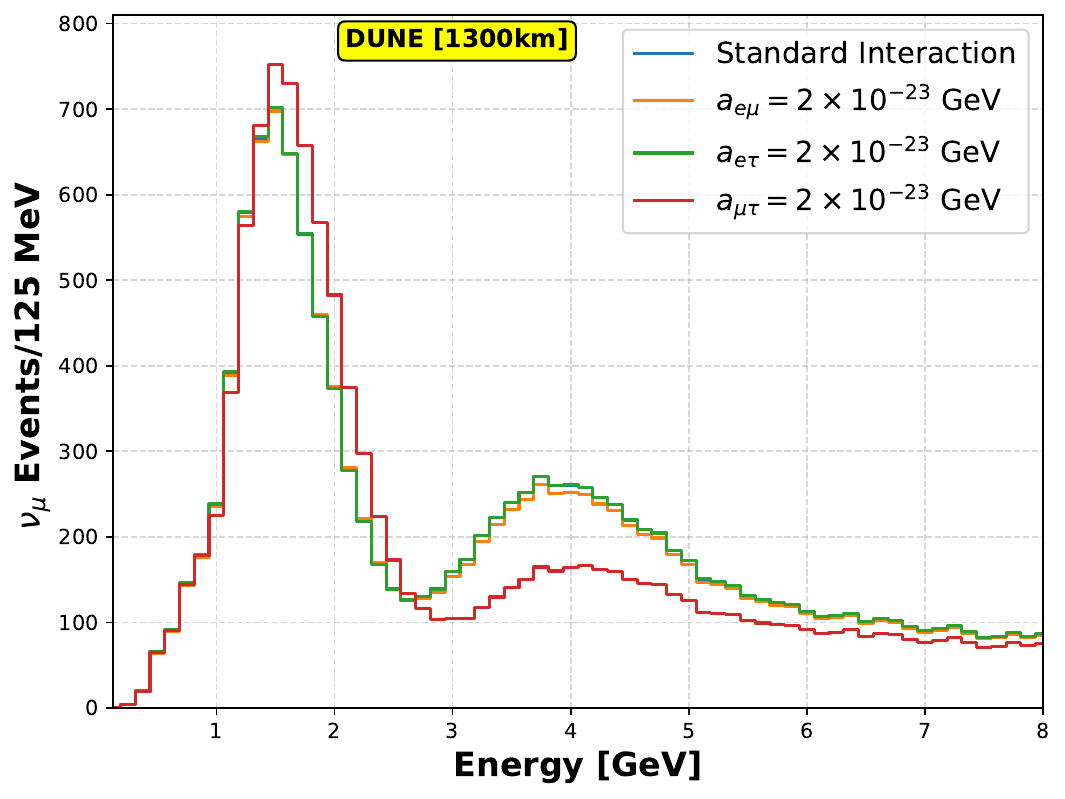}
    \caption{Neutrino disappearance ($\nu_e$) events in ESSnuSB (top [L=360km] and middle [L=540km] panel) and DUNE (Bottom panel) in presence of diagonal (left panel) and off-diagonal (right panel) LIV parameters. }
    \label{fig:event_nudisapp}
\end{figure}

The event spectra for the $\nu_{\mu}\rightarrow\nu_{\mu}$ disappearance channel are presented in Figure \ref{fig:event_nudisapp}. This channel provides a complementary probe to the appearance channel, exhibiting sensitivity to a distinct set of LIV parameters. Remarkably, the event spectrum for the ESSnuSB 540 km baseline (middle panels) exhibits a two-peak structure as opposed to the single-peak structure in the appearance channel. However, the 360 km baseline shows a single-peak structure in this channel, in contrast to the double-peak structure seen previously in the appearance channel. The DUNE 1300 km baseline (bottom panels) also shows two distinct peaks ($\sim 1.5$ GeV and $\sim 4$ GeV) and this characteristic dip ($\sim 2.5$ GeV) in this channel.

The left panels of Figure \ref{fig:event_nudisapp} demonstrate the impact of the diagonal LIV coefficients. For DUNE (bottom-left panel), the deviation is particularly visible around the oscillation dip ($\sim 2.5$ GeV), where the LIV terms shift the depth and position of the minimum. The effects of diagonal parameters are negligible, resulting in an event spectrum that is indistinguishable from the Standard Interaction (SI) case. The right panels of Figure 4 highlight the impact of the off-diagonal parameters. The event spectra for $a_{e\mu}$ (orange line) and $a_{e\tau}$ (green line) which induced large deviations in the $\nu_e$ appearance channel are essentially identical to the standard case here. In contrast, the parameter $a_{\mu\tau}$ (red line) induces a significant spectral distortion. For ESSnuSB 540km baseline (middle right panel), this manifests as a suppression of the event rate across the peak energies while for ESSnuSB 360 km baseline it manifests as an enhancement of the event rates. In the 1300 km DUNE baseline it has both suppression and enhancement effects to the event counts depending on the energy.

\subsubsection{Degeneracies in the $\nu_\mu \to \nu_e$ Appearance Channel}

To visually investigate the parameter degeneracies identified at the probability level, we next analyze the correlations at the event level. The heatmaps show the difference in the number of $\nu_{\mu}\rightarrow\nu_{e}$ appearance events between a scenario with LIV and the Standard Interaction (SI) case, calculated as $\Delta N_{events}$ =  $N_{events}(SI+LIV) - N_{events}(SI)$. This event difference is plotted across a two-dimensional plane, with a diagonal LIV parameter ($a_{\alpha\beta}$) on the x-axis and the atmospheric mixing angle $\theta_{23}$ or CP violating phase $\delta_{CP}$ on the y-axis. In these plots, red regions indicate an excess of events (more events than SI), while blue regions represent a suppression of event counts. The contour lines near the transition from red to blue (corresponding to the gray dashed line) mark the critical degeneracy region where the event difference is zero. Any point on this line represents a combination of a non-zero LIV parameter and a $\theta_{23}$ (or $\delta_{CP}$) value that would produce the same number of events as the Standard Interaction, making them indistinguishable in a simple event-counting analysis. The visualization is shown for the different peaks in the baselines (two peaks for ESSnuSB 360km, one peak for ESSnuSB 540km, and one peak for DUNE 1300km) at different energies (labeled E1, E2, E3 and E4) to show how these degeneracy patterns shift, illustrating the challenge they pose to individual measurements. Here, E1, E2, E3 and E4 correspond to energy values of 0.3 GeV, 0.6 GeV, 0.4 GeV and 2.5 GeV, respectively. All relevant plots in the $\nu_\mu \to \nu_e$ appearance channel appear in Appendix \ref{Appendix_A}. We have presented the parameter degeneracies only for the diagonal parameter $a_{\mu\mu}$ here for illustration.

\begin{figure}[H]
  \centering
  \begin{subfigure}[b]{0.47\textwidth}
    \centering
    \includegraphics[width=\textwidth]{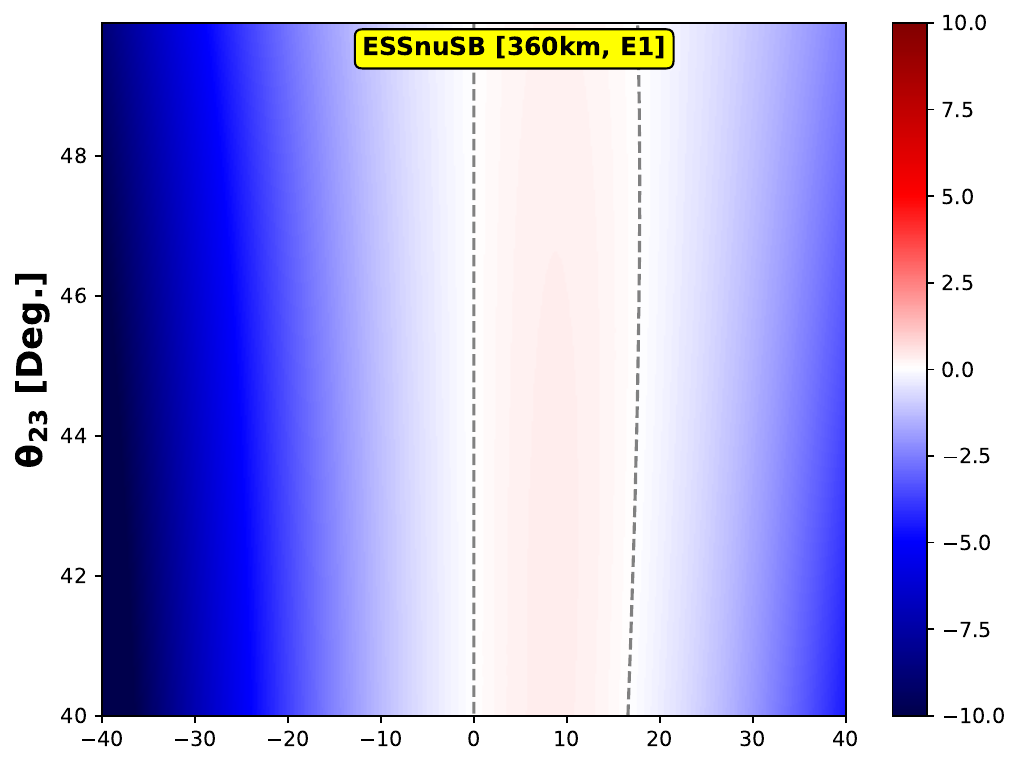}
    \label{fig:subA}
  \end{subfigure}
  \begin{subfigure}[b]{0.48\textwidth}
    \centering
    \includegraphics[width=\textwidth]{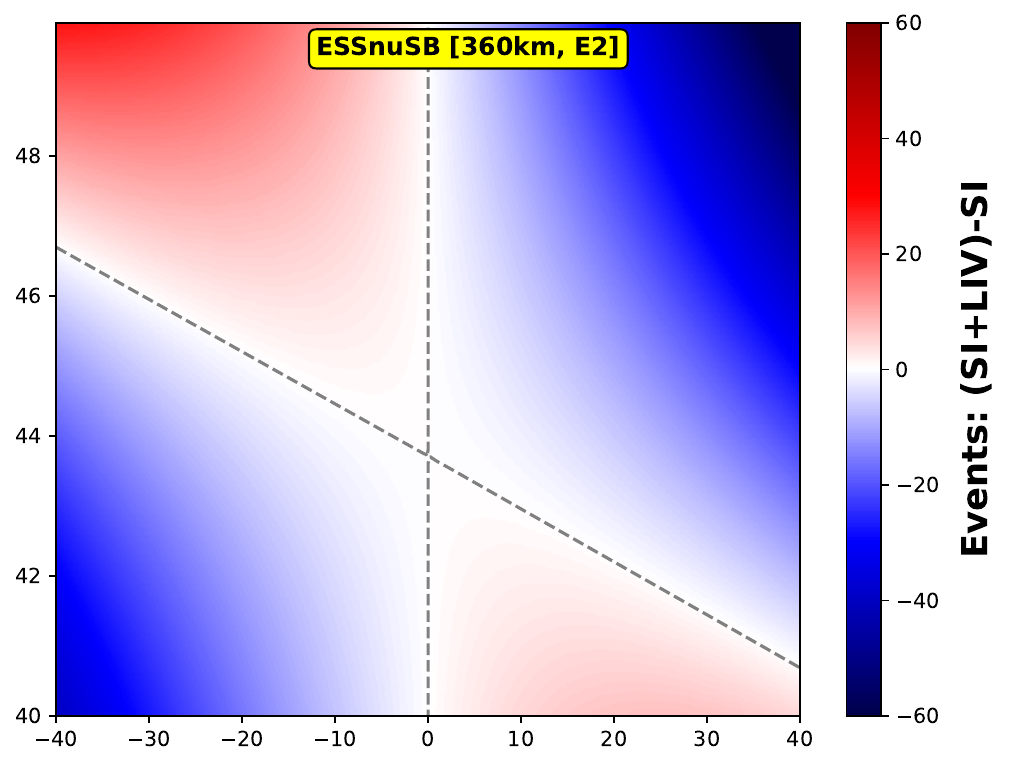}
    \label{fig:subB}
  \end{subfigure}
  \\[-4ex]  

  \begin{subfigure}[b]{0.47\textwidth}
    \centering
    \includegraphics[width=\textwidth]{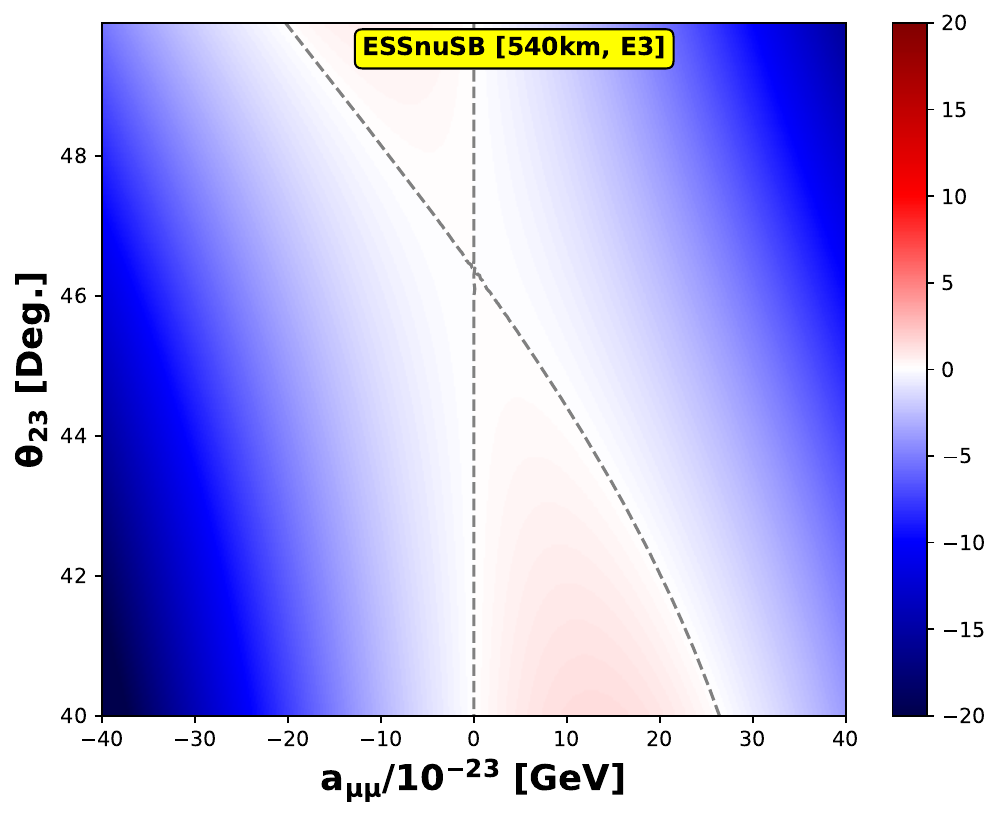}
    \label{fig:subC}
  \end{subfigure}
  \begin{subfigure}[b]{0.48\textwidth}
    \centering
    \includegraphics[width=\textwidth]{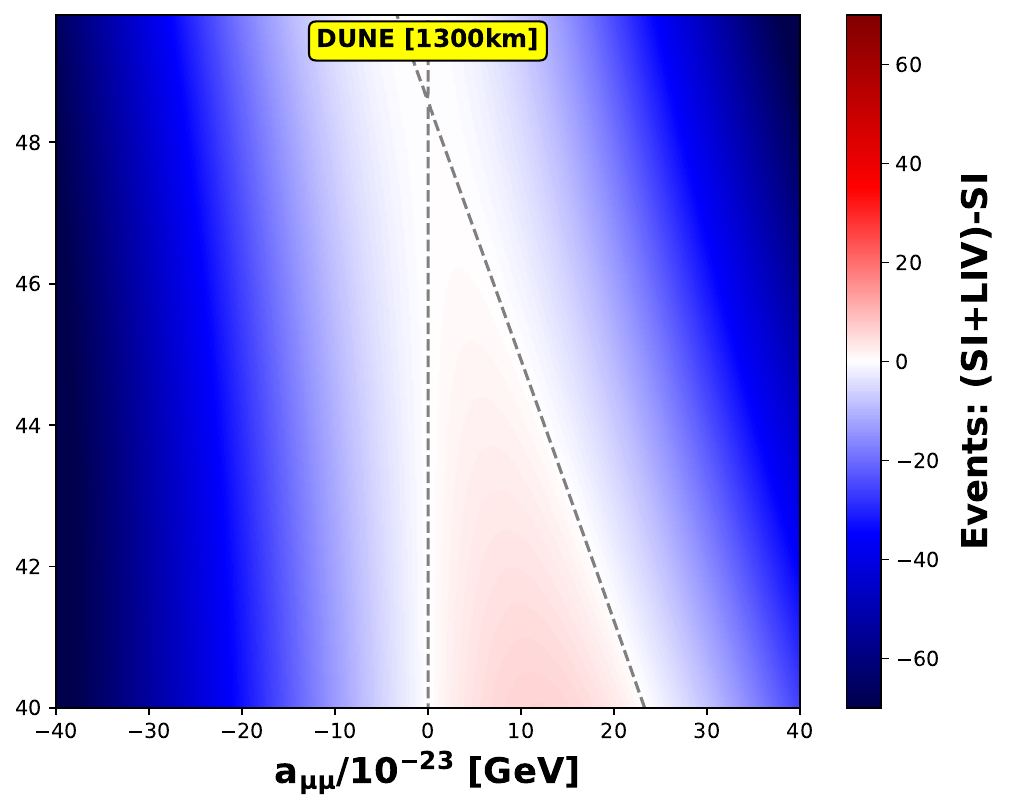}
    \label{fig:subD}
  \end{subfigure}
   \caption{The heatmap of event differences ($\Delta N_{events}$ =  $N_{events}(SI+LIV) - N_{events}(SI)$) in $a_{\alpha\beta}-\theta_{23}$ plane for the diagonal LIV parameter $a_{\mu\mu}$ at energies E1, E2, E3 and E4. The grey dashed line mark the critical degeneracy region where the event difference is zero }
   \label{fig:example_eventdiff_th23}
\end{figure}

A crucial observation across the panels in Figure \ref{fig:example_eventdiff_th23} is the difference in degeneracy patterns between different peaks in the ESSnuSB and DUNE baselines for the $a_{\mu\mu}$ parameter. Clear $\theta_{23}$–$a_{\alpha\beta}$ degeneracies appear in all panels highlighting the interplay between standard interactions and LIV effects. For the ESSnuSB 360 km (at E2) and 540 km (at E3) baselines, as well as the DUNE 1300 km baseline (at E4), the degeneracy contours exhibit a strong diagonal slope which indicates a strong negative correlation between $a_{\mu\mu}$ and $\theta_{23}$ in these regions. A positive shift in $a_{\mu\mu}$ can be masked by shifting $\theta_{23}$ to a lower value, and vice-versa. The red (excess events) and blue (suppressed events) regions are separated by this diagonal, showing that the LIV effect significantly distorts the event rate in a way that looks like a change in the mixing angle. The ESSnuSB heatmap (360 km) at E1 shows a distinctly different pattern. The degeneracy contour is much steeper (nearly vertical) compared to the other plots. This implies that at DUNE energies and baselines, the effect of $a_{\mu\mu}$ is less easily mimicked by simply shifting $\theta_{23}$ compared to ESSnuSB. The parameter $a_{\mu\mu}$ drives the event rate difference more independently of the mixing angle here in this channel.

Figure \ref{fig:dcp_2d_1} in the Appendix \ref{Appendix_A} shows the heatmaps for all diagonal parameters ($a_{ee}$ , $a_{\mu\mu}$ and $a_{\tau \tau}$) while Figure \ref{fig:dcp_2d_2} shows the heatmaps for the off-diagonal parameters ($a_{e\mu}$ $a_{e\tau}$ and $a_{\mu \tau}$) considering the associated LIV phase angle $\phi_{\alpha\beta}$ to be 0. Upon comparing the heatmaps, we see that, for $a_{\tau\tau}$ (right column), the event difference is nearly symmetric with respect to $a_{\tau\tau}=0$, showing broad regions of degeneracy that shift only slightly with $\theta_{23}$. $a_{ee}$ shows similar symmetric behaviour at E2 and E4 but nearly parallel vertical degenerate lines at E1 and E3.  In contrast, $a_{\mu\mu}$ (middle column) produce stronger, asymmetric changes and the degeneracy contours (gray dashed lines) for $a_{\mu\mu}$ exhibit a significant slope, indicating that a shift in the LIV parameter can be compensated by a shift in $\theta_{23}$ away from maximal mixing as discussed previously. 

The off-diagonal LIV parameters produce much stronger and more uniform changes in event rates across the $\theta_{23}$–$a_{\alpha\beta}$ plane, with the sign of $\Delta N_{events}$ mainly dictated by the LIV coefficient rather than $\theta_{23}$. The patterns remain consistent across energies and baselines, showing almost monotonic behaviour i.e. the E1, E2, E3 and E4 panels for both ESSnuSB and DUNE display similar colour gradients, indicating that off-diagonal terms affect the event rates more uniformly than the diagonal ones. The DUNE configuration shows the largest overall magnitude of $\Delta N_{events}$ among all setups, demonstrating that long baseline and higher energies enhance the sensitivity to off-diagonal LIV parameters even more prominently than for the diagonal case. Although no values of $\theta_{23}$ give rise to SI mimicking effects, as is evident from the absence of a grey dashed line in the heatmaps, yet varying  $\phi_{\alpha\beta}$ can still give rise to SI mimicking effects. 

\begin{figure}[h]
  \centering
  \begin{subfigure}[b]{0.47\textwidth}
    \centering
    \includegraphics[width=\textwidth]{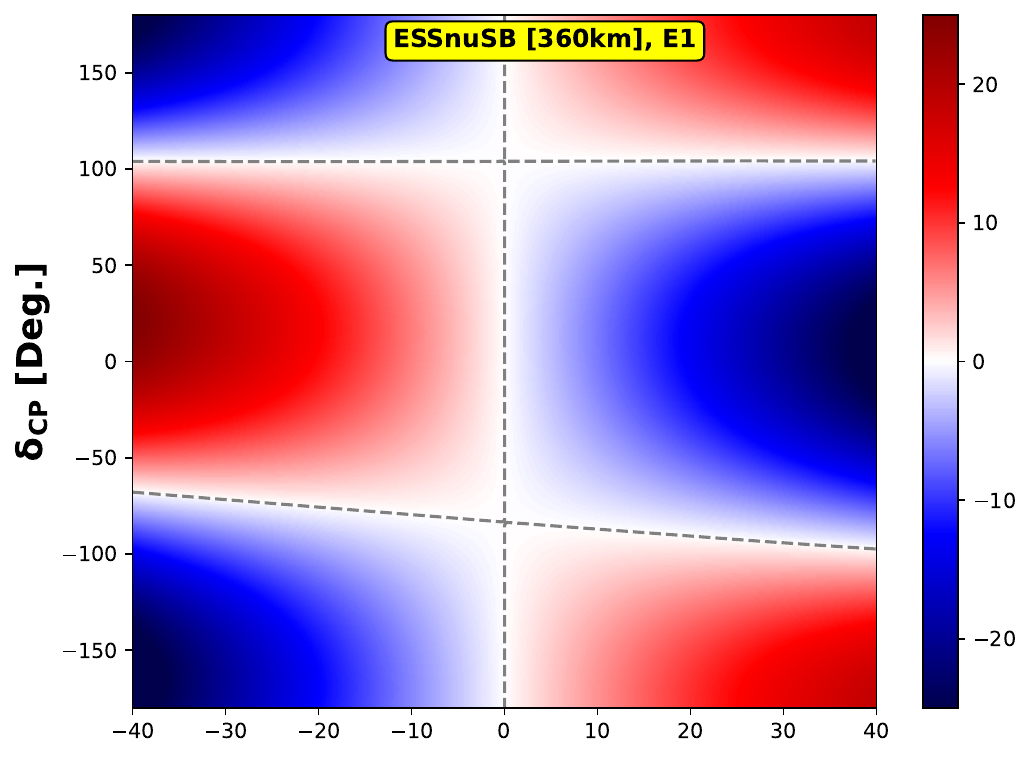}
    \label{fig:subA}
  \end{subfigure}
  \begin{subfigure}[b]{0.48\textwidth}
    \centering
    \includegraphics[width=\textwidth]{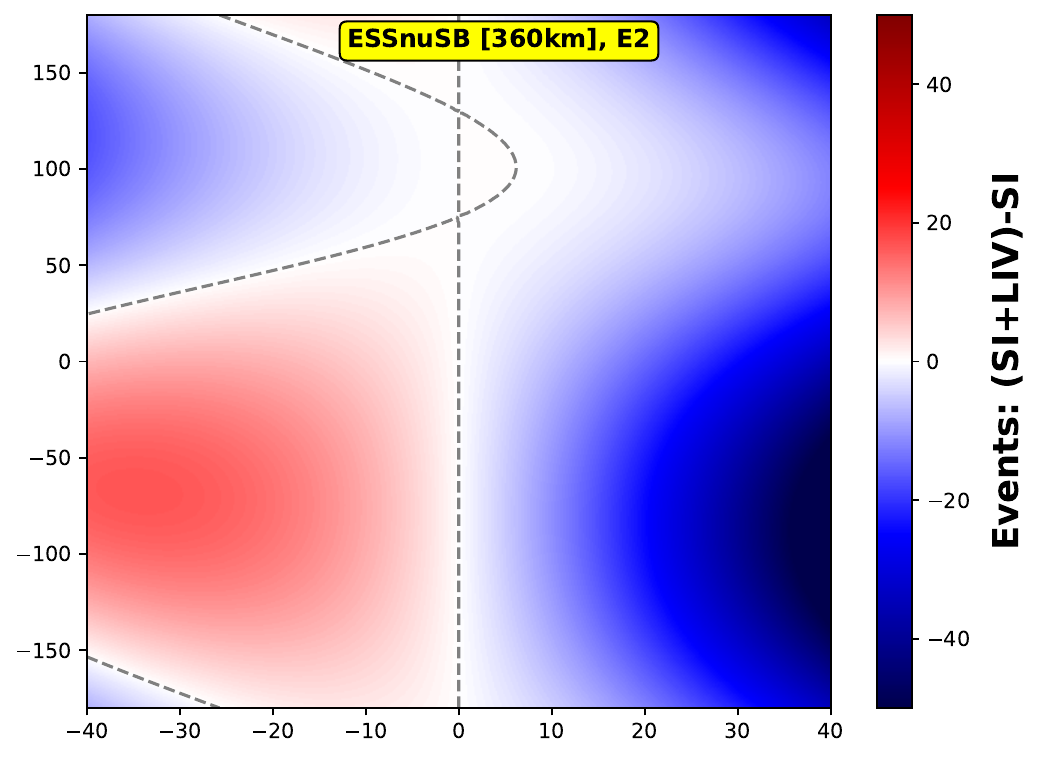}
    \label{fig:subB}
  \end{subfigure}
  \\[-4ex]  

  \begin{subfigure}[b]{0.47\textwidth}
    \centering
    \includegraphics[width=\textwidth]{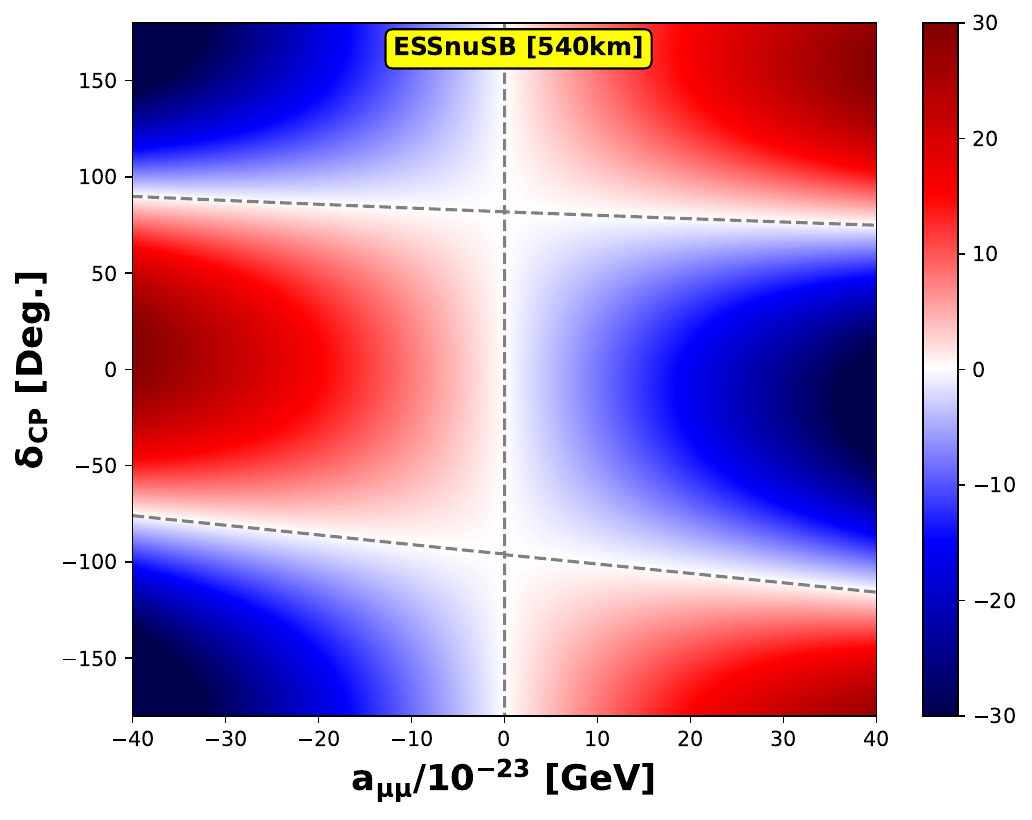}
    \label{fig:subC}
  \end{subfigure}
  \begin{subfigure}[b]{0.48\textwidth}
    \centering
    \includegraphics[width=\textwidth]{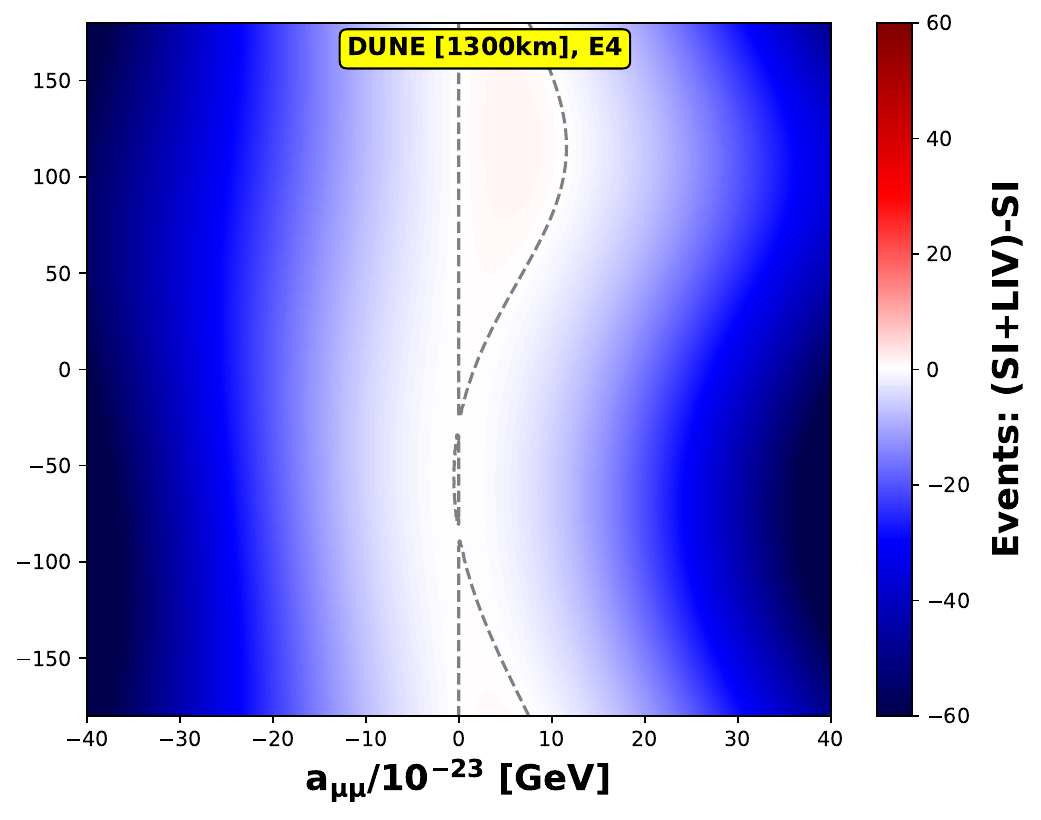}
    \label{fig:subD}
  \end{subfigure}
   \caption{The heatmap of event differences ($\Delta N_{events}$ =  $N_{events}(SI+LIV) - N_{events}(SI)$) in $a_{\alpha\beta}-\delta_{CP}$ plane for the diagonal LIV parameter $a_{\mu\mu}$ at energies E1, E2, E3 and E4. The grey dashed line mark the critical degeneracy region where the event difference is zero }
   \label{fig:example_eventdiff_dcp}
\end{figure}
 
Similarly, the heatmaps for $a_{\mu\mu}$ in Figure \ref{fig:example_eventdiff_dcp} exhibit a strong $\delta_{CP}$ dependence, characterized by alternating regions of positive and negative $\Delta N_{events}$. This confirms that diagonal LIV parameters significantly distort the CP-violation signature across the full range of $\delta_{CP}$. The bending of the degeneracy contours indicates that specific values of $a_{\alpha\alpha}$ can mimic different values of $\delta_{CP}$, creating ``fake" solutions.

The heatmaps for all diagonal parameters ($a_{ee}$ , $a_{\mu\mu}$ and $a_{\tau \tau}$) are presented in Figure \ref{fig:dcp_2d_3} the Appendix \ref{Appendix_A}. We see strong $\delta_{CP}$ dependence and oscillatory behaviour of the degeneracy curves giving us an alternating structures of red and blue regions. $a_{ee}$ and $a_{e\tau}$ shows similar degeneracy curves but opposite colours of the regions for all of E1, E2, E3 and E4 indicating the opposite effects of the parameters in the oscillation probabilities.

The heatmaps for all the off-diagonal parameters in Figure \ref{fig:dcp_2d_4} reveal that for all three off-diagonal parameters ($a_{e\mu}, a_{e\tau}, a_{\mu\tau}$) considering $\phi_{\alpha\beta} = 0$, the interplay between $\delta_{CP}$  and LIV produces characteristic areas of opposite signs. The bending of contours indicates degeneracy directions between $\delta_{CP}$ and each LIV coefficient. These parameters generate smooth, largely monotonic shifts in the event rate across the $\delta_{CP}$ space. Instead of the alternating quadrant structure seen with diagonal terms, we observe broad regions of uniform color. The $\delta_{CP}$ dependence is less oscillatory, implying that off-diagonal LIV terms introduce a bias to the CP measurement that is relatively uniform across the parameter space. As expected, the DUNE setup shows the greatest overall sensitivity, displaying the largest magnitude variations in $\Delta N_{events}$ across the $\delta_{CP}$--$a_{\alpha\beta}$ parameter space.

We have not explicitly presented the parameter degeneracy effects in the $\nu_\mu \to \nu_\mu$ disappearance channel, as it is evident from the above discussion that it is not possible to disentangle the LIV effects merely by counting events. We expect that a combined analysis across all energy ranges in all the available baselines leverages the non-overlapping degeneracy regions which will be able to break the correlations and constrain the true parameters.

\subsection{Sensitivity Analysis} \label{sec:sensitivity_analysis}

As described in simulation details section \ref{sec:simulation}, to determine the sensitivity of the ESSnuSB and DUNE experiments to the isotropic, CPT-violating LIV parameters, we perform a $\Delta\chi^2$ approach based on a Poissonian log-likelihood ratio, implemented within the GLoBES framework \cite{Huber:2004ka, HUBER2007439, Fogli:2002pt}. The true data is simulated assuming the Standard Model three-flavor oscillation paradigm (i.e., all LIV parameters are zero), while the test hypothesis includes non-zero LIV parameters as per the requirements (Table \ref{tab:marginalization_val}). 

During the $\chi^2$ analysis, the simulated data were generated using the latest NuFIT values for the oscillation parameters, except for $\delta_{CP}$. Throughout this work, the true value of $\delta_{CP}$ is fixed at $-90^\circ$, assuming a normal mass ordering and the higher octant for $\theta_{23}$. In the fit, along with the standard neutrino oscillation framework, we have incorporated the effects of Lorentz Invariance Violation (LIV). Accordingly, in the fit, $\theta_{23}$ and $\delta_{CP}$ are varied within their $3\sigma$ allowed ranges, while $\theta_{12}$, $\theta_{13}$, $\Delta m^2_{21}$ are varied within their $1\sigma$ allowed ranges as priors. The priors are estimated from the corresponding $3\sigma$ ranges using the formula $\frac{x_{u} - x_{l}}{3(x_{u} + x_{l})} \%,
$ where $x_{u}$ and $x_{l}$ denote the upper and lower limits of the $3\sigma$ allowed ranges, respectively. The prior on $\sin^2{\theta_{12}}$ is added from JUNO experiment as :
\begin{equation}
    \chi^2_{JUNO}(\sin^2{\theta_{12}})=(\frac{\sin^2{\theta_{12}}^{fit}-\sin^2{\theta_{12}}^{bf}}{0.0087})^{2}
\end{equation} To account for potential degeneracies arising from the wrong mass ordering, $\Delta m^2_{31}$ is varied across its $3\sigma$ range in both normal and inverted hierarchies.

The statistical analysis in this work follows the methodology detailed in Ref.~\cite{Fiza:2022xfw}, which employs a Poissonian log-likelihood ratio test to quantify the sensitivity of long-baseline experiments to Lorentz Invariance Violation. This method compares the event rates predicted under a ``test" hypothesis (including LIV) against the ``true" hypothesis (assuming standard three-flavor oscillations, i.e., $a_{\alpha\beta}=0$). The total $\Delta\chi^2$ function incorporates statistical fluctuations, systematic uncertainties (via the pull method), and prior constraints on known parameters:
\begin{equation}
\begin{split}
    \Delta\chi^{2}(p^{\text{true}}) = \min_{p^{\text{test}},\eta} \Bigg[ & 2\sum_{i,j,k}^{} \left\{ N_{ijk}^{\text{test}}(p^{\text{test}};\eta) - N_{ijk}^{\text{true}}(p^{\text{true}}) + N_{ijk}^{\text{true}}(p^{\text{true}}) \ln\frac{N_{ijk}^{\text{true}}(p^{\text{true}})}{N_{ijk}^{\text{test}}(p^{\text{test}};\eta)} \right\} \\
    & + \sum_{l} \frac{(p_{l}^{\text{true}} - p_{l}^{\text{test}})^{2}}{\sigma_{p_{l}}^{2}} + \sum_{m} \frac{\eta_{m}^{2}}{\sigma_{\eta m}^{2}} \Bigg]
\end{split}
\end{equation}
\begin{itemize}
    \item The first part of the equation is the Poissonian log-likelihood ratio. This sum is performed over all experimental configurations. $N^{true/test}_{ijk}$ are the expected event counts in energy bin $i$ for channel/mode $k$ under the true/test hypothesis where $i$ indicates the energy bins of the detector, $j$ indicates the oscillation channels ($\nu_e$ appearance, $\nu_\mu$ disappearance) and $k$ represents the beam operational modes (neutrino run and antineutrino run).
    \item  The Prior Term term $\sum_{l} \frac{(p_{l}^{\text{true}} - p_{l}^{\text{test}})^{2}}{\sigma_{p_{l}}^{2}}$ incorporates prior knowledge on standard oscillation parameters that are already measured with some precision. $p^{true/test}$ represents the sets of true/test oscillation and LIV parameters, and $\sigma_{pl}$ represents the 1$\sigma$ standad deviation for that parameter.  
    \item The Pull Term for Systematics $\sum_{m} \frac{\eta_{m}^{2}}{\sigma_{\eta m}^{2}}$ accounts for each systematic uncertainty (e.g., signal normalization, background normalization, matter density uncertainty) the pull method which are associated with a nuisance parameter $\eta_m$, which has a central value of zero and a 1$\sigma$ uncertainty of $\sigma_{\eta m}$.
\end{itemize}
The minimization is performed over all relevant test parameters ($p^{test}$) and nuisance parameters ($\eta$) to obtain conservative sensitivity limits. A combined analysis of both detectors at the two baselines will significantly improve the constraints on such new-physics scenarios. Furthermore, incorporating future data from DUNE could enhance this sensitivity even further.

\subsubsection{Constraints on LIV Parameters}\label{sec:constraints}

\begin{figure}[H]
  \centering
  \begin{subfigure}[b]{0.48\textwidth}
    \centering
    \includegraphics[width=\textwidth]{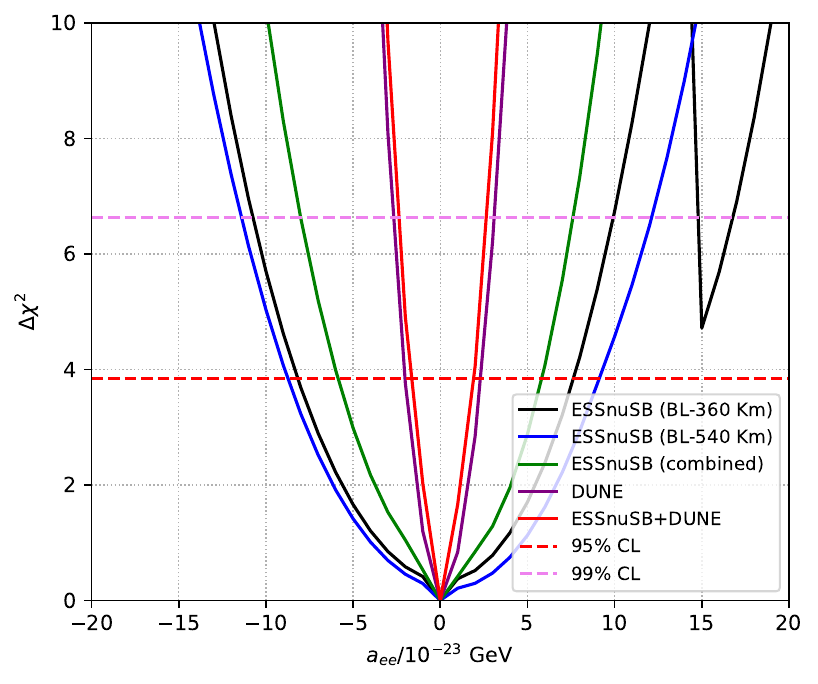}
    \label{fig:subA}
  \end{subfigure}
  \begin{subfigure}[b]{0.48\textwidth}
    \centering
    \includegraphics[width=\textwidth]{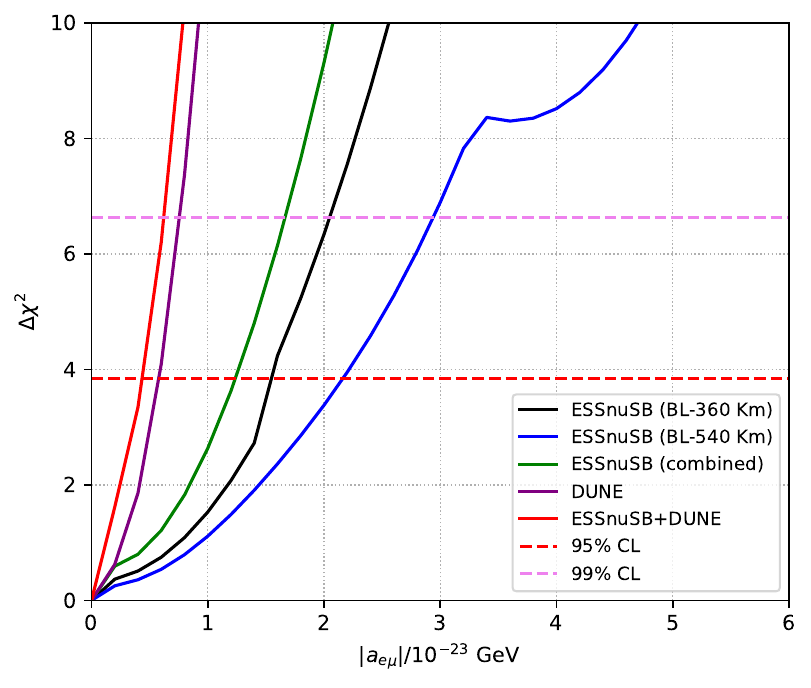}
    \label{fig:subB}
  \end{subfigure}
  \\[-4ex]  

  \begin{subfigure}[b]{0.48\textwidth}
    \centering
    \includegraphics[width=\textwidth]{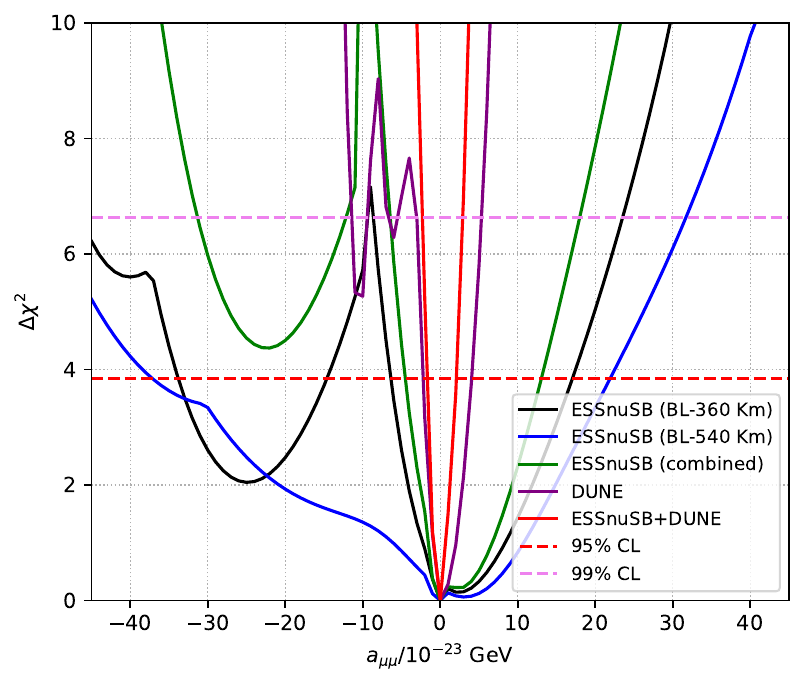}
    \label{fig:subC}
  \end{subfigure}
  \begin{subfigure}[b]{0.48\textwidth}
    \centering
    \includegraphics[width=\textwidth]{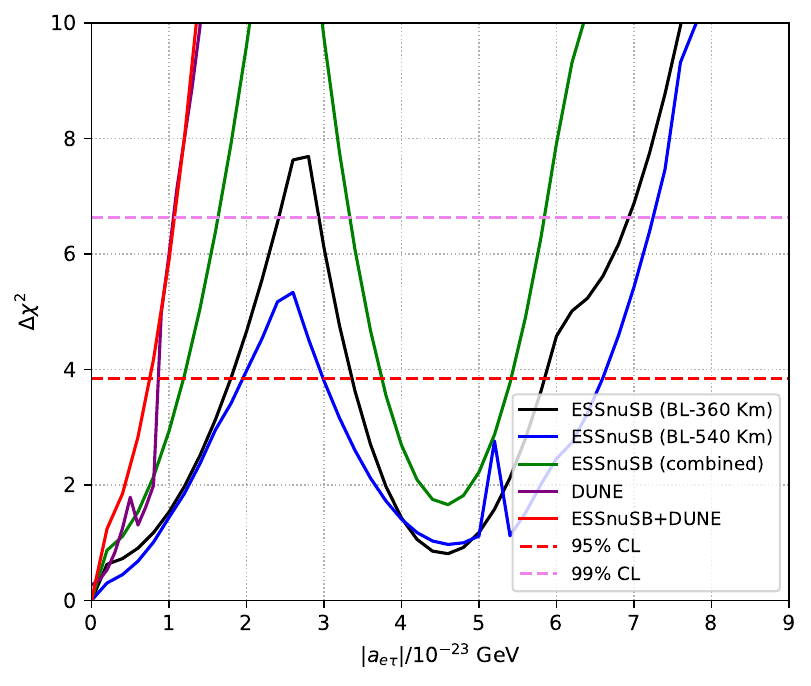}
    \label{fig:subD}
  \end{subfigure}
  \\[-4ex]

  \begin{subfigure}[b]{0.48\textwidth}
    \centering
    \includegraphics[width=\textwidth]{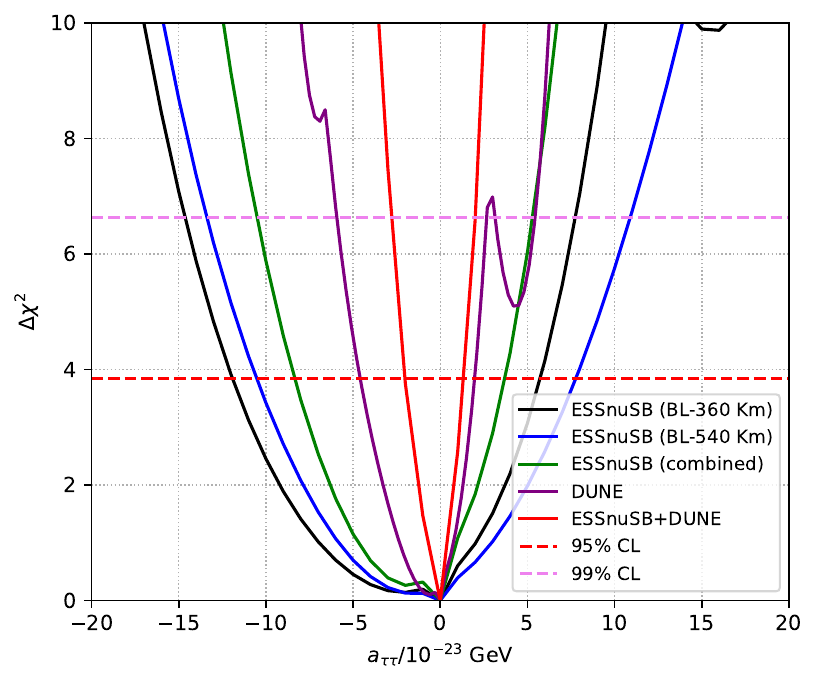}
    \label{fig:subE}
  \end{subfigure}
  \begin{subfigure}[b]{0.48\textwidth}
    \centering
    \includegraphics[width=\textwidth]{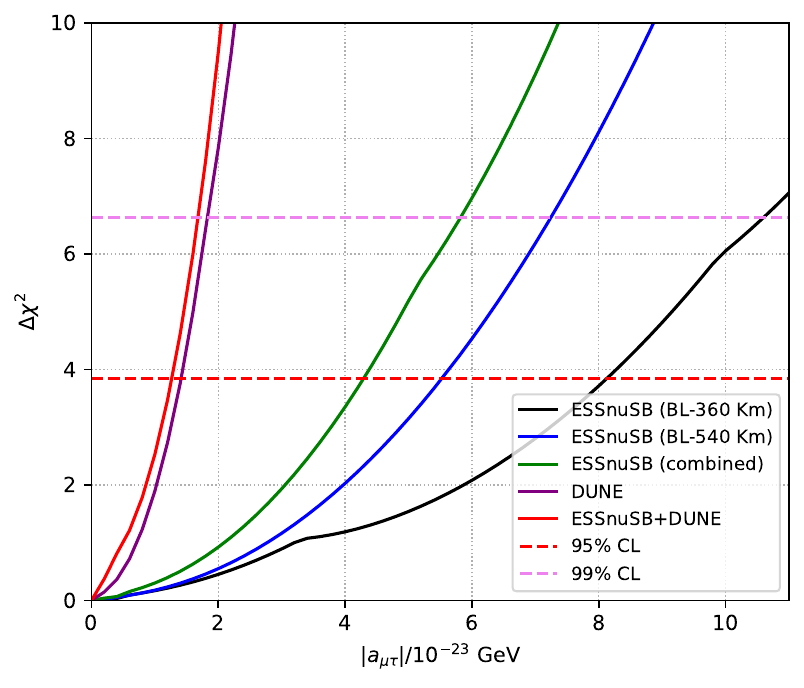}
    \label{fig:subF}
  \end{subfigure}

  \caption{One-dimensional $\Delta\chi^2$ profiles for the CPT-violating LIV parameters. The sensitivity is shown for ESSnuSB at BL=360 km (black), ESSnuSB at BL=540 km (blue), DUNE at BL=1300km (purple), the combination of both ESSnuSB setups (green), and the combined sensitivity of ESSnuSB and DUNE (red). The horizontal dashed lines indicate the 95\% (red) and 99\% (pink) confidence levels for one degree of freedom.}
  \label{fig:chi2}
\end{figure}

Figure \ref{fig:chi2} presents the one-dimensional limits on each of the six relevant LIV parameters for ESSnuSB at two different baselines (360 km and 540 km), their combined setup, DUNE, and the combined analysis of ESSnuSB and DUNE. The results for the diagonal LIV parameters are obtained by marginalising over $\theta_{23}$, $\delta_{CP}$, and $\Delta m^2_{31}$ as described earlier. For the off-diagonal parameters, denoted as $a_{\alpha\beta}$, the analysis is done by marginalising over the corresponding phase $\phi_{\alpha\beta}$ in the range $[-\pi, \pi]$. 
On the left panel, we have the results for the diagonal LIV parameters, while on the right, we have the results for the off-diagonal absolute parameters. 
Across all three diagonal parameters, $a_{ee}$, $a_{\mu\mu}$ and $a_{\tau\tau}$, combining DUNE and ESSnuSB gives the strongest constraint, meaning this setup can rule out the widest range of values with high confidence. DUNE alone also performs very well, showing a steep rise in $\Delta \chi^2$ and a narrow curve around zero. The combined ESSnuSB setup improves over its individual baselines, thanks to more data and complementary coverage. Interestingly, the 360 km baseline gives better constraints than the 540 km one, which might be due to better energy resolution or optimised detector performance. The 540 km baseline proves to be the weakest, showing that a longer baseline doesn’t always yield better sensitivity. 

In the analysis of the \( a_{\mu\mu} \) parameter, the ESSnuSB setup at 360~km exhibits a pronounced degeneracy, characterized by a secondary local minimum in the \( \Delta \chi^2 \) curve around \( a_{\mu\mu} \approx -25 \times 10^{-23}~\mathrm{GeV} \). The combined ESSnuSB configuration also displays a milder version of this degeneracy. Although DUNE offers greater sensitivity, it is not entirely free from such degeneracies; the \( \Delta \chi^2 \) curve for DUNE reveals shallow dips near \( a_{\mu\mu} \approx -10 \times 10^{-23}~\mathrm{GeV} \) and \( a_{\mu\mu} \approx -6 \times 10^{-23}~\mathrm{GeV} \). A similar feature is observed for \( a_{\tau\tau} \) at DUNE, with a dip around \( a_{\tau\tau} \approx 4.9 \times 10^{-23}~\mathrm{GeV} \). However, when DUNE is combined with ESSnuSB, all such degeneracies are effectively resolved. The resulting \( \Delta \chi^2 \) curve exhibits a single, sharp minimum at the true value, underscoring the enhanced resolving power of the combined experimental setup.

On the right panel, the results for the off-diagonal parameters show a similar trend, except in the case of $|a_{\mu\tau}|$. The combination DUNE+ESSnuSB performs the best and puts tight constraints on all three off-diagonal parameters. DUNE alone provides stronger bounds than the combined ESSnuSB, while the combined ESSnuSB performs better than its individual baselines. For $|a_{\mu\tau}|$, the detector at the 540 km baseline performs better than that at 360 km, yielding tighter constraints on $|a_{\mu\tau}|$.

From this analysis, we observe that combining the two ESSnuSB baselines at 
$360~\text{km}$ and $540~\text{km}$ significantly enhances sensitivity to the 
LIV parameters compared to each baseline individually. Moreover, the inclusion 
DUNE further strengthens the constraints, effectively excluding large portions of the previously allowed parameter space and thereby improving the overall precision of the measurements. In the case of $|a_{e\tau}|$, the degeneracy observed around $|a_{\mu\tau}| = 4.5~\text{GeV}$ can be resolved through the addition of DUNE. These degeneracies are discussed in detail in the preceding sections within the context of oscillation parameter correlations. The constraints derived in this work can be found in table \ref{tab:liv_bounds_95cl}.

\begin{table}[h!]
\centering
\renewcommand{\arraystretch}{1.5} 
\begin{tabular}{|p{.8cm}|p{2.9cm}|p{2.4cm}|p{2.4cm}|p{2.4cm}|p{2.3cm}|}
\hline
\textbf{$a_{\alpha\beta}$} & \textbf{ESSnuSB (360 km)} & \textbf{ESSnuSB (540 km)} & \textbf{ESSnuSB (comb.)} & \textbf{DUNE (1300 km)} & \textbf{ESSnuSB (comb.) + DUNE} \\
 & \textbf{$[10^{-23} \text{ GeV}]$} & \textbf{$[10^{-23} \text{ GeV}]$} & \textbf{$[10^{-23} \text{ GeV}]$} & \textbf{$[10^{-23} \text{ GeV}]$} & \textbf{$[10^{-23} \text{ GeV}]$} \\
\hline \hline
\multirow{2}{*}{$a_{ee}$} &$[-8.16, 7.63]$ & $[-8.73, 9.12]$ & $[-5.85, 5.81]$ & $[-2.02, 2.30]$ & $[-1.63, 1.91]$\\ 
 & & & & & \\ 
\hline
\multirow{2}{*}{$a_{\mu\mu}$} & $[-33.79, -14.64] \cup$ & $[-37.12, 21.91]$ & $[-4.50, 13.04]$ & $[-2.20, 4.05]$ & $[-1.66, 2.08]$\\ 
 &$[-6.36, 17.05]$ & & & & \\ 
\hline
\multirow{2}{*}{$a_{\tau\tau}$} & $[-11.91, 5.71]$ & $[-10.51, 7.77]$ & $[-8.33, 3.69]$ & $[-4.59, 1.97]$ & $[-2.02, 1.32]$\\ 
 & & & & & \\ 
\hline
\multirow{2}{*}{$|a_{e\mu}|$} & $[0, 1.55]$ & $[0, 2.16]$ & $[0, 1.24]$ & $[0, 0.58]$ & $[0, 0.43]$\\ 
 & & & & & \\ 
\hline
\multirow{2}{*}{$|a_{e\tau}|$} & $[0, 1.80] \cup$ & $[0, 1.95] \cup$ & $[0, 1.19] \cup$ & $[0, 0.86]$ & $[0, 0.75]$\\ 
 & $[3.36, 5.85]$ & $[2.99, 6.59]$ & $[3.75, 5.41]$ & & \\ 
\hline
\multirow{2}{*}{$|a_{\mu\tau}|$} & $[0, 8.12]$ & $[0, 5.52]$ & $[0, 4.29]$ & $[0, 1.41]$ & $[0, 1.26]$\\ 
 & & & & & \\ 
  \hline
\end{tabular}
\renewcommand{\arraystretch}{1.0} 
\caption{Projected bounds on the CPT-violating LIV parameters at 95\% C.L. derived from 1D $\Delta\chi^2$ profiles}
\label{tab:liv_bounds_95cl}
\end{table}

\subsubsection{Resolving $\theta_{23}$ and $\delta_{CP}$ Degeneracies }\label{sec:degeneracy}

A critical challenge in the search for new physics is the potential degeneracies between new physics parameters and the standard oscillation parameters. Such degeneracies can obscure the true values of the parameters and lead to incorrect physical interpretations. In this section, we analyse the correlation between the CPT-violating LIV parameters and the atmospheric mixing angle, $\theta_{23}$ and the standard CP-violating phase $\delta_{CP}$, to assess how effectively different experimental configurations can resolve these degeneracies. 

\begin{figure}[!h]
  \centering
  \begin{subfigure}[b]{0.48\textwidth}
    \centering
    \includegraphics[width=\textwidth]{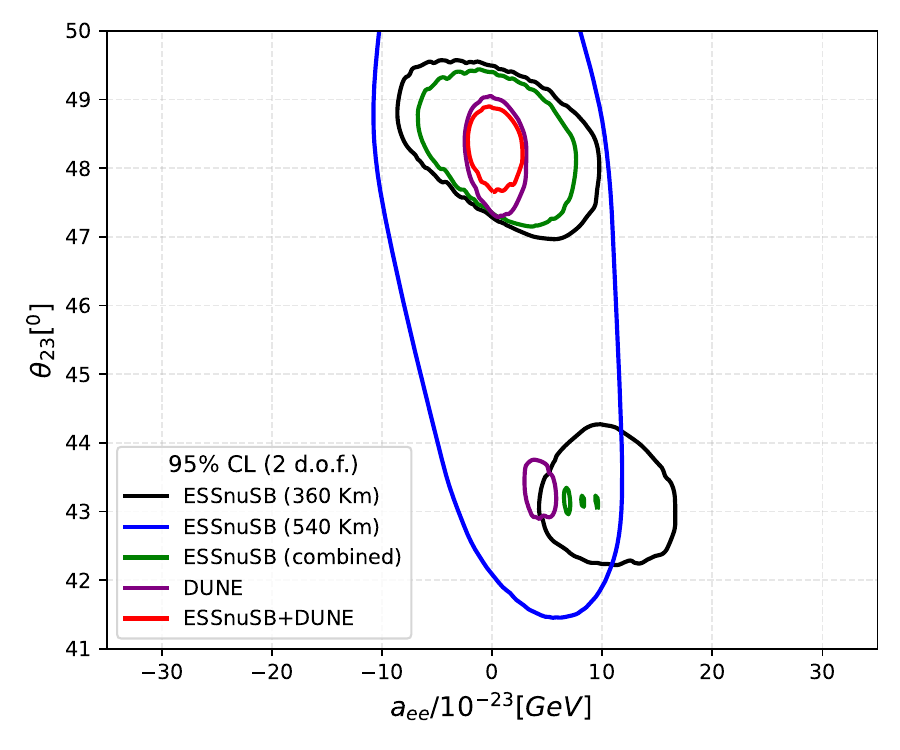}
    \label{fig:subA}
  \end{subfigure}
  \begin{subfigure}[b]{0.48\textwidth}
    \centering
    \includegraphics[width=\textwidth]{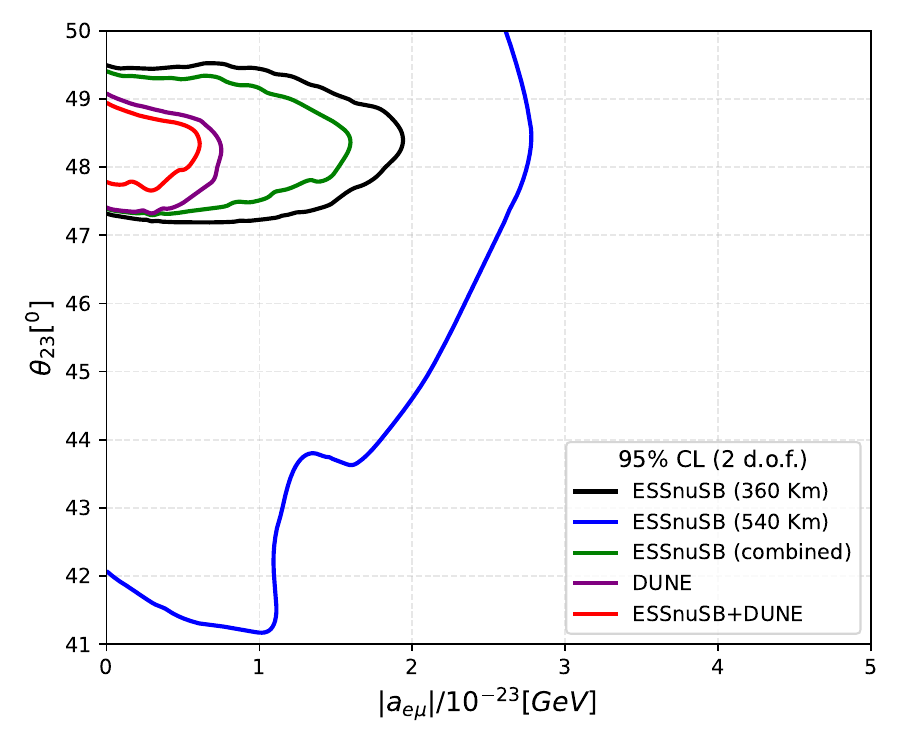}
    \label{fig:subB}
  \end{subfigure}
  \\[-4ex]

  \begin{subfigure}[b]{0.48\textwidth}
    \centering
    \includegraphics[width=\textwidth]{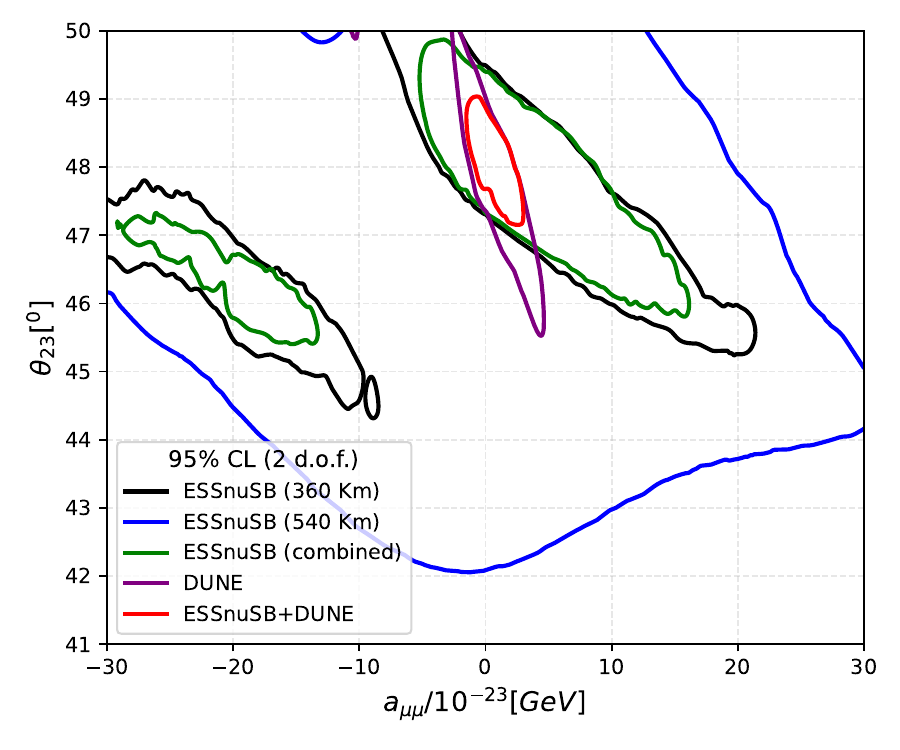}
    \label{fig:subC}
  \end{subfigure}
  \begin{subfigure}[b]{0.48\textwidth}
    \centering
    \includegraphics[width=\textwidth]{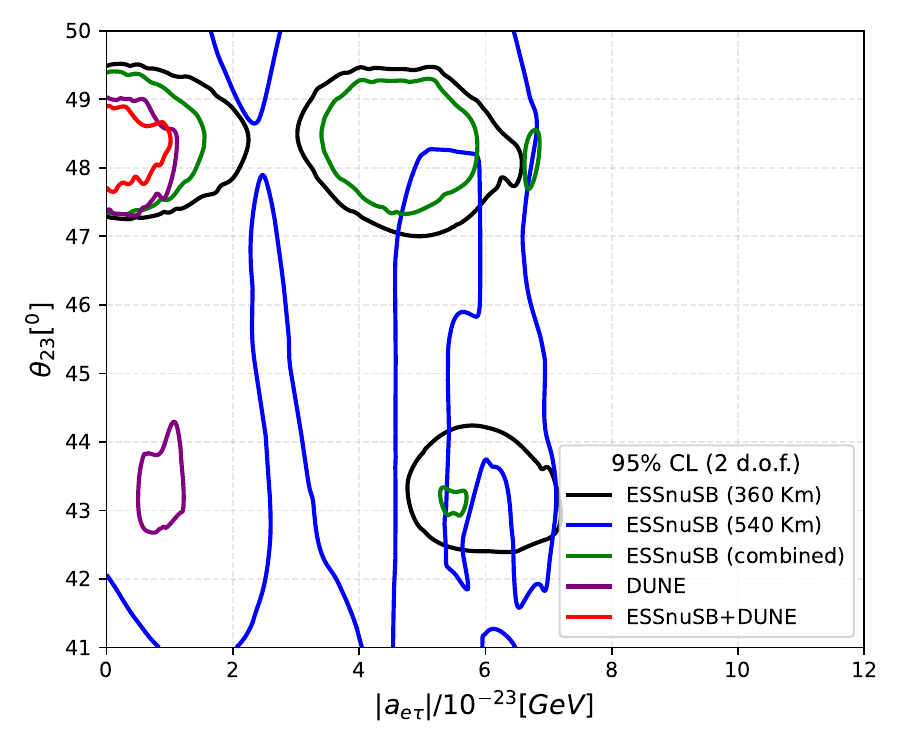}
    \label{fig:subD}
  \end{subfigure}
  \\[-4ex]

  \begin{subfigure}[b]{0.48\textwidth}
    \centering
    \includegraphics[width=\textwidth]{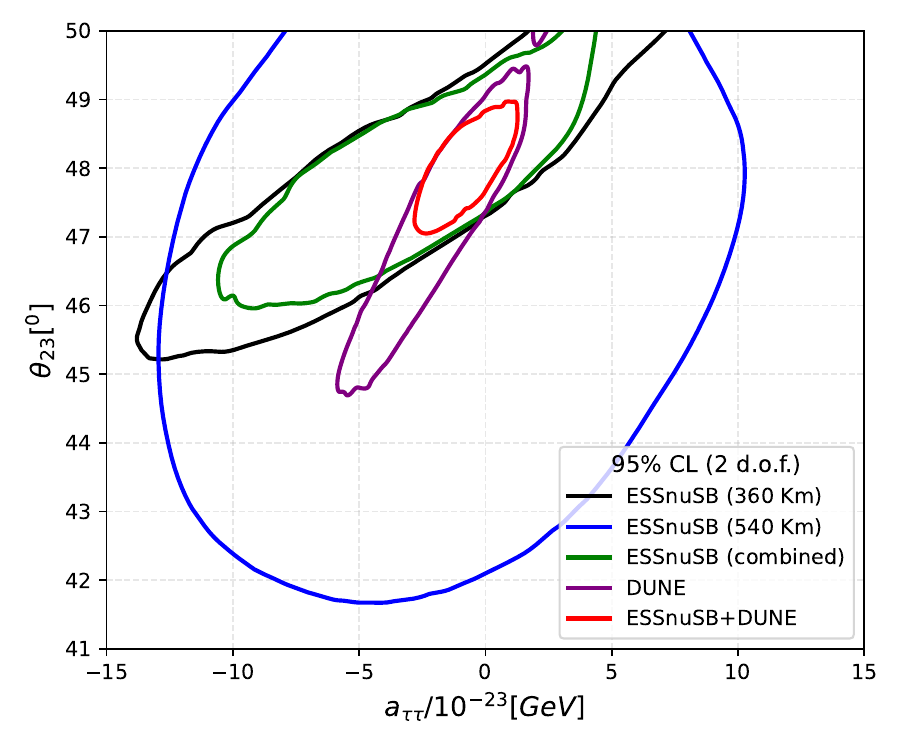}
    \label{fig:subE}
  \end{subfigure}
  \begin{subfigure}[b]{0.48\textwidth}
    \centering
    \includegraphics[width=\textwidth]{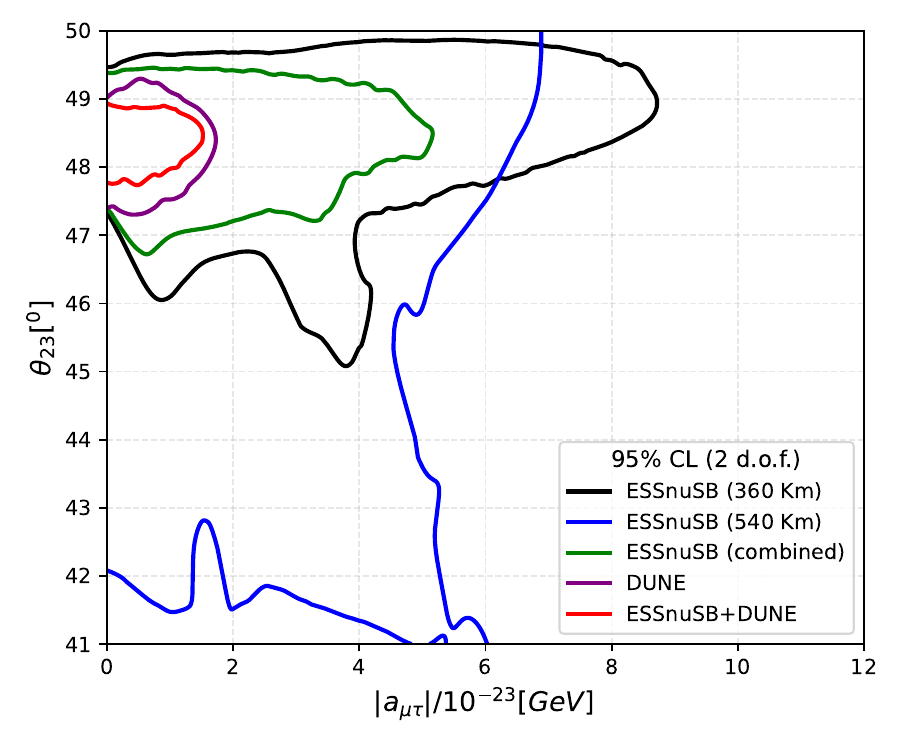}
    \label{fig:subF}
  \end{subfigure}
  \\[-4ex]

  \caption{Allowed regions at 95\% C.L. (2 d.o.f.) in the plane of each LIV parameter versus $\theta_{23}$. The contours show the sensitivity for ESSnuSB at BL=360 km (black), ESSnuSB at BL=540 km (blue), the combined ESSnuSB setup (green), DUNE (purple), and the combined analysis of ESSnuSB and DUNE (red). The true value of all LIV parameters is assumed to be zero.}
  \label{fig:theta23_degeneracy}
\end{figure}

\begin{figure}[!h]
  \centering
  \begin{subfigure}[b]{0.48\textwidth}
    \centering
    \includegraphics[width=\textwidth]{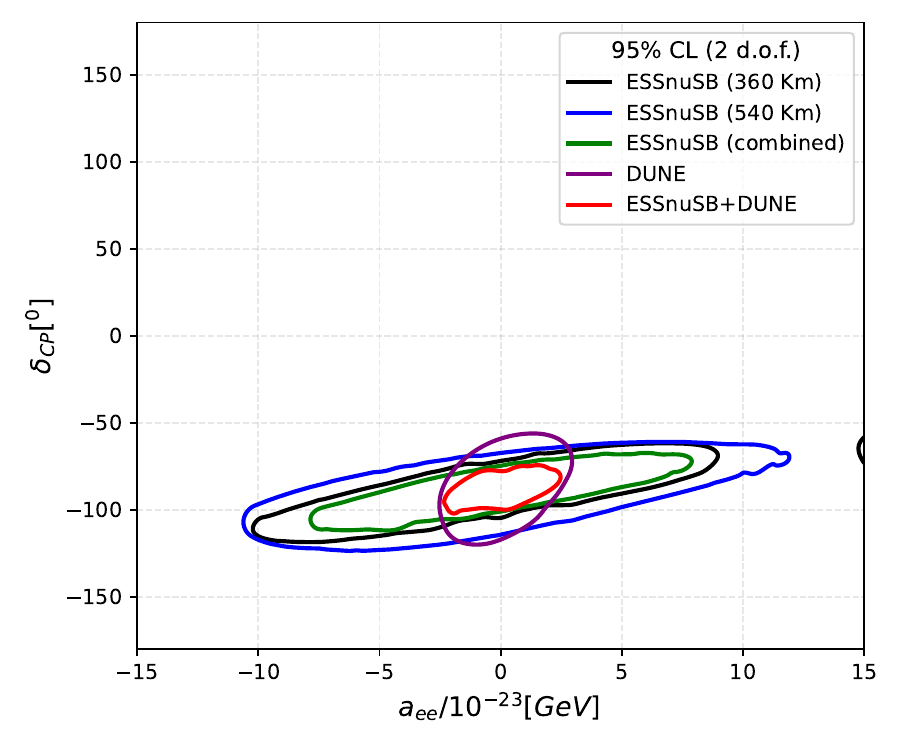}
    \label{fig:subA}
  \end{subfigure}
  \begin{subfigure}[b]{0.48\textwidth}
    \centering
    \includegraphics[width=\textwidth]{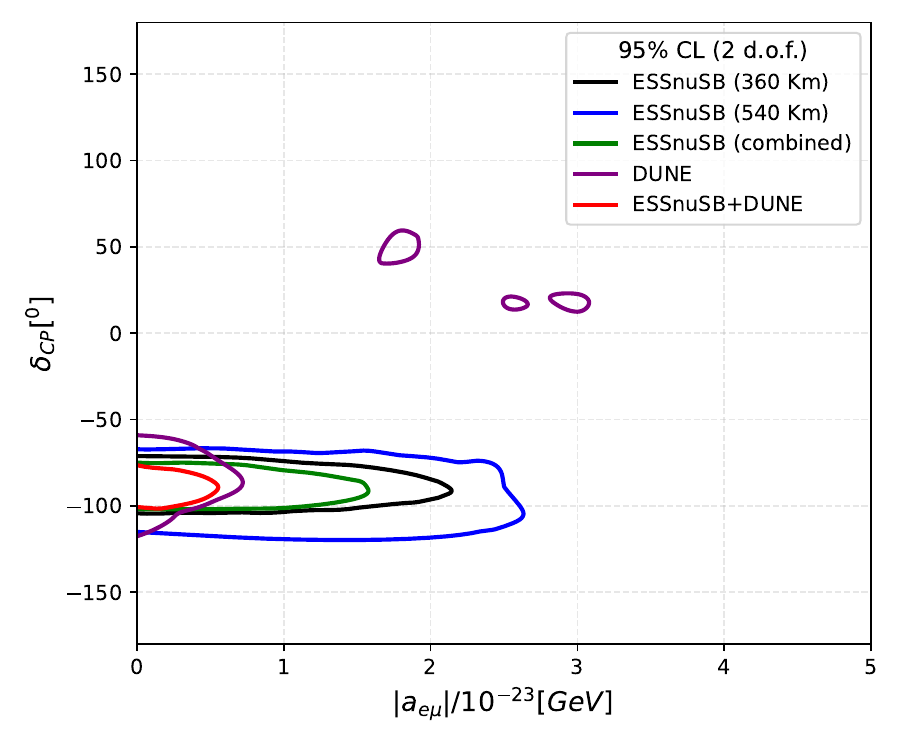}
    \label{fig:subB}
  \end{subfigure}
  \\[-4ex]

  \begin{subfigure}[b]{0.48\textwidth}
    \centering
    \includegraphics[width=\textwidth]{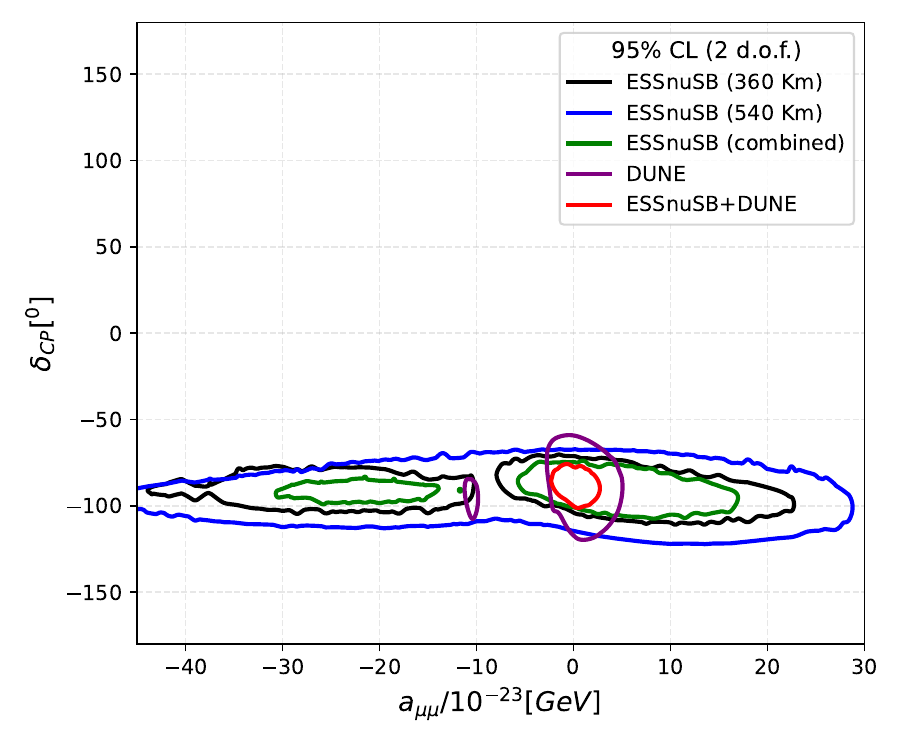}
    \label{fig:subC}
  \end{subfigure}
  \begin{subfigure}[b]{0.48\textwidth}
    \centering
    \includegraphics[width=\textwidth]{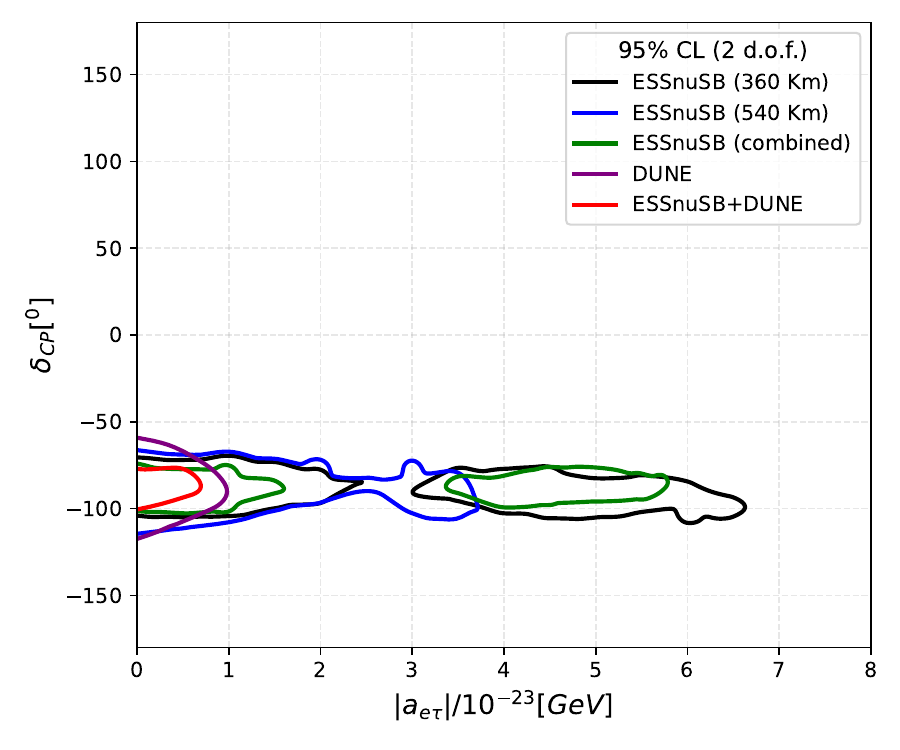}
    \label{fig:subD}
  \end{subfigure}
  \\[-4ex]

  \begin{subfigure}[b]{0.48\textwidth}
    \centering
    \includegraphics[width=\textwidth]{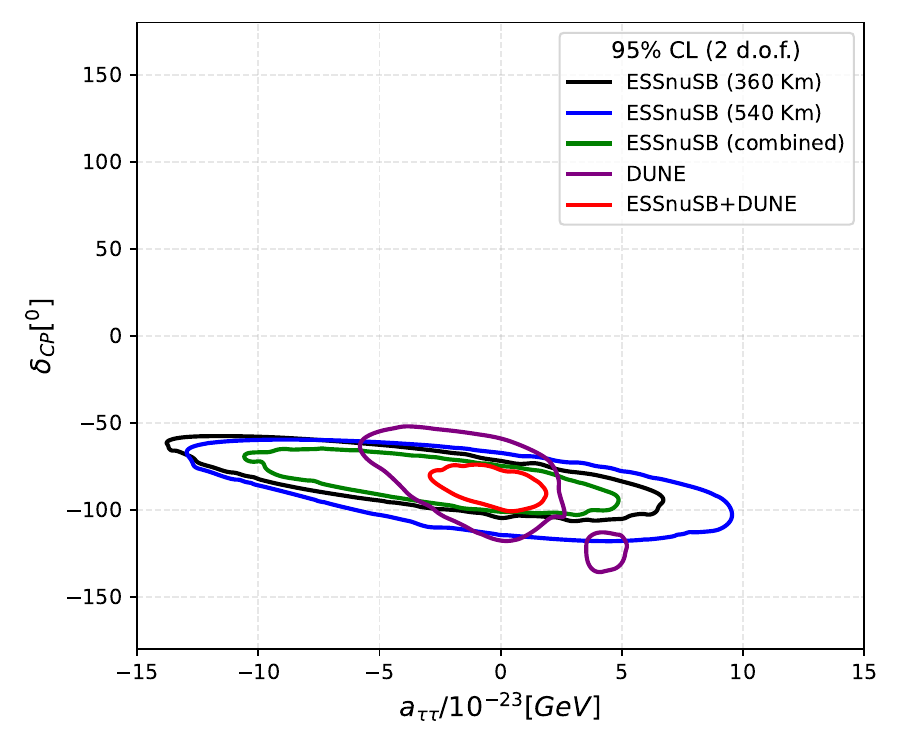}
    \label{fig:subE}
  \end{subfigure}
  \begin{subfigure}[b]{0.48\textwidth}
    \centering
    \includegraphics[width=\textwidth]{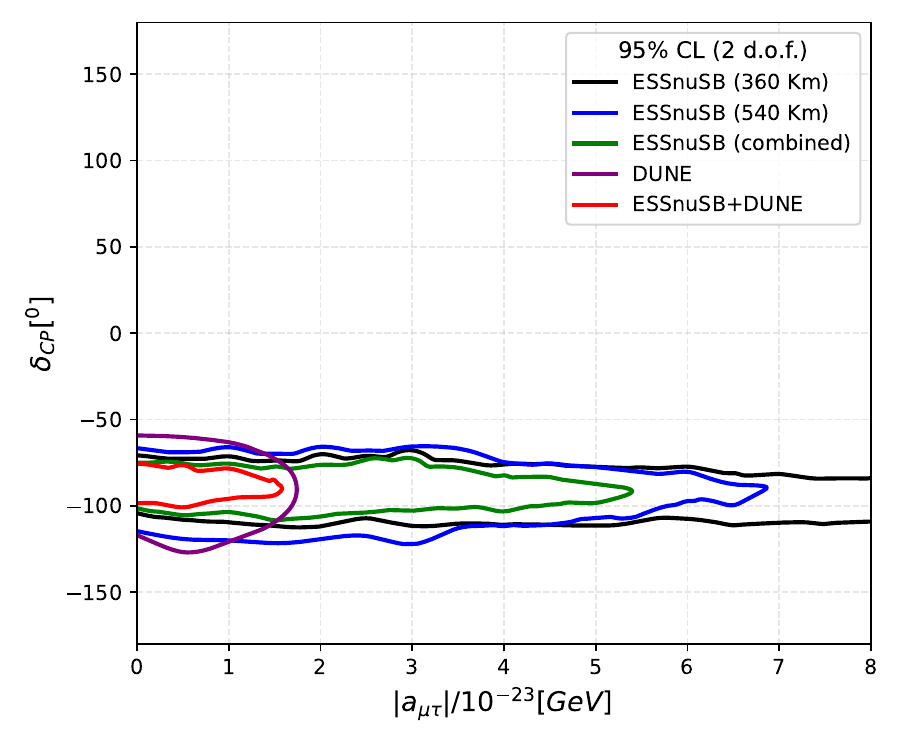}
    \label{fig:subF}
  \end{subfigure}
  \\[-4ex]

  \caption{Allowed regions at 95\% C.L. (2 d.o.f.) in the plane of each LIV parameter versus the standard CP phase $\delta_{CP}$. The contours represent the sensitivity for ESSnuSB at BL=360 km (black), ESSnuSB at BL=540 km (blue), the combined ESSnuSB setup (green), DUNE (purple), and the combined analysis of ESSnuSB and DUNE (red). The true value of all LIV parameters is assumed to be zero, with a true value of $\delta_{CP}$ near maximal CP violation.}
  \label{fig:deltaCP_degeneracy}
\end{figure}

Figure~\ref{fig:theta23_degeneracy} and Figure~\ref{fig:deltaCP_degeneracy} 
show the 95\% exclusion regions in the two-dimensional parameter space of each 
LIV parameter versus $\theta_{23}$ and the standard $\delta_{CP}$, respectively, after collecting data for the projected periods described in the experimental details section. The left panel presents the results for the diagonal LIV parameters, while the right panel displays the results for the absolute values of the off-diagonal LIV parameters with $\theta_{23}$ and the standard $\delta_{CP}$, respectively. It is important to note that smaller and more localised allowed regions indicate a stronger ability to constrain both parameters and to resolve 
potential ambiguities simultaneously.

While generating these plots, we vary one LIV parameter at a time. For the off-diagonal LIV parameters, we marginalize over the corresponding phase without restriction, in addition to the standard oscillation parameters discussed earlier. In Figures~\ref{fig:theta23_degeneracy} and \ref{fig:deltaCP_degeneracy}, when examining the correlation of $\theta_{23}$ ($\delta_{CP}$) with the LIV parameters, we marginalize over $\delta_{CP}$ ($\theta_{23}$) within its allowed $3\sigma$ range, along with the other parameters. 

\paragraph{$\theta_{23}$ vs LIV Parameters:} Figure \ref{fig:theta23_degeneracy} presents the 95\% confidence level (C.L.) allowed regions in the two-dimensional parameter space of each LIV parameter versus $\theta_{23}$ for 2 degrees of freedom (d.o.f.). Smaller and more localized allowed regions indicate a better ability to simultaneously constrain both parameters and resolve potential ambiguities.

From Figure~\ref{fig:theta23_degeneracy}, we observe that the 540 km baseline at ESSnuSB covers a larger portion of the parameter space compared to the 360 km baseline. Across all six correlations studied between LIV parameters and $\theta_{23}$, the 540 km baseline accommodates nearly all values of $\theta_{23}$ within the $3\sigma$ range. In contrast, the 360 km baseline excludes lower-octant values of $\theta_{23}$ for $|a_{e\mu}|$, $|a_{\mu\tau}|$, and $a_{\tau\tau}$. For $|a_{e\tau}|$ and $a_{ee}$, although regions around the true $\theta_{23}=48.5^{\circ}$ remain allowed, additional spurious regions appear in the wrong octant, complicating the identification of the true solution. Consequently, for $|a_{e\tau}| \approx (5.5-6)\times 10^{-23}$ GeV and $a_{ee} \approx (5-10)\times 10^{-23}$ GeV, allowed values of $\theta_{23}$ exist in both octants, making it difficult to determine the correct octant and the absolute value of $\theta_{23}$. Similarly, in the case of $|a_{e\tau}|$ and $a_{\mu\mu}$, for a given $\theta_{23}$ we find two allowed regions corresponding to different values of $|a_{e\tau}|$ and $a_{\mu\mu}$, preventing a unique determination of their true values. A combination of both baselines at ESSnuSB can further restrict the parameter space, but cannot resolve the degeneracies observed with the 360 km baseline. DUNE performs better than the combined ESSnuSB setup, though it still suffers from $\theta_{23}$ degeneracies. In particular, for $|a_{e\tau}|$ and $a_{ee}$, degenerate solutions arise where $\theta_{23}$ lies in the opposite octant. Crucially, the combined use of the two baselines at ESSnuSB together with DUNE resolves all observed degeneracies, enabling the true regions to be identified definitively. Moreover, at the 95$\%$ confidence level, this synergy excludes most of the parameter space, thereby yielding significantly improved precision in the measurements. Table \ref{tab:liv_theta23_corr_theta} presents the results derived from figure \ref{fig:theta23_degeneracy}, displaying the allowed ranges of $\theta_{23}$ obtained from the fit when the data are generated under exact Lorentz invariance, with the true value of $\theta_{23}$ fixed at $48.5^{\circ}$.

\clearpage

\begin{table}[htbp]
\centering
\renewcommand{\arraystretch}{1.3}
\begin{tabular}{|p{1.2cm}|p{2.4cm}|p{2.4cm}|p{2.4cm}|p{2.4cm}|p{2.4cm}|}
\hline
\textbf{Corr. w/} & \textbf{ESSnuSB (360 km)} & \textbf{ESSnuSB (540 km)} & \textbf{ESSnuSB (comb.)} & \textbf{DUNE (1300 km)} & \textbf{ESSnuSB (comb.) + DUNE} \\
\textbf{Param} & \textbf{$\theta_{23}$ [deg]} & \textbf{$\theta_{23}$ [deg]} & \textbf{$\theta_{23}$ [deg]} & \textbf{$\theta_{23}$ [deg]} & \textbf{$\theta_{23}$ [deg]} \\
\hline \hline
\multirow{2}{*}{$a_{ee}$} & $[42.2, 44.3] \cup$ & $[41.5, 50.0]$ & $[43.0, 43.3] \cup$ & $[42.9, 43.8] \cup$ & $[47.6, 48.9]$\\
 & $[47.0, 49.6]$ & & $[47.1, 49.4]$ & $[47.3, 49.0]$ & \\
\hline
\multirow{2}{*}{$a_{\mu\mu}$} & $[44.3, 50.0]$ & $[42.1, 50.0]$ & $[45.4, 49.9]$ & $[45.5, 50.0]$ & $[47.2, 49.0]$\\
 & & & & & \\
\hline
\multirow{2}{*}{$a_{\tau\tau}$} & $[45.2, 50.0]$ & $[41.7, 50.0]$ & $[46.0, 50.0]$ & $[44.7, 49.5] \cup$ & $[47.0, 49.0]$\\
 & & & & $[49.8, 50.0]$ & \\
\hline
\multirow{2}{*}{$|a_{e\mu}|$} & $[47.2, 49.5]$ & $[41.2, 50.0]$ & $[47.3, 49.4]$ & $[47.3, 49.1]$ & $[47.7, 48.9]$\\
 & & & & & \\
\hline
\multirow{2}{*}{$|a_{e\tau}|$} & $[42.4, 44.2] \cup$ & $[41.0, 50.0]$ & $[42.9, 43.3] \cup$ & $[42.7, 44.3] \cup$ & $[47.7, 48.9]$\\
 & $[47.0, 49.5]$ & & $[47.3, 49.4]$ & $[47.3, 49.0]$ & \\
\hline
\multirow{2}{*}{$|a_{\mu\tau}|$} & $[45.1, 49.9]$ & $[41.0, 50.0]$ & $[46.7, 49.5]$ & $[47.3, 49.3]$ & $[47.7, 48.9]$\\
 & & & & & \\
  \hline
\end{tabular}
\renewcommand{\arraystretch}{1.0}
\caption{Projected 95\% C.L. allowed ranges for $\theta_{23}$ in the presence of various LIV parameters. The combined analysis consistently restricts $\theta_{23}$ to the higher octant region ($47.6^\circ - 49.0^\circ$), successfully breaking the octant degeneracy observed in individual experiments.}
\label{tab:liv_theta23_corr_theta}
\end{table}

\paragraph{$\delta_{CP}$ vs LIV Parameters:} Figure \ref{fig:deltaCP_degeneracy} presents the 95\% confidence level (C.L.) allowed regions in the two-dimensional parameter space of each LIV parameter versus $\delta_{CP}$ for 2 degrees of freedom (d.o.f.). 

In Figure~\ref{fig:deltaCP_degeneracy}, the parameter space appears cleaner compared to Figure~\ref{fig:theta23_degeneracy}. Except for $|a_{e\tau}|$ and $|a_{\mu\tau}|$, the longer baseline of 560 km at ESSnuSB allows a broader parameter space than the 360 km baseline. Across all six correlations studied, we find that the measurement of the standard $\delta_{CP}$ is tightly constrained around its true value, $\delta_{CP} = -90^{\circ}$. Notably, the combination of the two baselines at ESSnuSB provides a better constraint on the allowed values of $\delta_{CP}$ compared to DUNE. However, in terms of excluding the overall parameter space spanned by $\delta_{CP}$ and the LIV parameters, DUNE outperforms both the individual baselines and their combination. For $|a_{e\tau}|$ and $|a_{\mu\tau}|$, we observe that degenerate solutions appear near the true value of $\delta_{CP}$ at the 360 km baseline as well as in the combined ESSnuSB case, hindering precise determination of the LIV parameters. Crucially, combining ESSnuSB with DUNE resolves all such ambiguities, making this synergy particularly effective for eliminating these new degeneracies and enabling precise measurements of the LIV parameters.

\begin{table}[H]
\centering
\renewcommand{\arraystretch}{1.5}
\begin{tabular}{|p{.8cm}|p{2.5cm}|p{2.5cm}|p{2.5cm}|p{2.5cm}|p{2.5cm}|}
\hline
\textbf{Corr. w/} & \textbf{ESSnuSB (360 km)} & \textbf{ESSnuSB (540 km)} & \textbf{ESSnuSB (comb.)} & \textbf{DUNE (1300 km)} & \textbf{ESSnuSB (comb.) + DUNE} \\
 & \textbf{$\delta_{CP}$ [deg]} & \textbf{$\delta_{CP}$ [deg]} & \textbf{$\delta_{CP}$ [deg]} & \textbf{$\delta_{CP}$ [deg]} & \textbf{$\delta_{CP}$ [deg]} \\
\hline \hline
\multirow{2}{*}{$a_{ee}$} & $[-118.6, -58.3]$ & $[-123.7, -61.0]$ & $[-112.0, -67.4]$ & $[-120.0, -56.2]$ & $[-102.1, -74.5]$ \\ 
 & & & & & \\ 
\hline
\multirow{2}{*}{$a_{\mu\mu}$} & $[-110.7, -70.2]$ & $[-122.0, -67.2]$ & $[-107.3, -73.8]$ & $[-119.6, -59.0]$ & $[-101.1, -75.7]$ \\ 
 & & & & & \\ 
\hline
\multirow{2}{*}{$a_{\tau\tau}$} & $[-106.5, -57.7]$ & $[-118.1, -59.7]$ & $[-103.0, -64.9]$ & $[-135.8, -52.1]$ & $[-100.8, -74.1]$ \\ 
 & & & & & \\ 
\hline
\multirow{2}{*}{$|a_{e\mu}|$} & $[-104.4, -71.1]$ & $[-119.7, -66.6]$ & $[-101.9, -74.9]$ & $[-117.5, -59.0]$ & $[-101.5, -76.4]$ \\ 
 & & & & & \\ 
\hline
\multirow{2}{*}{$|a_{e\tau}|$} & $[-108.3, -69.6]$ & $[-114.5, -66.4]$ & $[-102.7, -74.1]$ & $[-117.5, -59.2]$ & $[-100.5, -76.5]$ \\ 
 & & & & & \\ 
\hline
\multirow{2}{*}{$|a_{\mu\tau}|$} & $[-112.4, -67.6]$ & $[-122.2, -65.4]$ & $[-108.3, -72.3]$ & $[-126.9, -59.2]$ & $[-100.8, -75.6]$ \\ 
 & & & & & \\ 
  \hline
\end{tabular}
\renewcommand{\arraystretch}{1.0} 
\caption{Projected 95\% C.L. allowed ranges for $\delta_{CP}$ in the presence of LIV parameters.}
\label{tab:liv_dcp_corr_dcp_final_cubic}
\end{table}


We can see a striking feature emerging from the correlation analysis about the superior performance of the ESSnuSB 360 km baseline in preserving the precision of the $\delta_{CP}$ measurement against LIV-induced distortions. While DUNE generally provides tighter constraints on the magnitude of the LIV parameters due to its long baseline and matter effects, ESSnuSB consistently yields narrower allowed intervals for the phase $\delta_{CP}$. For instance, in the presence of the parameter $a_{\tau\tau}$, the ESSnuSB (360 km) setup restricts $\delta_{CP}$ to an interval of approximately $46^\circ$ width ($[-105.6^\circ, -59.0^\circ]$), whereas the DUNE interval spans roughly $66^\circ$ ($[-121.7^\circ, -55.6^\circ]$). A similar trend is observed for $a_{ee}$ and $a_{\mu\mu}$. This may be attributed to the fact that ESSnuSB operates at the second oscillation maximum, where the intrinsic CP asymmetry is approximately three times larger than at the first maximum probed by DUNE. This amplified CP signal makes the measurement more strict against new physics effects, ensuring that even if LIV parameters are non-zero, they are less likely to mimic a false CP value at ESSnuSB than at DUNE. Table \ref{tab:liv_dcp_corr_dcp_final_cubic} summarises the results from figure \ref{fig:deltaCP_degeneracy}, showing the allowed ranges of $\delta_{CP}$, when the data are generated assuming exact Lorentz invariance, with the true value of $\delta_{CP} = -90^{\circ}$.

\begin{figure}[!h]
  \centering
  \begin{subfigure}[b]{0.48\textwidth}
    \centering
    \includegraphics[width=\textwidth]{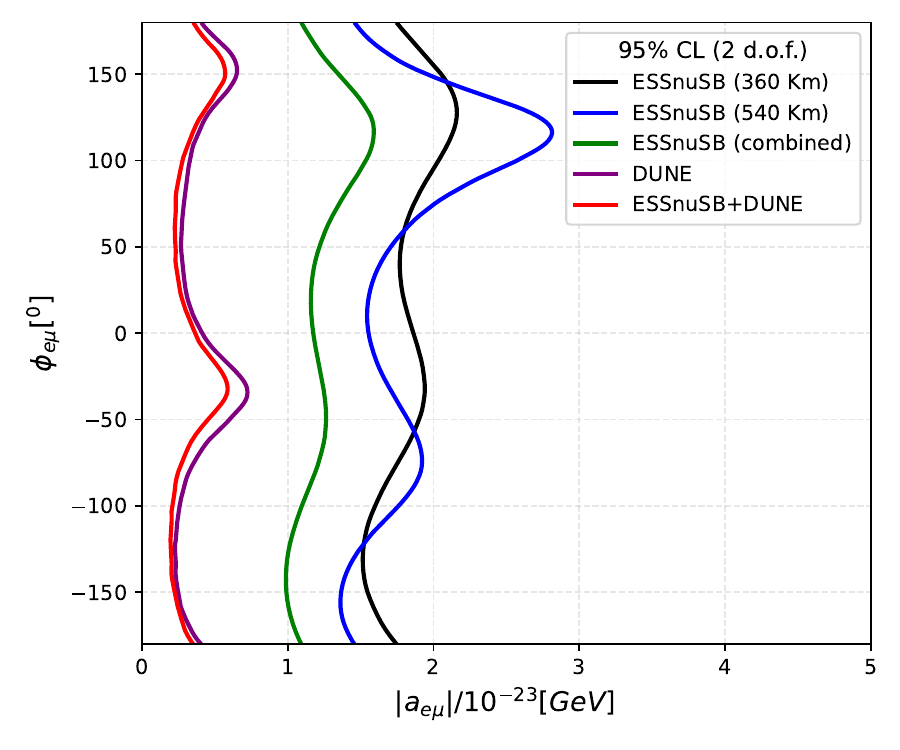}
    \label{fig:subA}
  \end{subfigure}
  \begin{subfigure}[b]{0.48\textwidth}
    \centering
    \includegraphics[width=\textwidth]{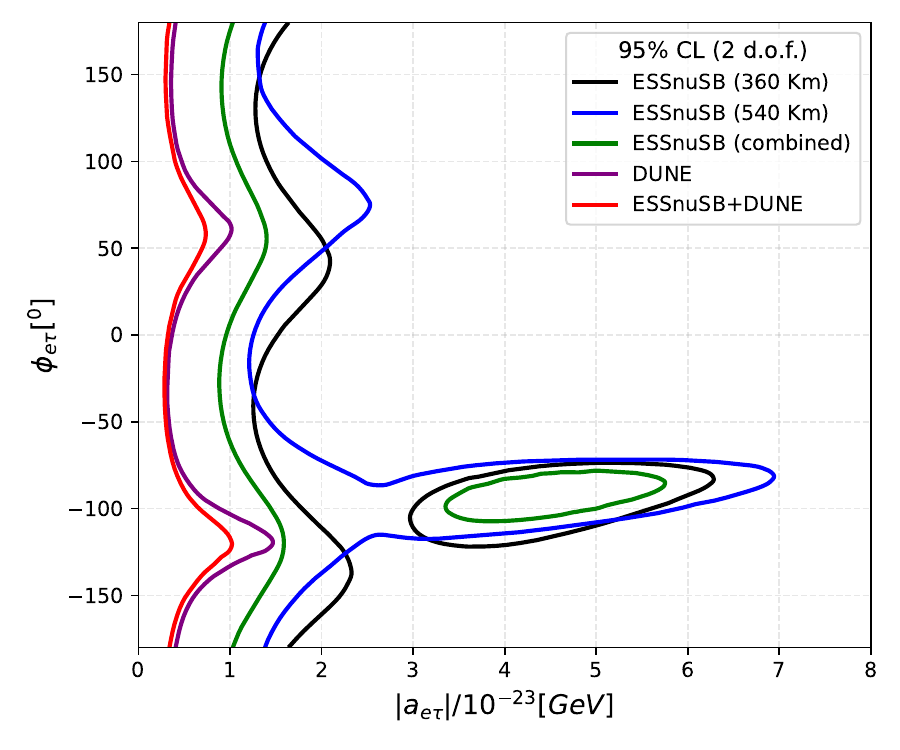}
    \label{fig:subB}
  \end{subfigure}
  \\[-4ex]

  \begin{subfigure}[b]{0.48\textwidth}
    \centering
    \includegraphics[width=\textwidth]{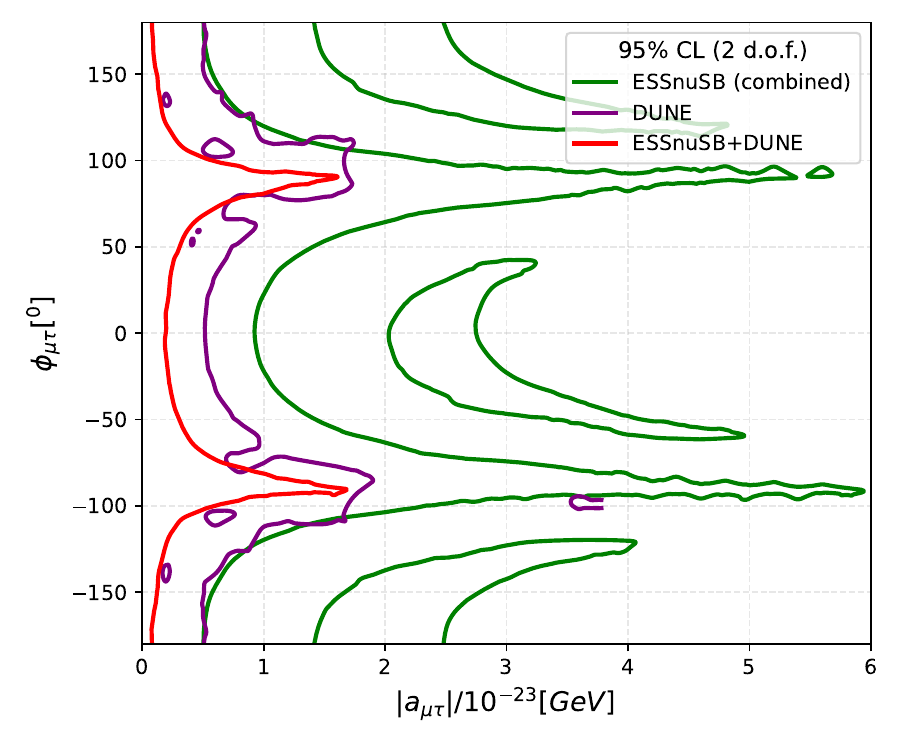}
    \label{fig:subC}
  \end{subfigure}

  \caption{Allowed regions at 95\% C.L. (2 d.o.f.) in the plane of each LIV parameter versus the ne LIV phase $\phi_{\alpha\beta}$. The contours represent the sensitivity for ESSnuSB at BL=360 km (black), ESSnuSB at BL=540 km (blue), the combined ESSnuSB setup (green), DUNE (purple), and the combined analysis of ESSnuSB and DUNE (red). The true value of all LIV parameters is assumed to be zero, with a true value of $\delta_{CP}$ near maximal CP violation.}
  \label{fig:livphase_degeneracy}
\end{figure}

In Figure \ref{fig:livphase_degeneracy}, we present the sensitivities to the off-diagonal LIV parameters, shown at 95$\%$ C.L. (2 d.o.f) as functions of their respective phases. Throughout the analysis, the true values of the standard oscillation parameters are kept fixed, with $\delta_{CP} = -90^\circ$ and normal mass ordering. We find that the combined ESSnuSB configuration provides noticeably better sensitivity than the individual baselines for $|a_{e\mu}|$ and $|a_{e\tau}|$. DUNE exhibits even stronger sensitivity compared to the combined ESSnuSB setup, and thus their combined analysis shows a significant improvement over ESSnuSB combined. For $|a_{\mu\tau}|$, the sensitivities of ESSnuSB at 360 km and 540 km are too weak to be presented. The appearance of the island structured regions at ESSnuSB around $|a_{e\tau}| = 4.5 \times 10^{-23}$ GeV and $\phi_{e\tau} = -90^\circ$ can be explained from the local minima observed in figure~\ref{fig:chi2} at $|a_{e\tau}| = 4.5 \times 10^{-23}$ GeV.

\newpage
\section{Summary \& Concluding Remarks}\label{sec:conclusion}

In this work, we analyzed the individual potential of the ESSnuSB long-baseline experiment's two proposed baselines (360 km to Zinkgruvan and 540 km to Garpenberg) to constrain CPT-violating isotropic Lorentz Invariance Violation (LIV) parameters $a_{\alpha\beta}$ and explored the synergetic potential of the two when combined with the longer 1300 km DUNE baseline. Table \ref{table:all-exp} compares our findings with the limits on LIV parameters ($a_{\alpha\beta}$) from available literatures. Since the non-zero LIV parameters critically introduce parameter degeneracies leading to effects that can mimic the signatures of standard oscillation parameters, we may find ``fake solutions" where a true signal of new physics could be misinterpreted as a standard parameter measurement, or conversely, where standard physics could mask a LIV signal.

In our investigation, we found that the resolution of these degeneracies through the synergy between DUNE and ESSnuSB is possible. It is largely due to the fact that DUNE, with its 1300 km baseline and broadband beam centered at the first oscillation maximum, provides high statistics and strong sensitivity to matter effects while ESSnuSB, operating at shorter baselines (360 km and 540 km) near the second oscillation maximum, offers a complementary probe with a different $L/E$ dependence and reduced matter effects. Since the LIV contribution to the Hamiltonian is energy-independent while the standard mass term scales as $1/E$, the combination effectively disentangles the two effects. We saw that the combined analysis successfully eliminates the ``fake" octant solutions for $\theta_{23}$ that appeared in the individual ESSnuSB analysis (e.g., for $a_{\mu\mu}$). No octant degeneracy apeared for any LIV parameter when a synergetic approach was taken by combining all the available baselines from ESSnuSB and DUNE. 

A remarkable finding from our correlation analysis is the superior performance of ESSnuSB in preserving the precision of the $\delta_{CP}$ measurement over DUNE. As detailed in Table \ref{tab:liv_dcp_corr_dcp_final_cubic}, the ESSnuSB 360 km baseline consistently provides tighter constraints on $\delta_{CP}$ compared to DUNE for all considered LIV parameters. For instance, in the presence of $a_{\mu\mu}$, the allowed $\delta_{CP}$ range for ESSnuSB is nearly half the width of that for DUNE. This is also a direct consequence of ESSnuSB operating at the second oscillation maximum, where the intrinsic CP asymmetry is significantly larger than at the first maximum. This confirms that although high-energy, long-baseline data (DUNE) is essential for constraining matter-dependent LIV effects, the high-intensity data at the second maximum (ESSnuSB) is indispensable for eliminating the CP phase against ``fake" solutions.

While deriving the projected upper bounds on the LIV coefficients at 95$\%$ C.L., we found that combined setup gave us limits of order $\mathcal{O}(10^{-24})$ GeV for off-diagonal parameters (e.g., $|a_{e\mu}| < 0.5 \times 10^{-23}$ GeV) and $\mathcal{O}(10^{-23})$ GeV for diagonal parameters. These projected constraints are competitive with, and in some cases superior to, current limits from atmospheric neutrino experiments (Super-Kamiokande, IceCube) as highlighted in \ref{table:all-exp}. It is also seen that adding the 540km baseline to our analysis and marginalization over a broader range of standard oscillation parameter values gave us slightly better constraints on LIV parameters over the similar setup where only a single baseline (360 km) from ESSnuSB was considered \cite{Delgadillo:2025sme}. The inclusion of the 540km baseline in the analysis significantly improved the constraint on $a_{e\mu}$ making it competitive with the DUNE+P2O setup and ICAL+T2HK+DUNE combined setup.

\section*{Acknowledgement}
DD thanks Monojit Ghosh for providing the GLoBES files for the ESSnuSB detector on behalf of the ESSnuSB collaboration. DD also acknowledges INSA for awarding the IASc–INSA–NASI Focus Area Science Technology Summer Fellowship, which enabled a visit to the Physical Research Laboratory, Ahmedabad, where part of this work was carried out. 

\clearpage
\begin{table*}[htbp]
\centering
\renewcommand{\arraystretch}{0.95} 
\setlength{\tabcolsep}{6pt} 
\scalebox{0.9}{
\begin{tabular}{|p{9.5cm}|p{1.5cm}|p{4.0cm}|} 
\hline
\textbf{Experiment Details} & \textbf{Param.} & \textbf{95\% C.L. Limit ($10^{-23}$ GeV)} \\
\hline \hline

\multirow{3}{=}{SK (Atmospheric) \cite{Super-Kamiokande:2014exs}} 
    & $a_{e\mu}$     & $1.8$  \\
    & $a_{e\tau}$    & $2.8$  \\
    & $a_{\mu\tau}$  & $0.51$  \\
\hline

\multirow{2}{=}{IceCube   \cite{IceCube:2017qyp}} 
    & $a_{\mu\tau}$  & $0.2 ^{\dagger}$ \\ 
    & $a_{\tau\tau}$ & $0.002 ^{\dagger}$ \\
\hline

\multirow{3}{=}{ICAL \\ (Atmospheric neutrino simulated, 1--25 GeV range) \cite{2110}} 
    & $a_{e\mu}$     & $1.34$ \\
    & $a_{e\tau}$    & $1.58$ \\
    & $a_{\mu\tau}$  & $0.22$ \\
\hline

\multirow{5}{=}{DUNE \\ (Long-baseline neutrino simulated, 0.2--10 GeV range) \cite{Barenboim:2018ctx}} 
    & $a_{e\mu}$     & $0.7$ \\
    & $a_{e\tau}$    & $1.0$ \\
    & $a_{\mu\tau}$  & $1.7$ \\
    & $a_{ee}$      & $(-2.5,3.2)$ \\
    & $a_{\mu\mu}$    & $(-3.7,2.8)$ \\
\hline

\multirow{6}{=}{T2K + NOVA \\ (Simulation for LIV sensitivity) \cite{Majhi:2019tfi}} 
    & $a_{e\mu}$     & $3.6$ \\
    & $a_{e\tau}$    & $10.8$ \\
    & $a_{\mu\tau}$  & $8.0$ \\
    & $a_{ee}$       & $(-55, 34)$ \\
    & $a_{\mu\mu}$   & $(-10.7, 11.8)$ \\\
    & $a_{\tau\tau}$ & $(-11.2, 9.0)$ \\
\hline

\multirow{5}{=}{DUNE + P2O \\ (Long-baseline simulation for LIV) \cite{Fiza:2022xfw}} 
    & $a_{e\mu}$     & $0.47$ \\
    & $a_{e\tau}$    & $0.6$ \\
    & $a_{\mu\tau}$  & $1.3$ \\
    & $a_{ee}$       & $(-2.6, 3.3)$ \\
    & $a_{\mu\mu}$   & $(-1.5, 1.6)$ \\
\hline

\multirow{6}{=}{ICAL + T2HK + DUNE \\ (Combined atmospheric and LBL) \cite{Raikwal:2023lzk}} 
    & $a_{e\mu}$     & $0.4$  \\
    & $a_{e\tau}$    & $0.6$  \\
    & $a_{\mu\tau}$  & $0.95$  \\
    & $a_{ee}$       & $(-2.1,2.1)$ \\
    & $a_{\mu\mu}$   & $(-1.8,1.9)$ \\
    & $a_{\mu\mu}$   & $(-1.5,1.5)$ \\
\hline

\multirow{6}{=}{DUNE + ESSnuSB \\ (Combined analysis, 1300km DUNE + 360 km ESSnuSB) \cite{Delgadillo:2025sme}} 
    & $a_{e\mu}$     & $0.9$ \\
    & $a_{e\tau}$    & $1.6$ \\
    & $a_{\mu\tau}$  & $0.8$ \\
    & $a_{ee}$       & $(-1.3, 1.4)$ \\
    & $a_{\mu\mu}$   & $(-1.1, 0.9)$ \\
    & $a_{\tau\tau}$ & $(-0.9, 0.9)$ \\
\hline

\multirow{6}{=}{\textbf{DUNE + ESSnuSB (Full Setup)} \\ (Combined DUNE + 360 km + 540 km, Global marginalization) (\textbf{This Work})} 
    & $a_{e\mu}$     & $0.43$ \\
    & $a_{e\tau}$    & $0.75$ \\
    & $a_{\mu\tau}$  & $1.26$ \\
    & $a_{ee}$       & $(-1.63, 1.91)$ \\
    & $a_{\mu\mu}$   & $(-1.66, 2.08)$ \\
    & $a_{\tau\tau}$ & $(-2.02, 1.32)$ \\
    
\hline
\end{tabular}
}
\caption{95\% C.L. limits on LIV parameters ($a_{\alpha\beta}$) from available literature and this work. All values are scaled to units of $10^{-23}$ GeV.  $^{\dagger}$ IceCube bounds are at 90\% C.L.}
\label{table:all-exp}
\end{table*}

\begin{appendix}

\section{Additional Plots} \label{Appendix_A}

\begin{figure}[H]
    \centering
    \includegraphics[width=0.3298\linewidth]{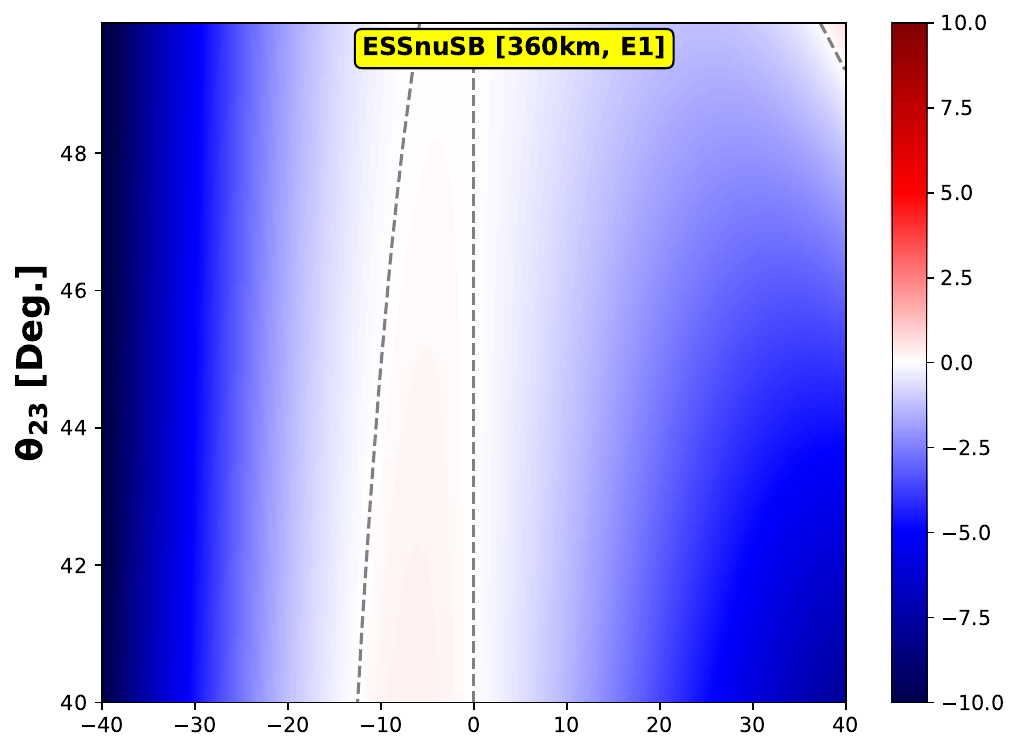}
    \includegraphics[width=0.3141\linewidth]{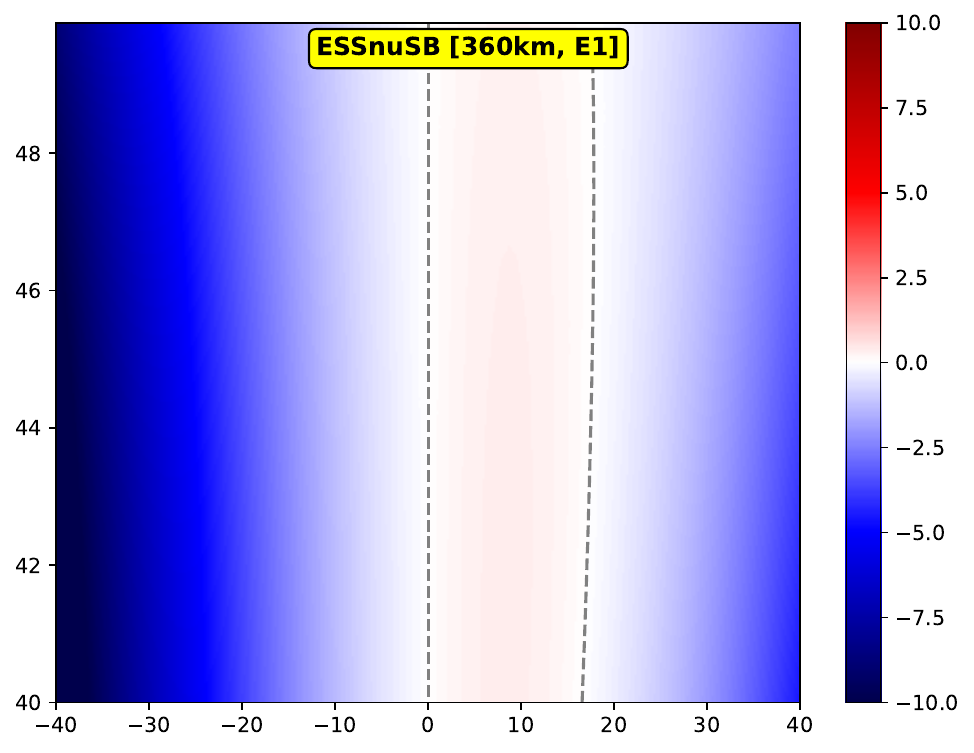}
    \includegraphics[width=0.3259\linewidth]{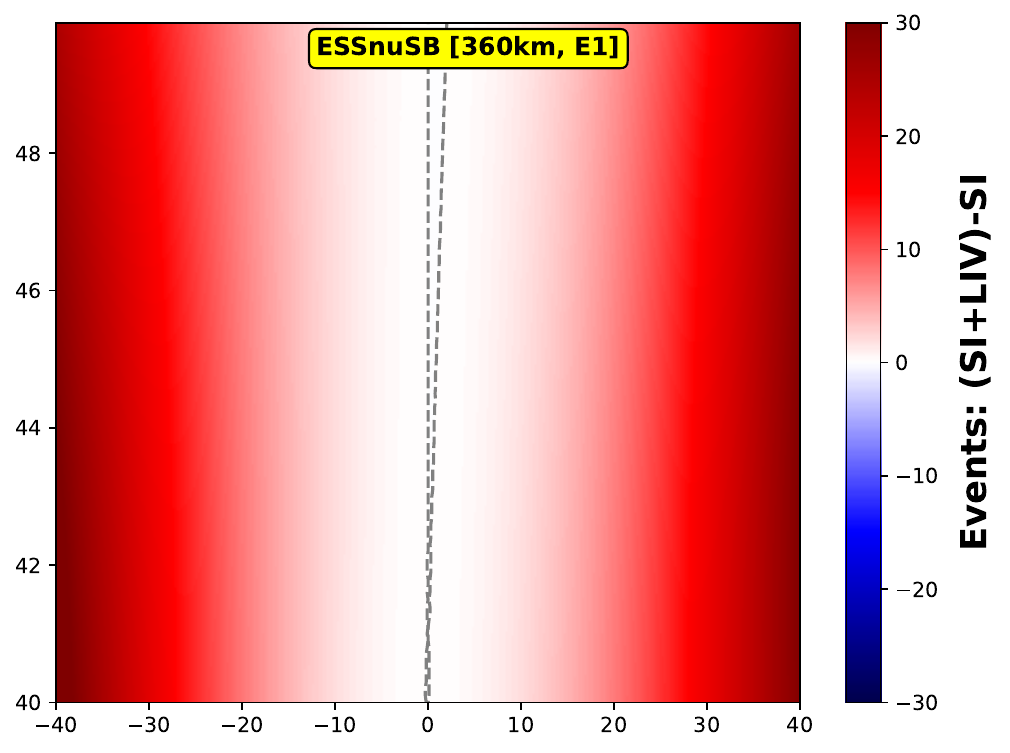}

    \includegraphics[width=0.3298\linewidth]{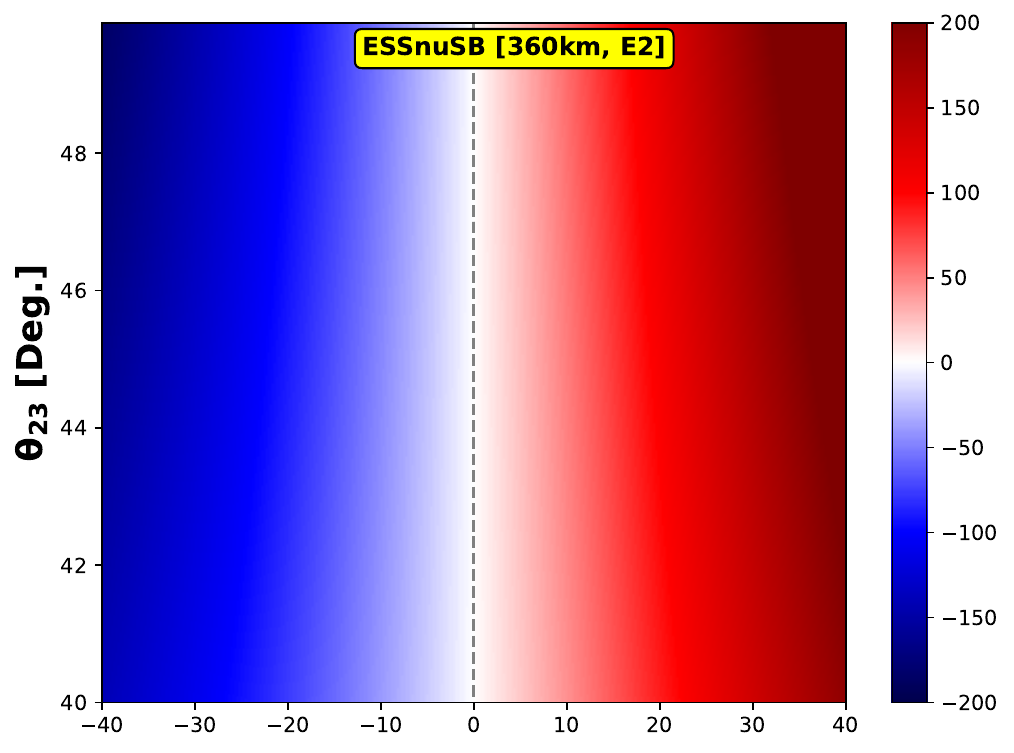}
    \includegraphics[width=0.3141\linewidth]{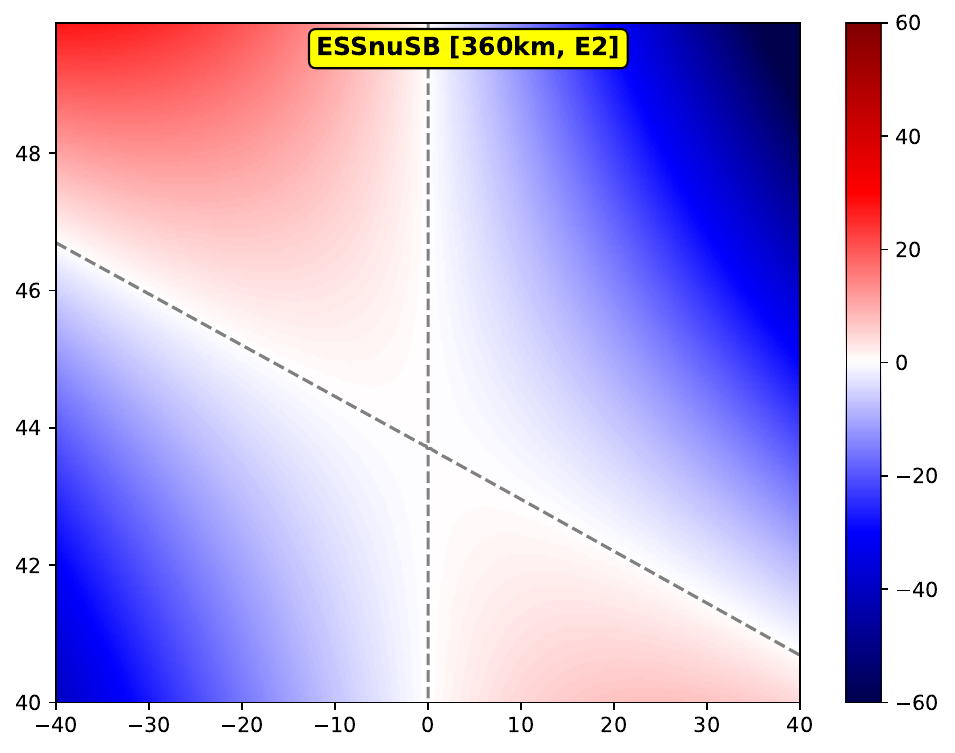}
    \includegraphics[width=0.3259\linewidth]{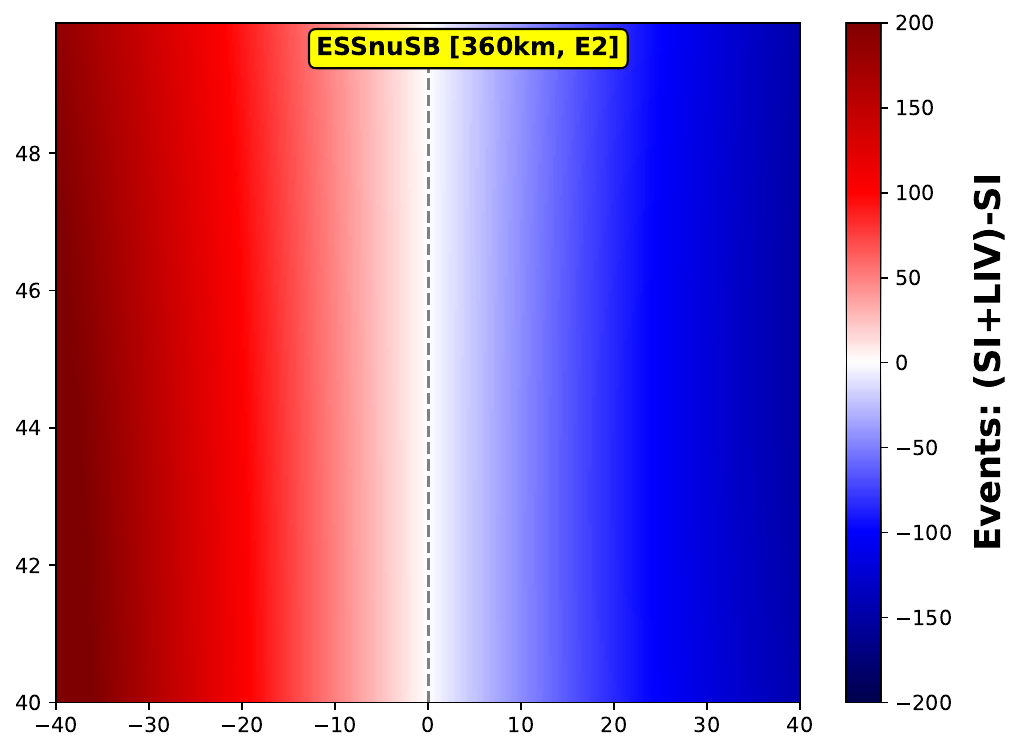}

    \includegraphics[width=0.3298\linewidth]{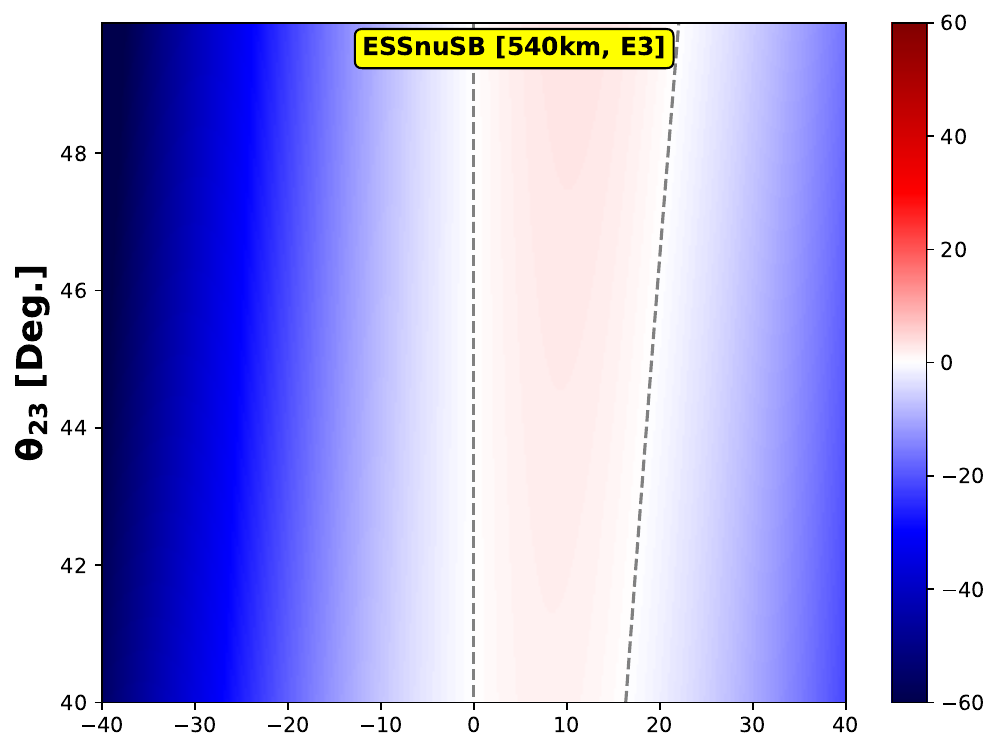}
    \includegraphics[width=0.3141\linewidth]{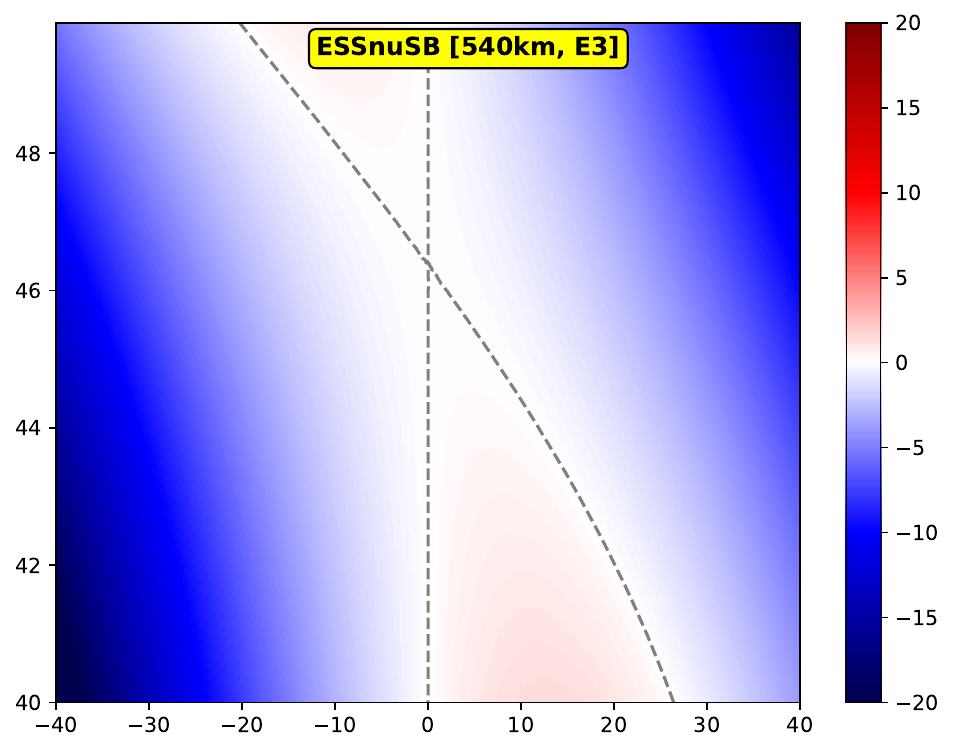}
    \includegraphics[width=0.3259\linewidth]{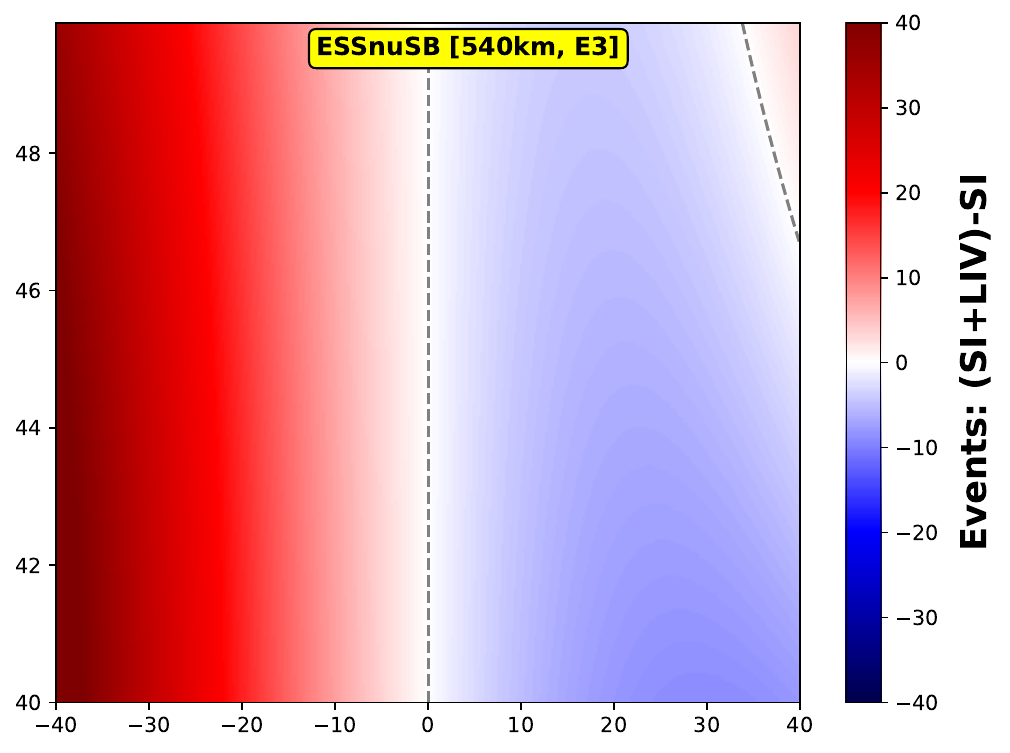}

    \includegraphics[width=0.3298\linewidth]{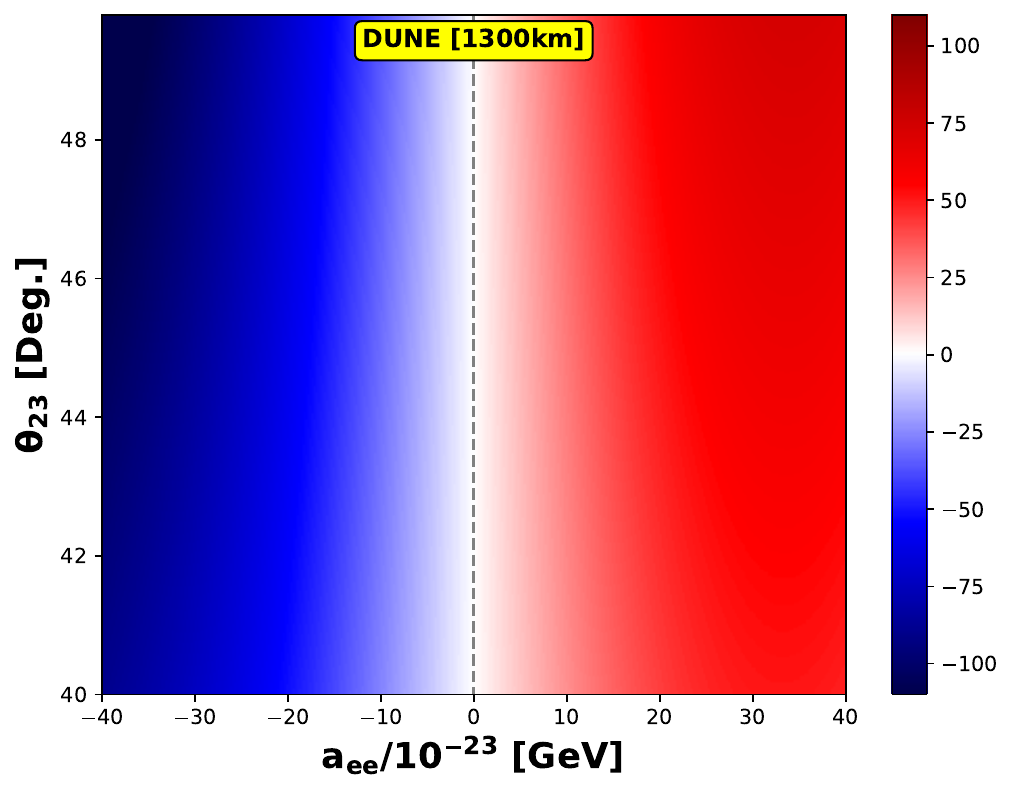}
    \includegraphics[width=0.3141\linewidth]{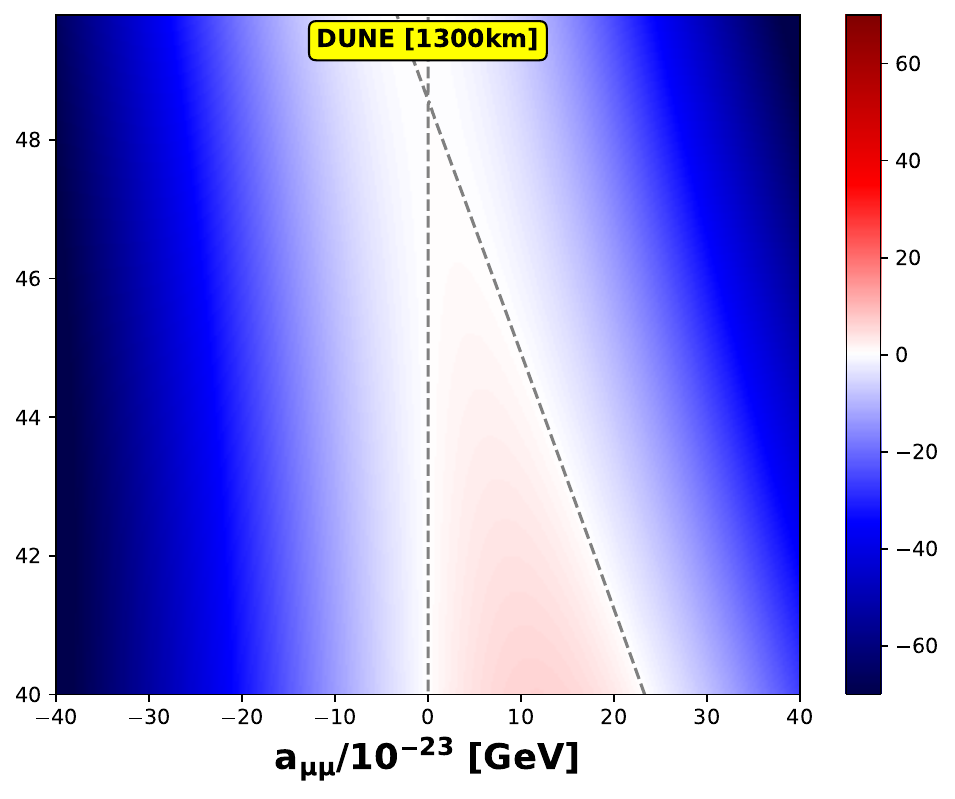}
    \includegraphics[width=0.3259\linewidth]{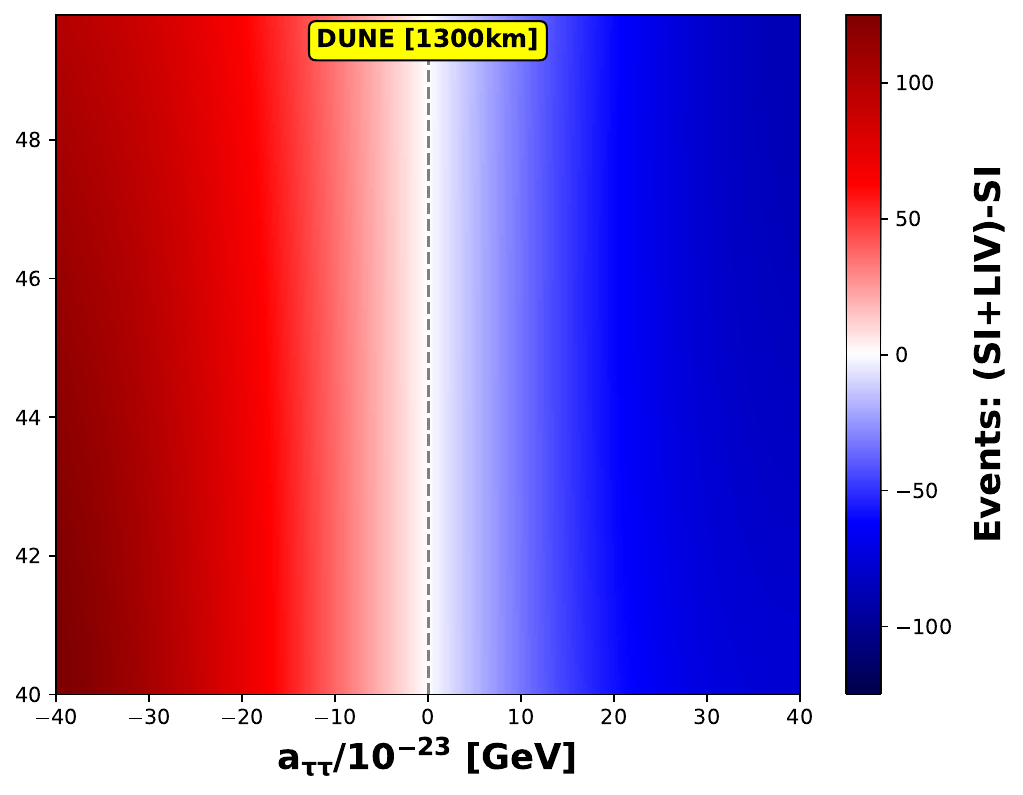}
    
    \caption{The heatmap of change in events ($\Delta N_{events}$ =  $N_{events}(SI+LIV) - N_{events}(SI)$) in $\theta_{23}$--$a_{\alpha\beta}$ plane for the diagonal LIV parameters. The top three panels are for ESSnuSB and the bottom panel is for the DUNE experiment. The E1, E2 and E3 correspond to energy values of 0.3 GeV, 0.6 GeV, and 2.5 GeV, respectively, which represent the peak values for the ESSnuSB.}
    \label{fig:dcp_2d_1}
 \end{figure}

 \begin{figure}[H]
    \centering
    \includegraphics[width=0.3298\linewidth]{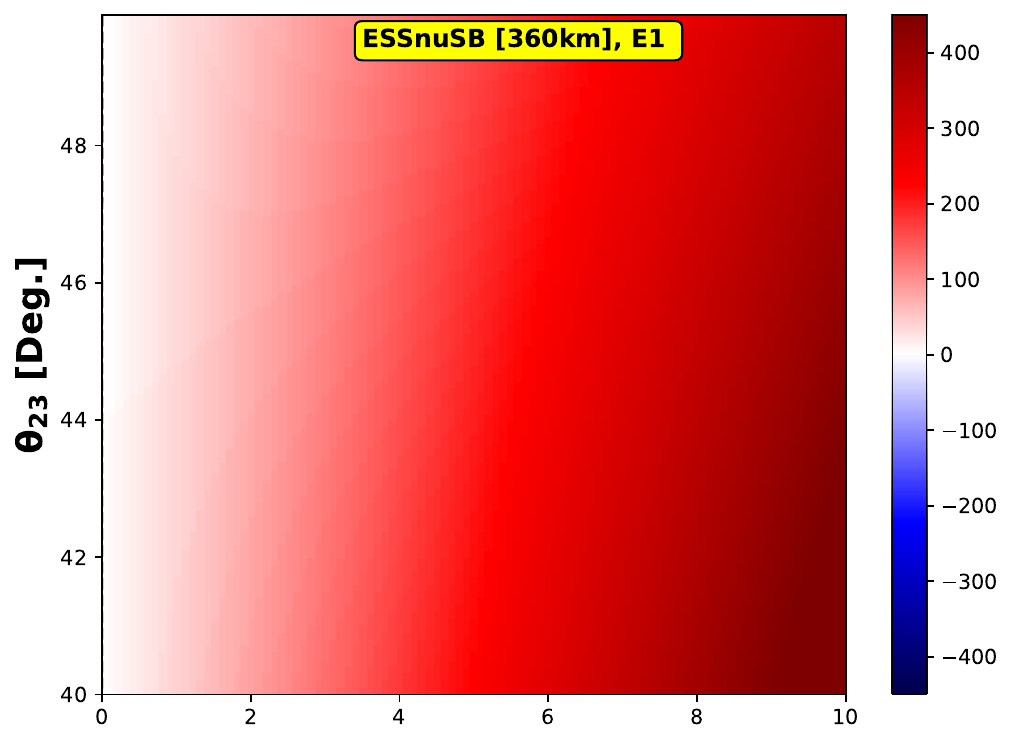}
    \includegraphics[width=0.3141\linewidth]{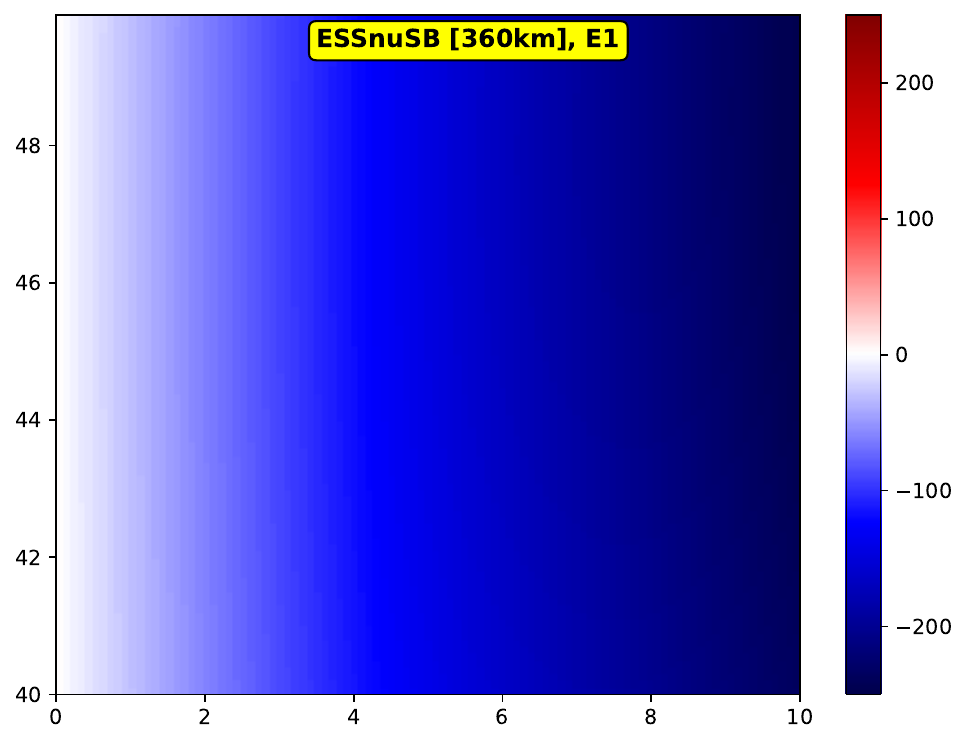}
    \includegraphics[width=0.3259\linewidth]{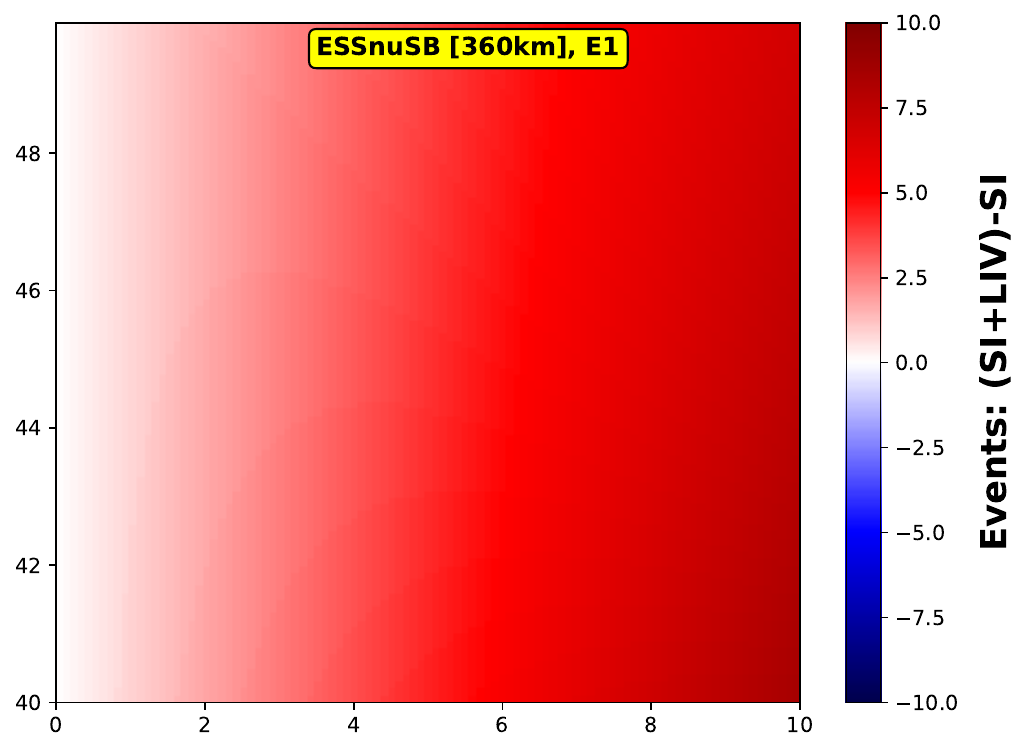}

    \includegraphics[width=0.3298\linewidth]{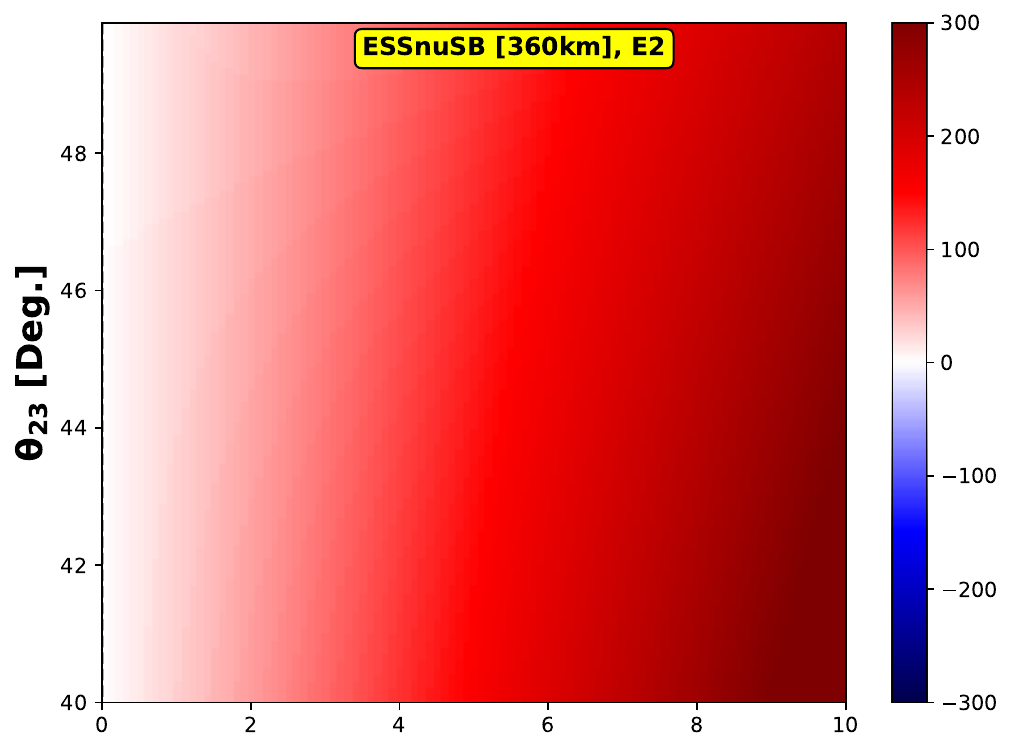}
    \includegraphics[width=0.3141\linewidth]{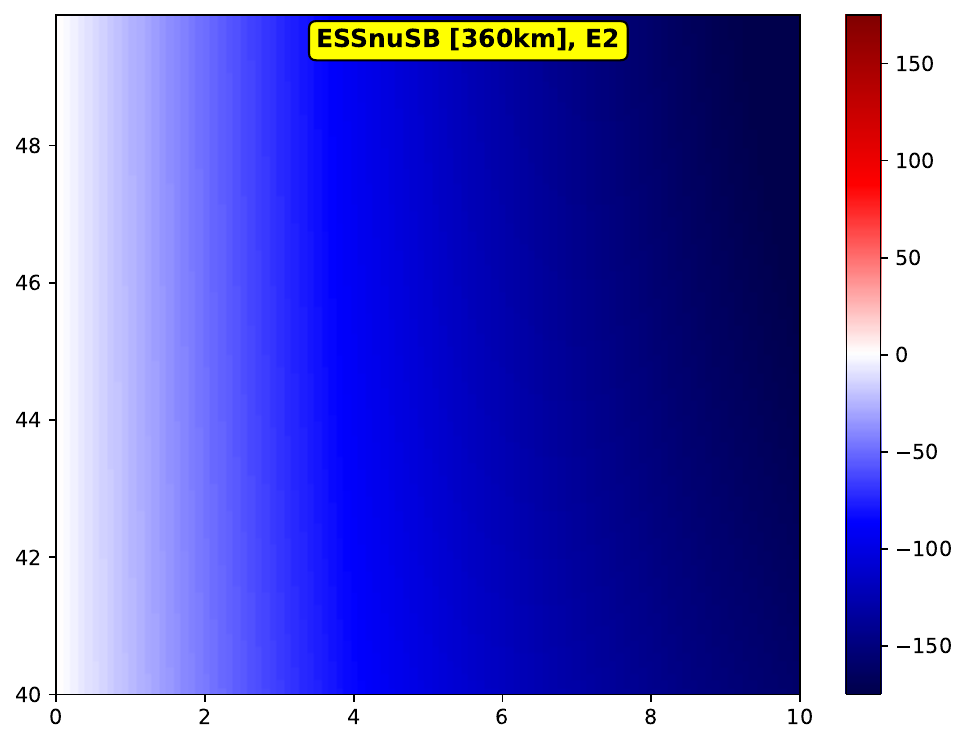}
    \includegraphics[width=0.3259\linewidth]{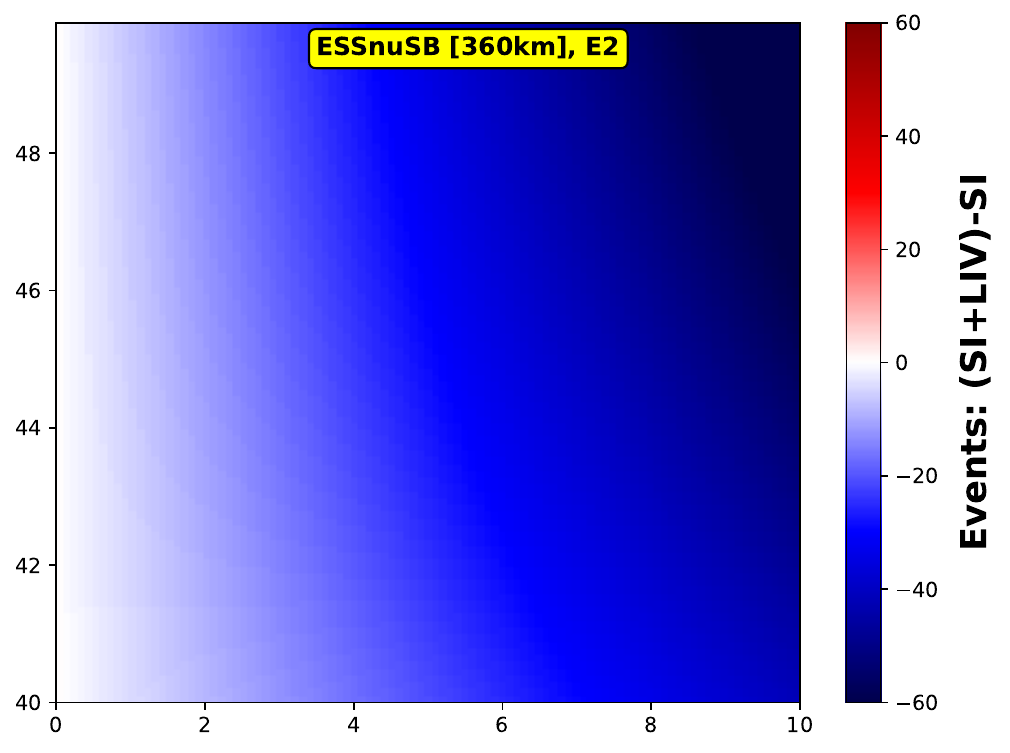}

    \includegraphics[width=0.3298\linewidth]{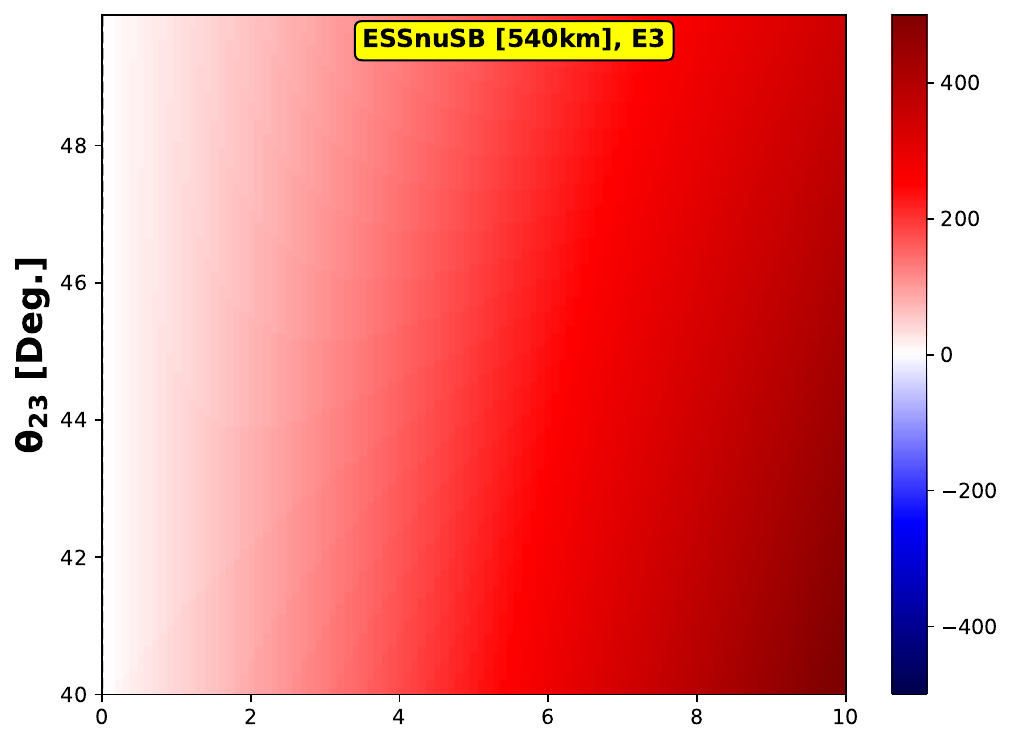}
    \includegraphics[width=0.3141\linewidth]{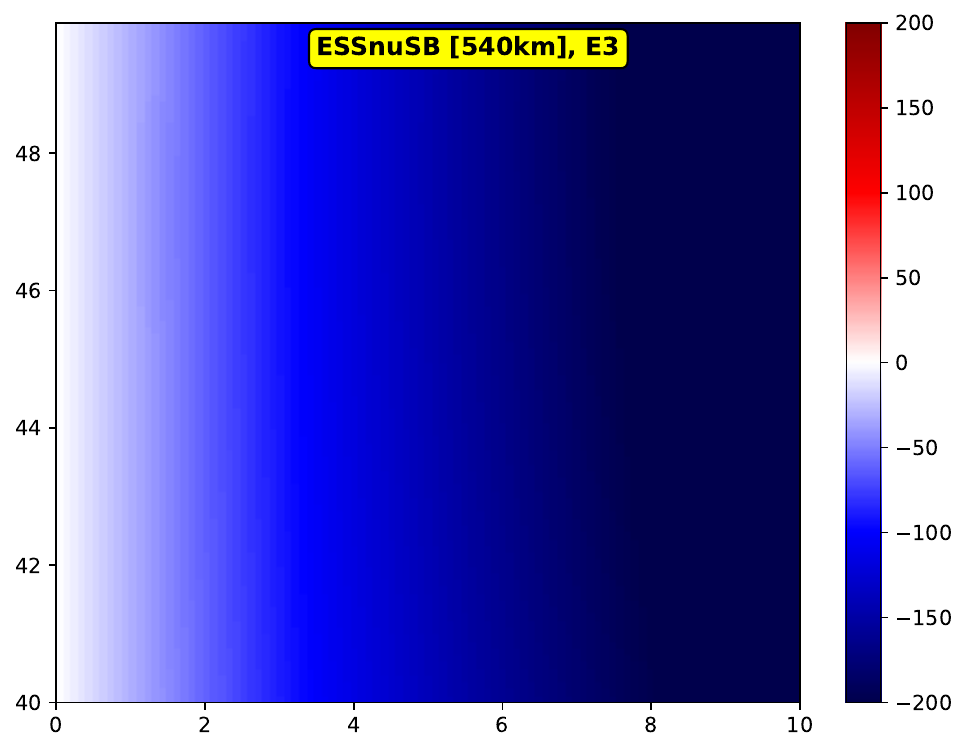}
    \includegraphics[width=0.3259\linewidth]{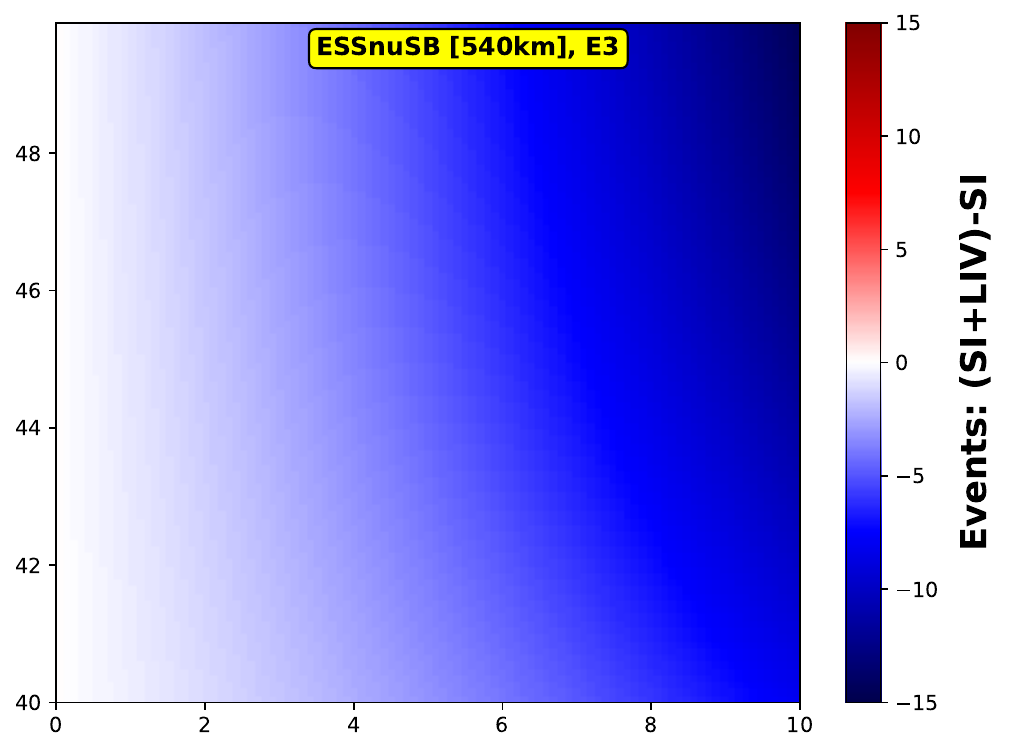}

    \includegraphics[width=0.3298\linewidth]{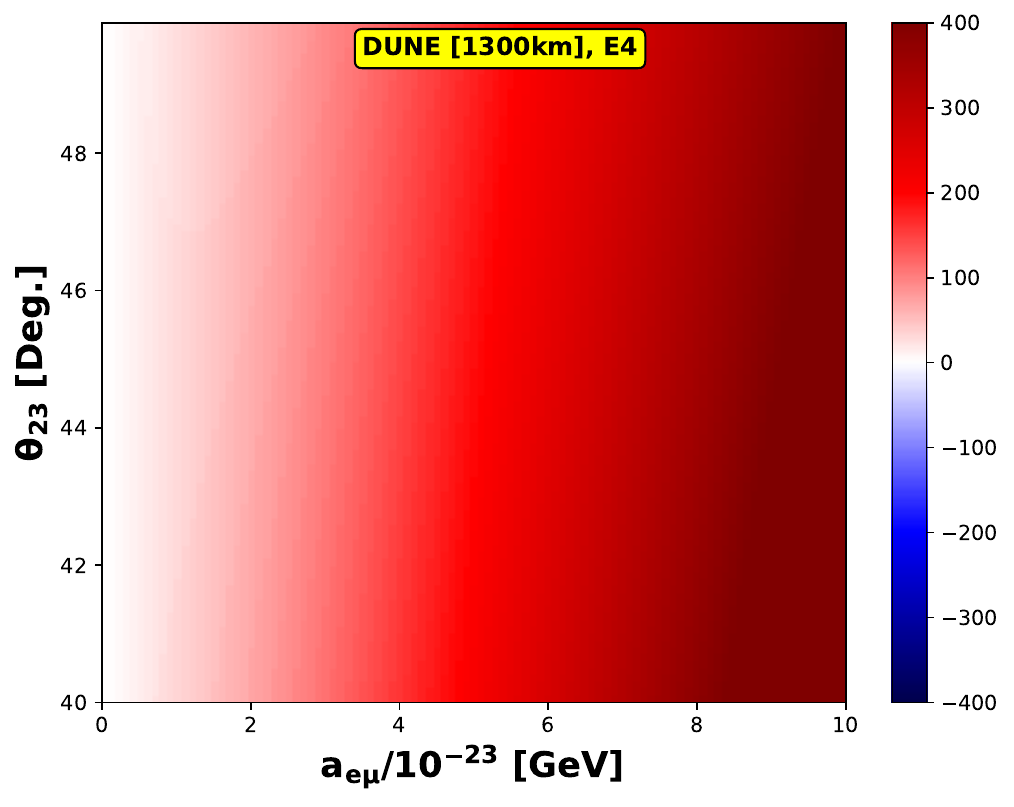}
    \includegraphics[width=0.3141\linewidth]{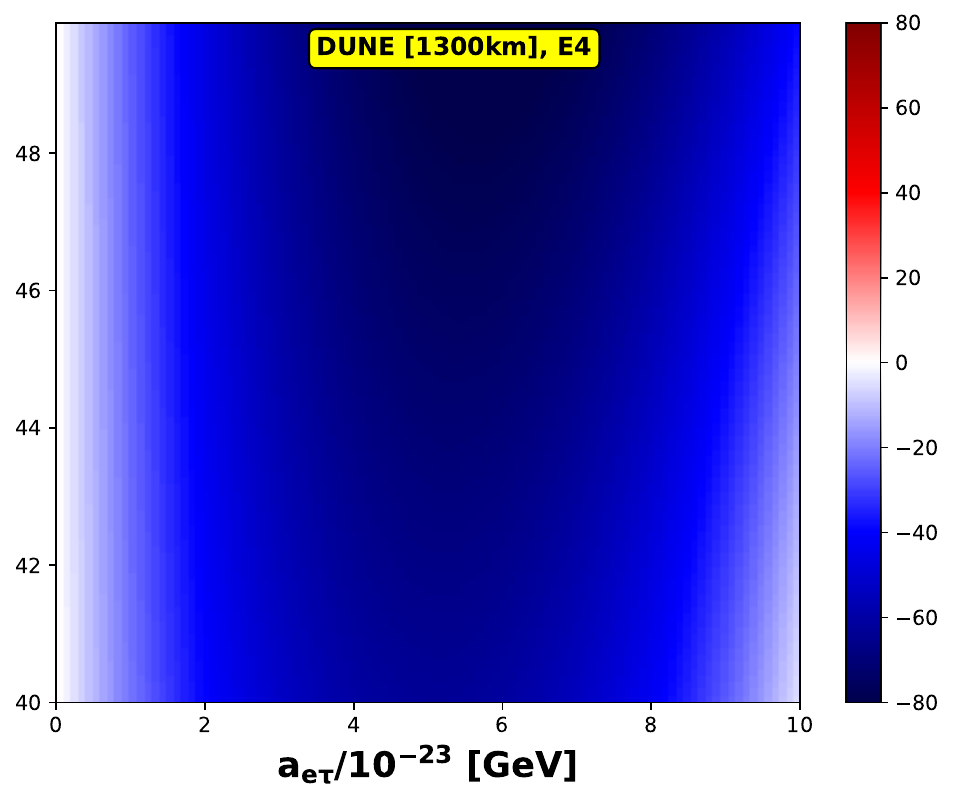}
    \includegraphics[width=0.3259\linewidth]{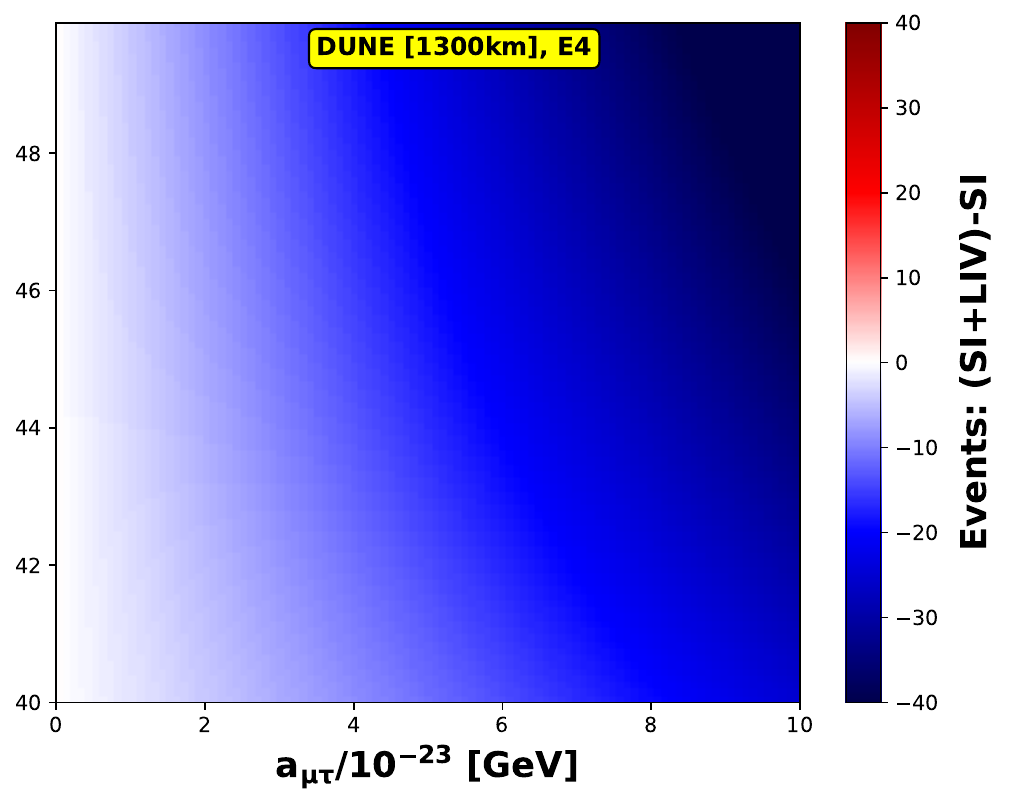}
    
    \caption{The heatmap of change in events ($\Delta N_{events}$ =  $N_{events}(SI+LIV) - N_{events}(SI)$) in $\theta_{23}$--$a_{\alpha\beta}$ plane for the off diagonal LIV parameters. The top three panels are for ESSnuSB and the bottom panel is for the DUNE experiment. The E1, E2 and E3 correspond to energy values of 0.3 GeV, 0.6 GeV, and 2.5 GeV, respectively, which represent the peak values for the ESSnuSB.}
    \label{fig:dcp_2d_2}
\end{figure}

\begin{figure}[H]
    \centering
    \includegraphics[width=0.3298\linewidth]{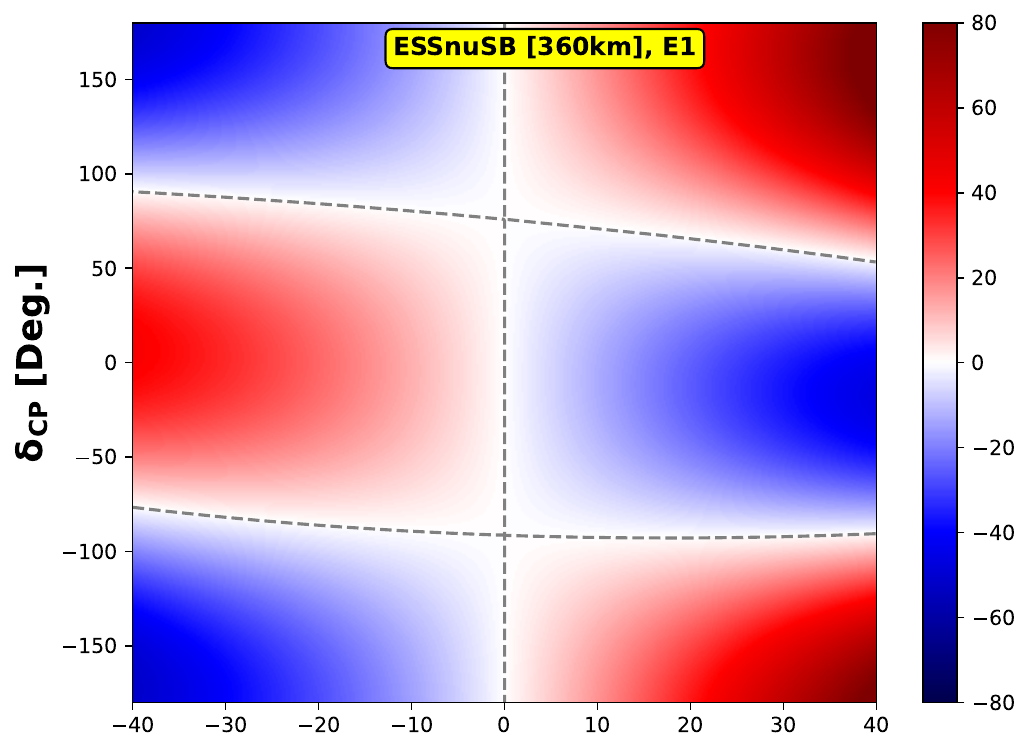}
    \includegraphics[width=0.3141\linewidth]{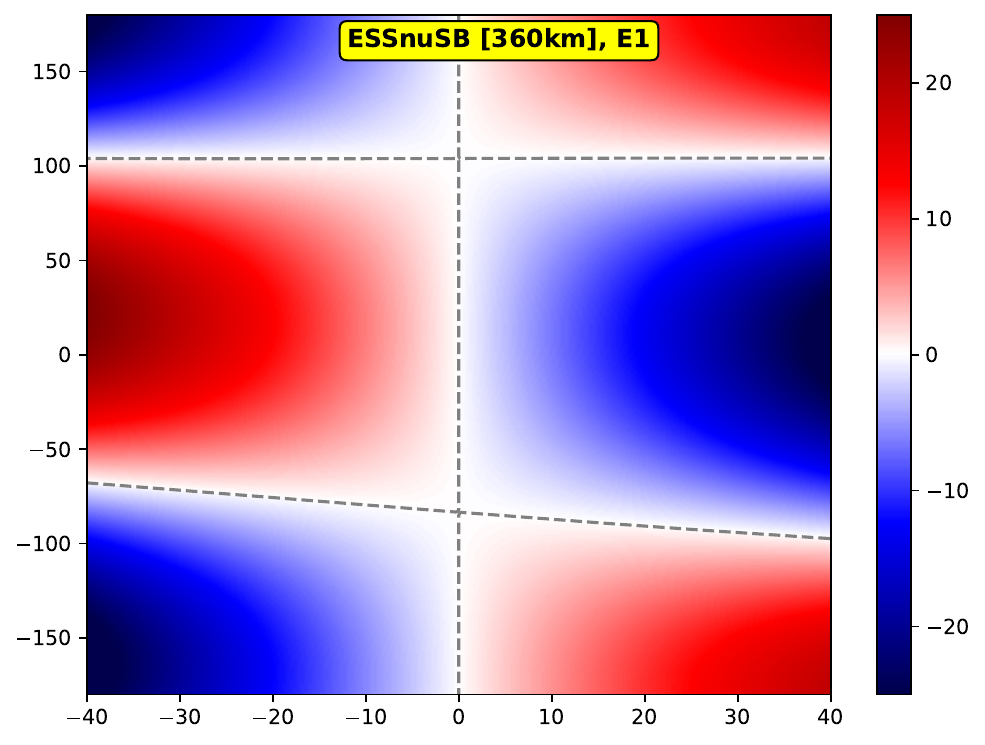}
    \includegraphics[width=0.3259\linewidth]{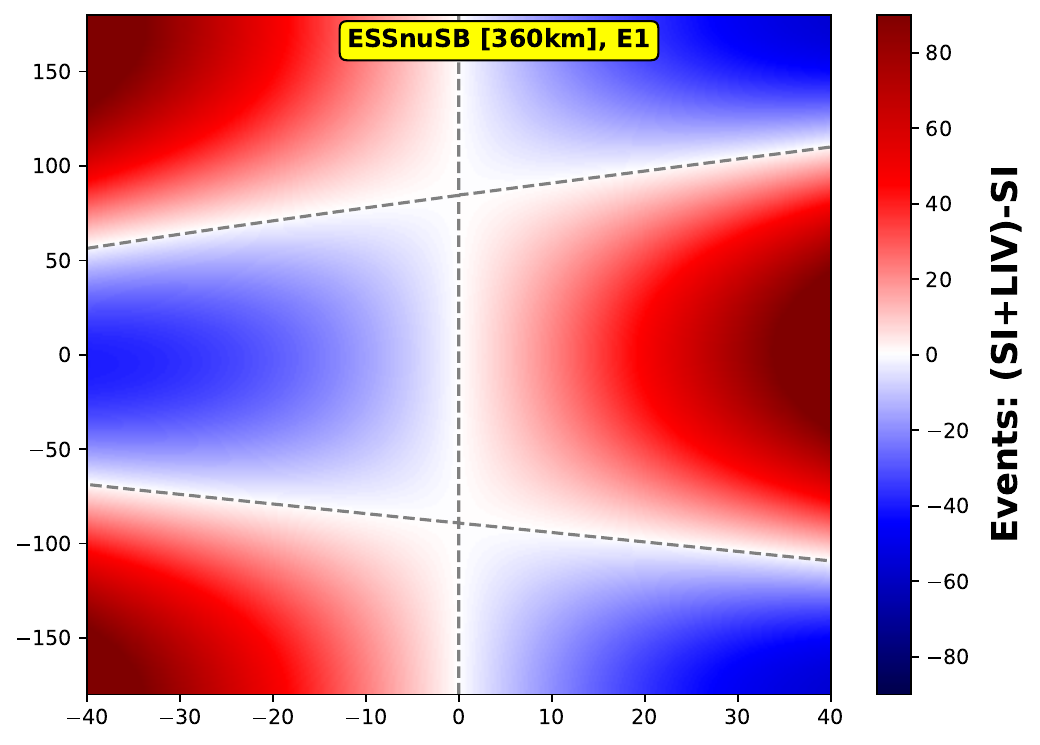}

    \includegraphics[width=0.3298\linewidth]{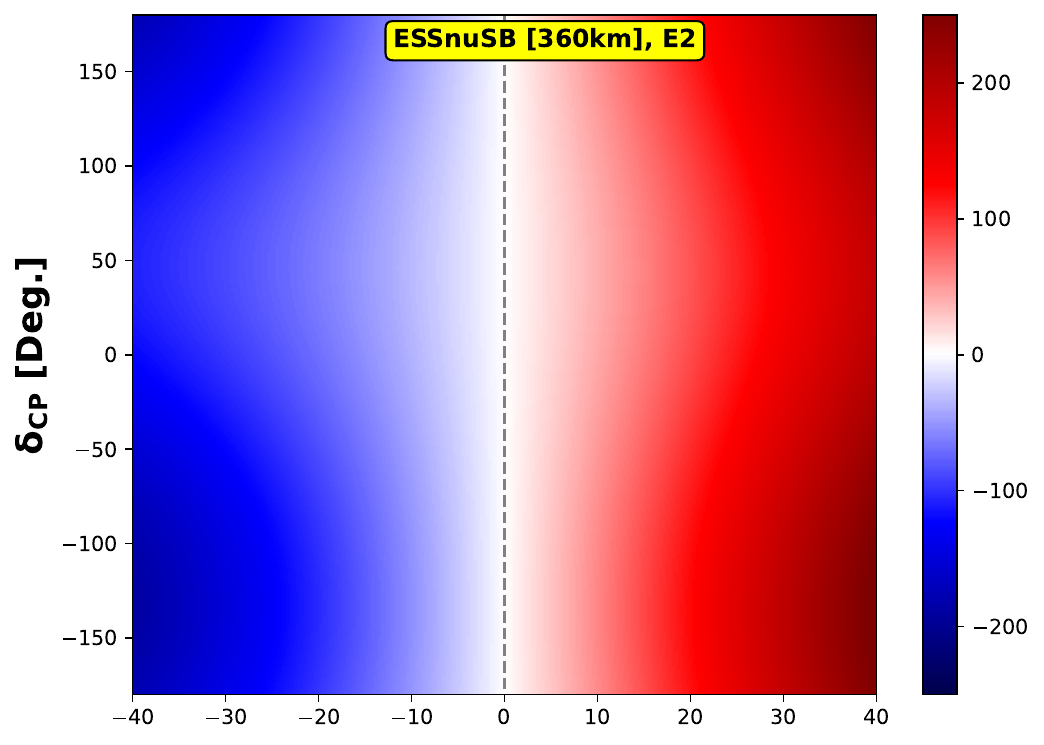}
    \includegraphics[width=0.3141\linewidth]{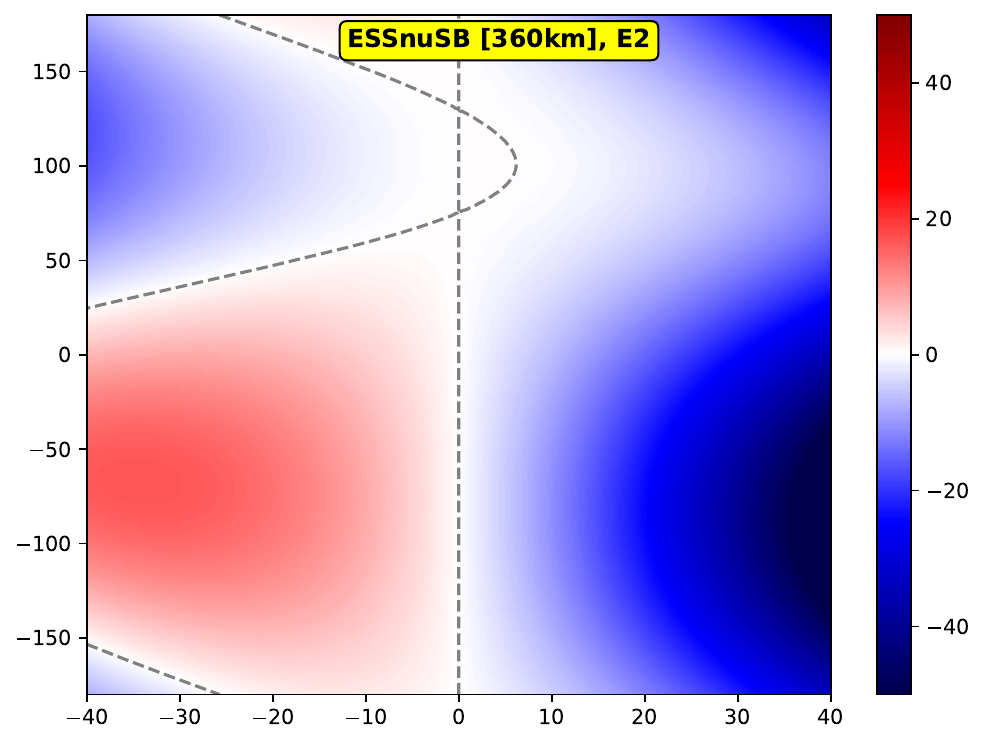}
    \includegraphics[width=0.3259\linewidth]{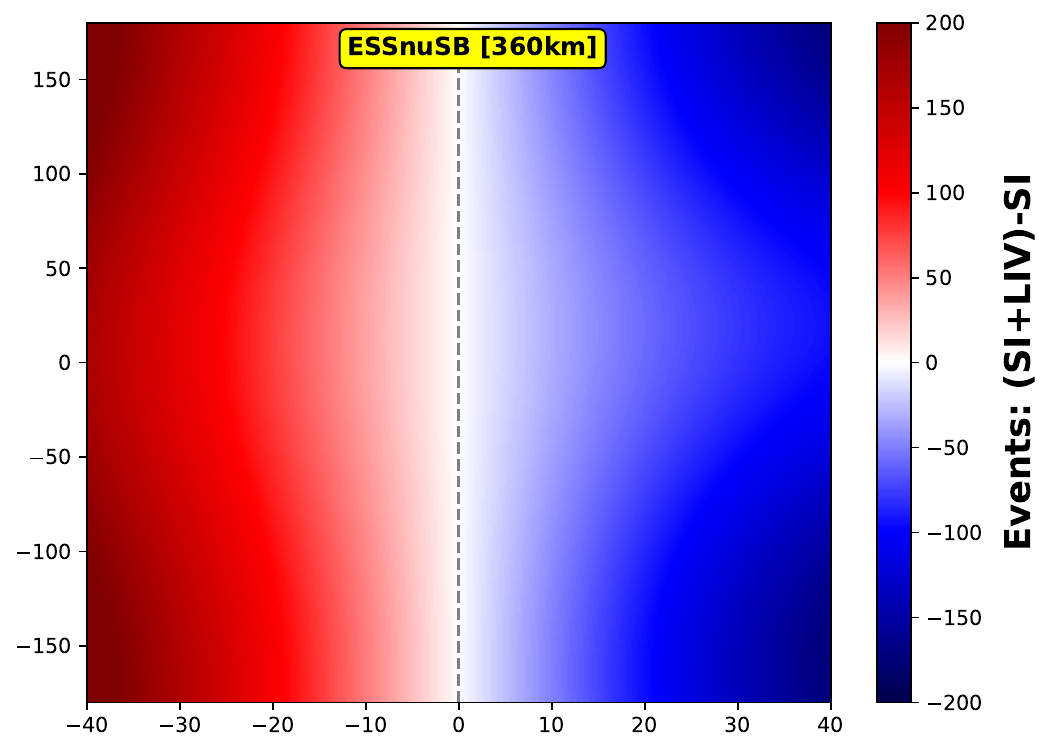}

    \includegraphics[width=0.3298\linewidth]{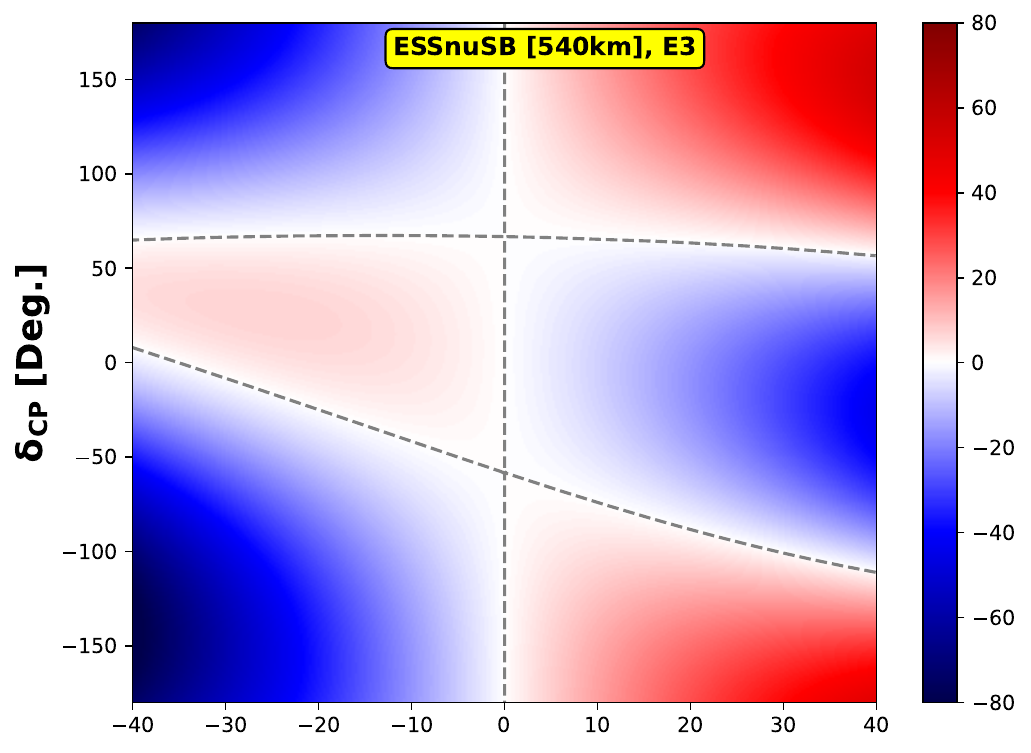}
    \includegraphics[width=0.3141\linewidth]{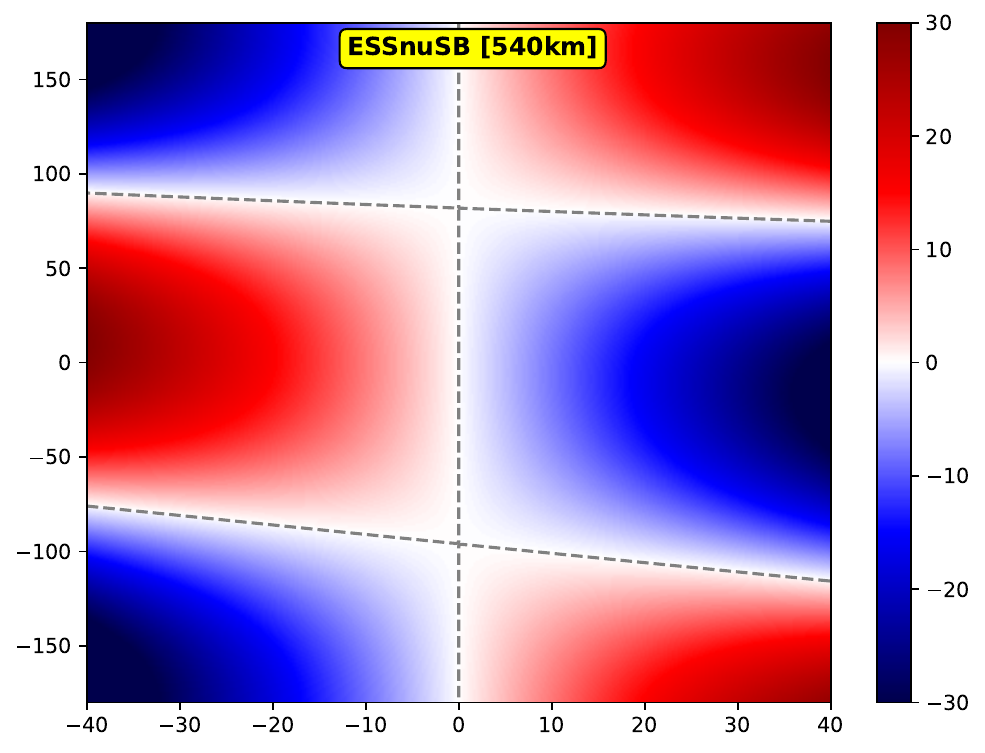}
    \includegraphics[width=0.3259\linewidth]{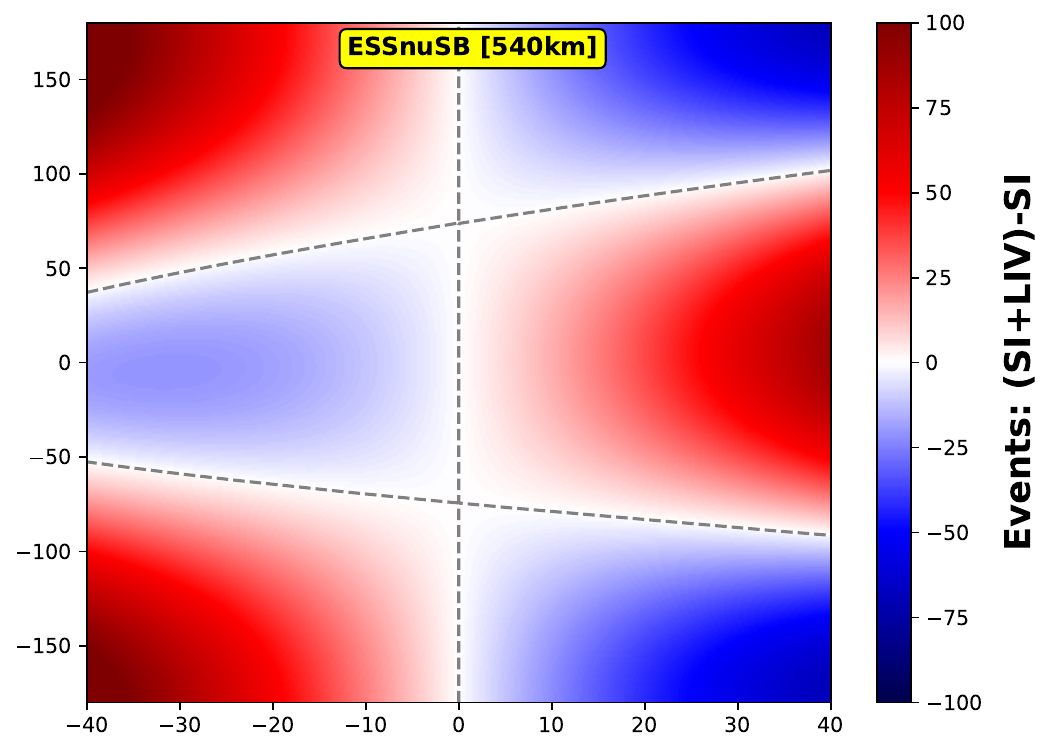}

    \includegraphics[width=0.3298\linewidth]{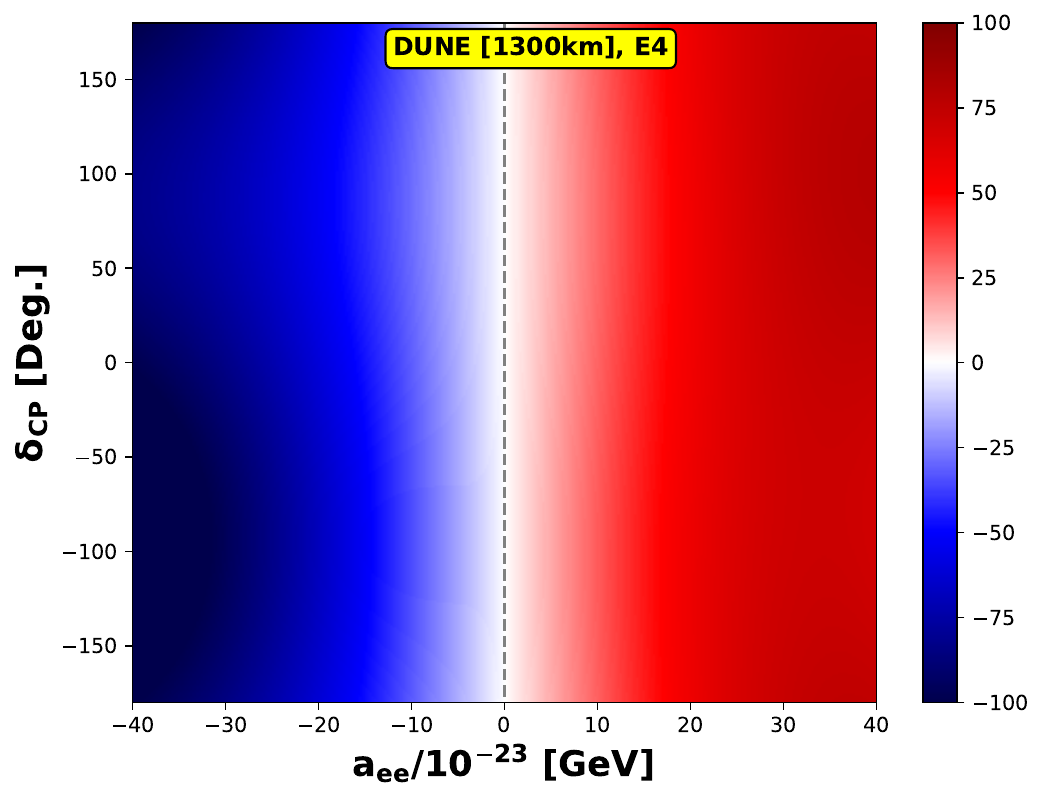}
    \includegraphics[width=0.3141\linewidth]{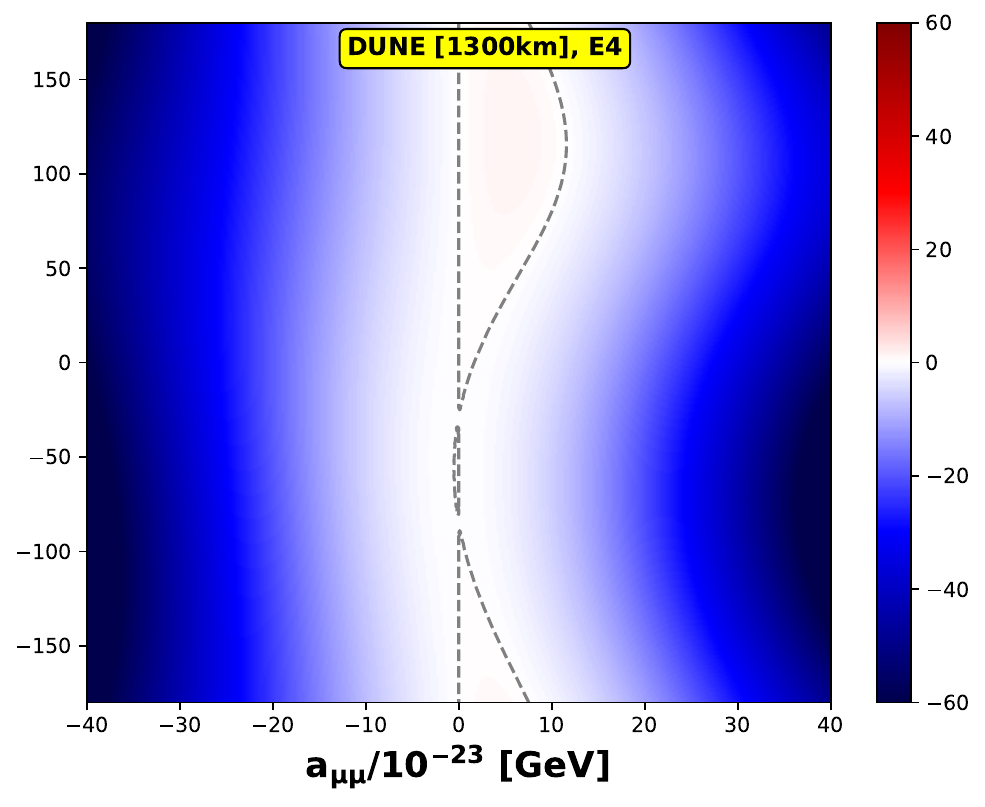}
    \includegraphics[width=0.3259\linewidth]{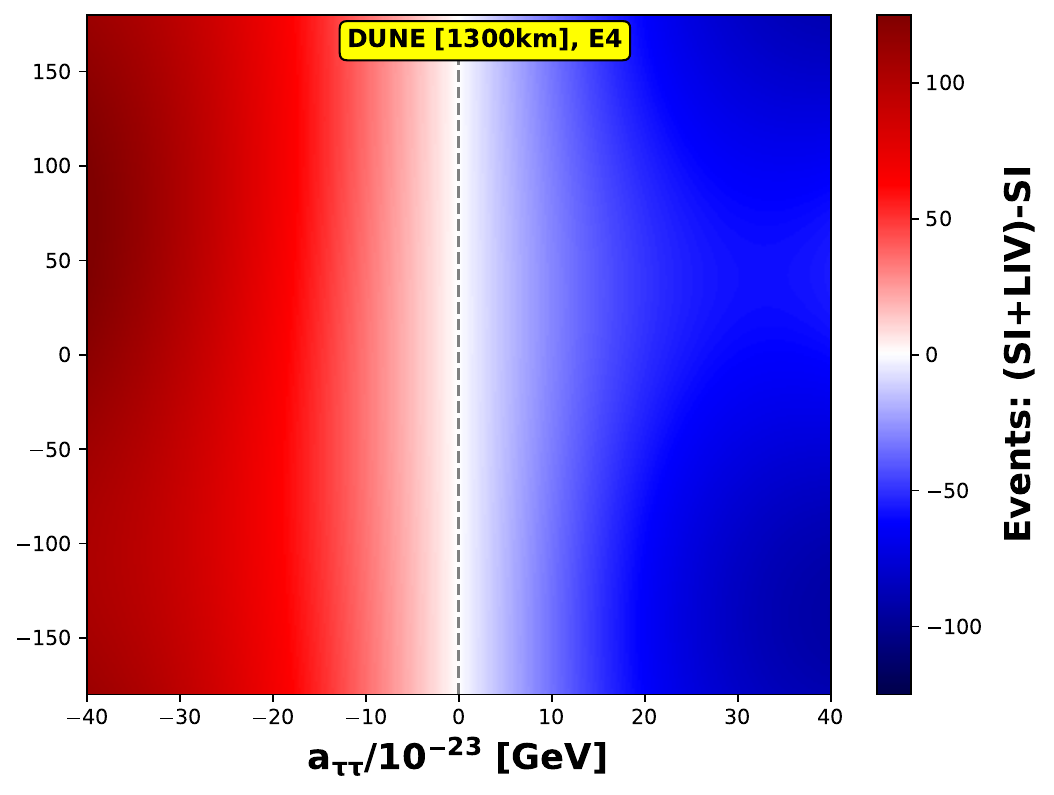}
    
    \caption{The heatmap of change in events ($\Delta N_{events}$ =  $N_{events}(SI+LIV) - N_{events}(SI)$) in $\delta_{CP}$--$a_{\alpha\beta}$ plane for the diagonal LIV parameters. The top three panels are for ESSnuSB and the bottom panel is for the DUNE experiment. The E1, E2 and E3 correspond to energy values of 0.3 GeV, 0.6 GeV, and 2.5 GeV, respectively, which represent the peak values for the ESSnuSB.}
    \label{fig:dcp_2d_3}
\end{figure}

\begin{figure}[H]
    \centering
    \includegraphics[width=0.3298\linewidth]{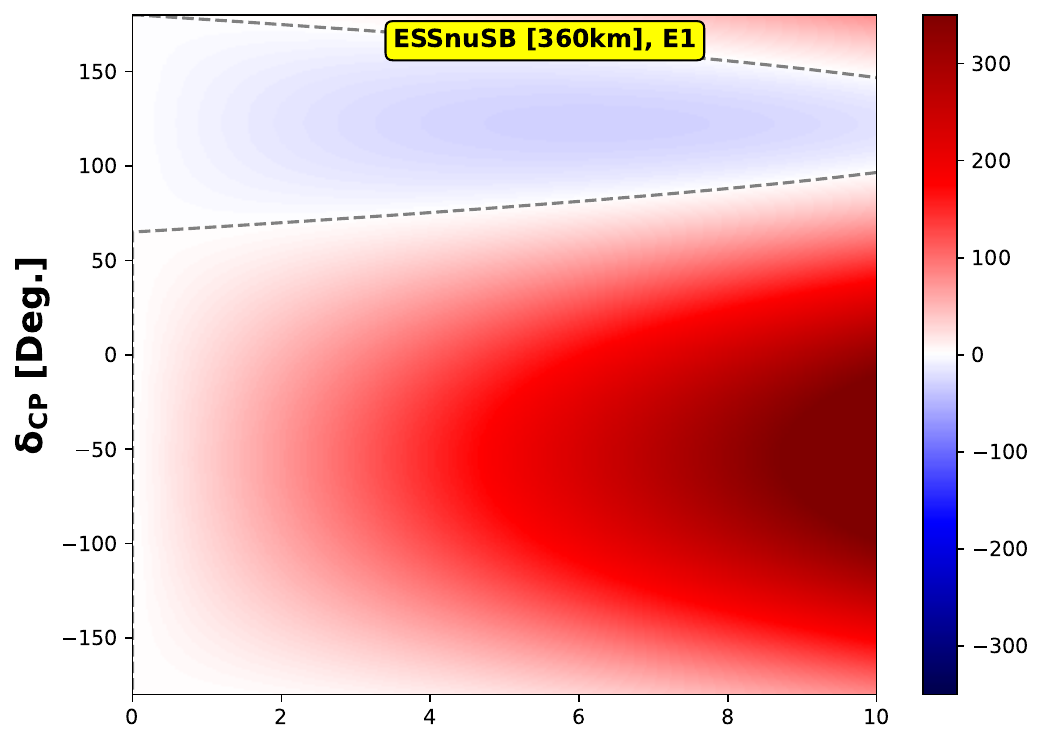}
    \includegraphics[width=0.3141\linewidth]{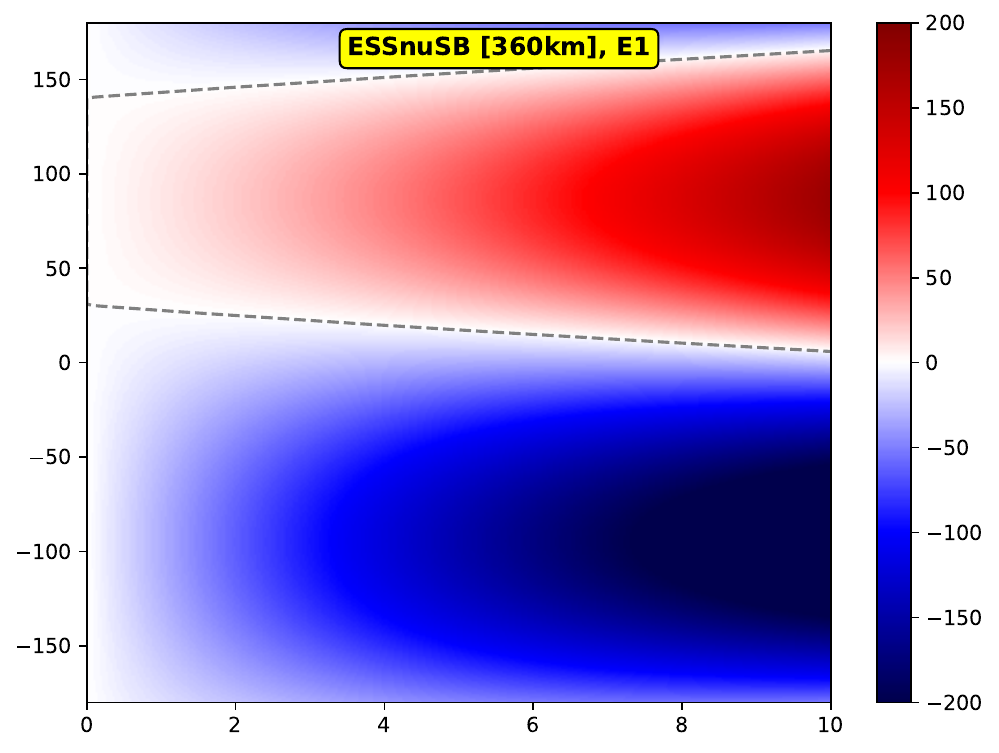}
    \includegraphics[width=0.3259\linewidth]{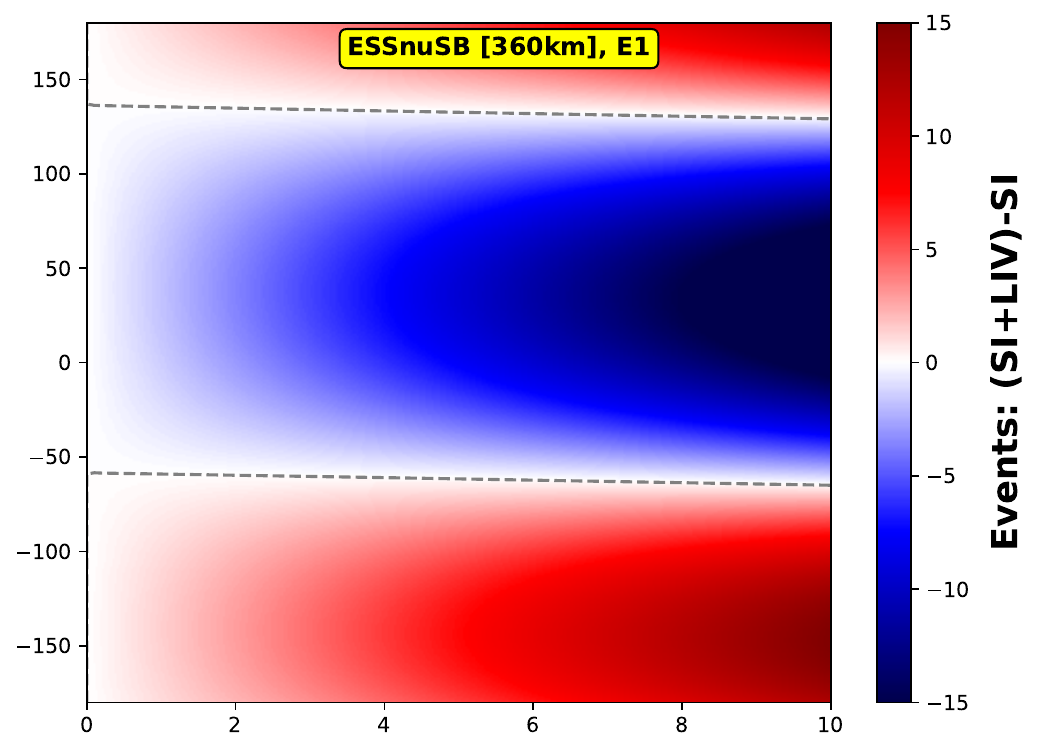}

    \includegraphics[width=0.3298\linewidth]{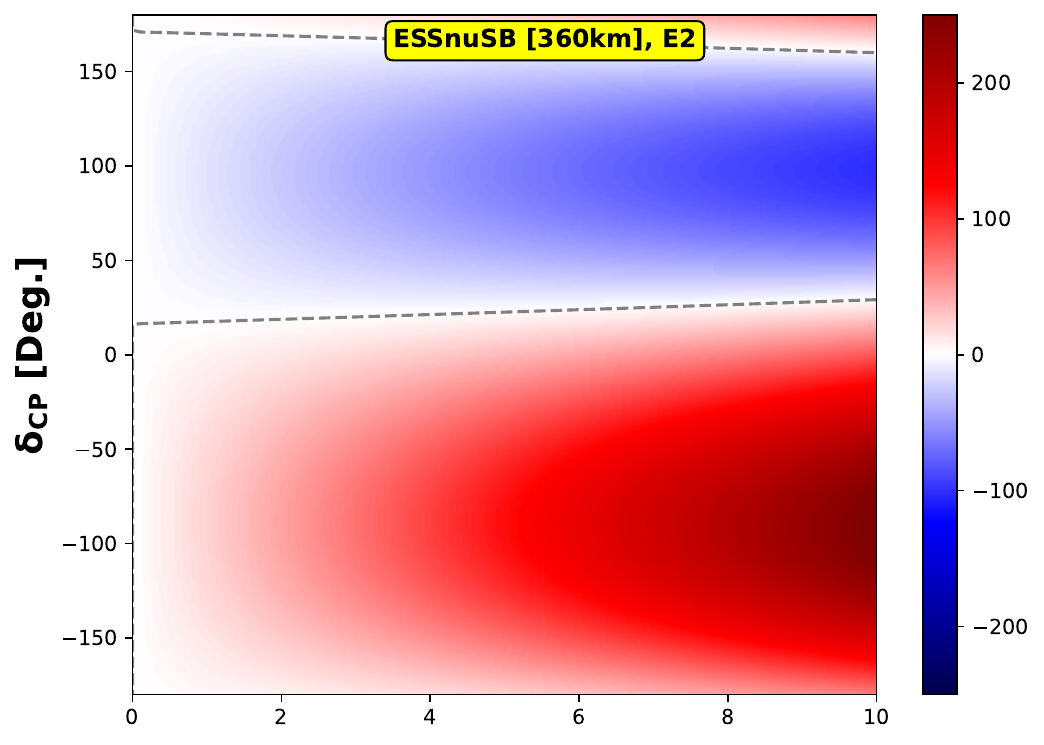}
    \includegraphics[width=0.3141\linewidth]{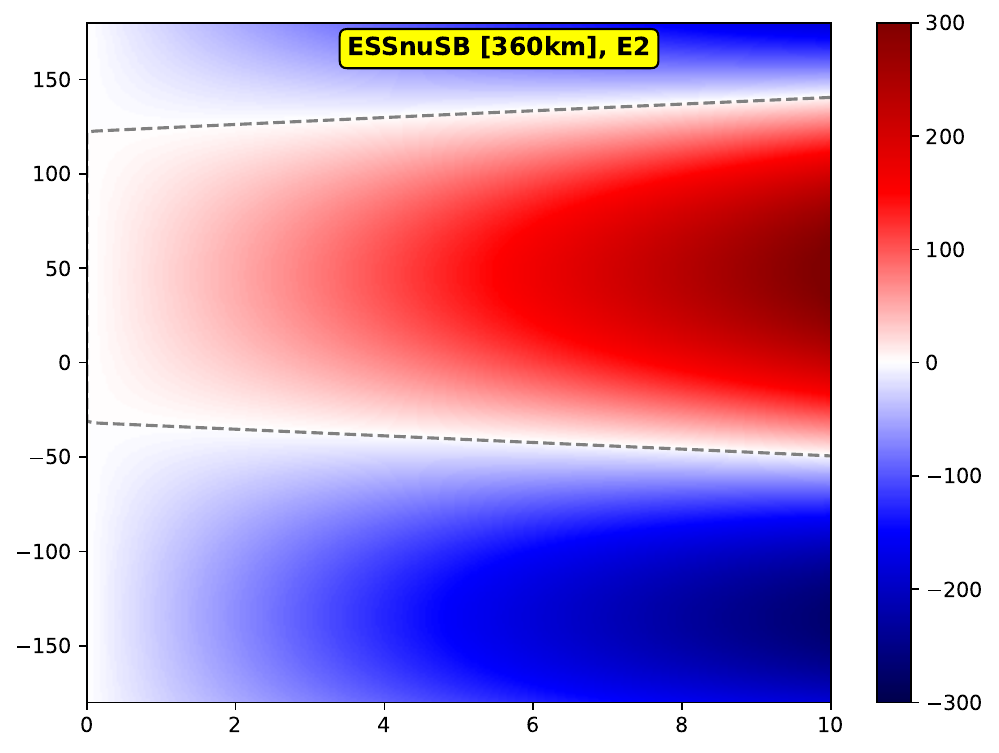}
    \includegraphics[width=0.3259\linewidth]{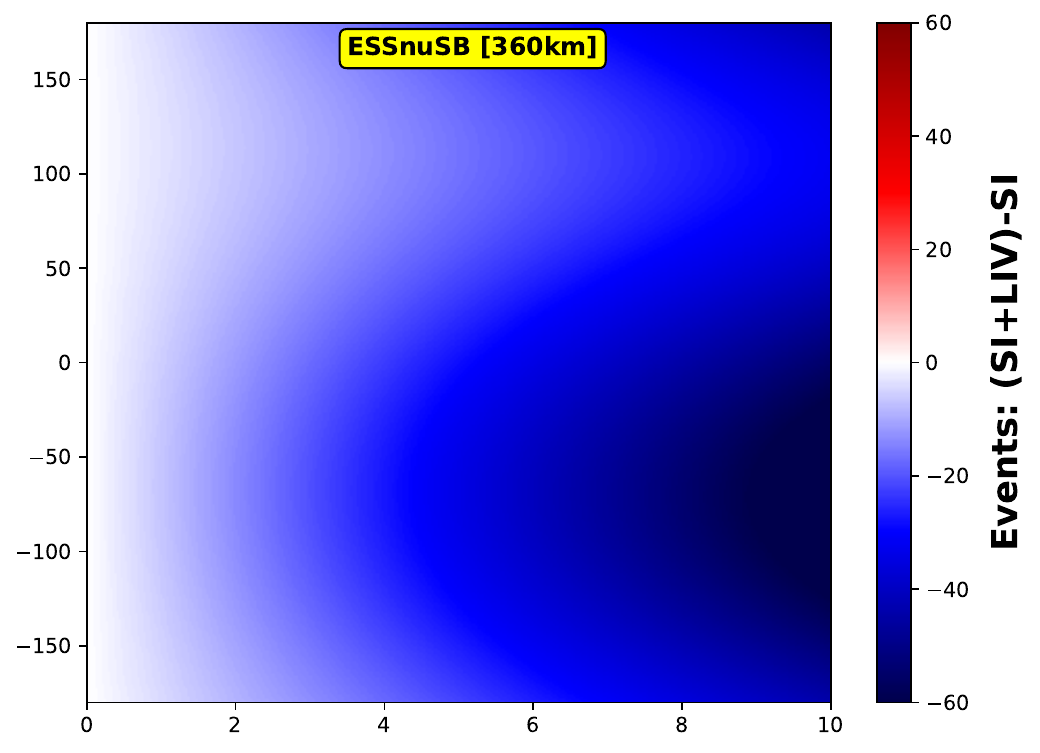}

    \includegraphics[width=0.3298\linewidth]{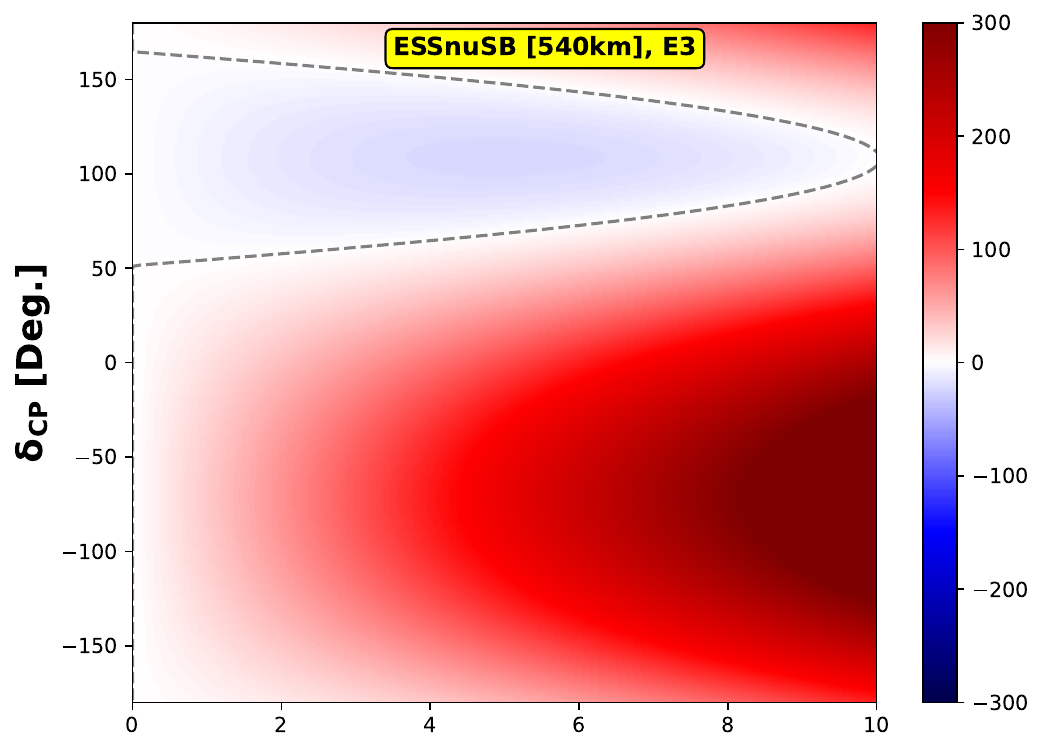}
    \includegraphics[width=0.3141\linewidth]{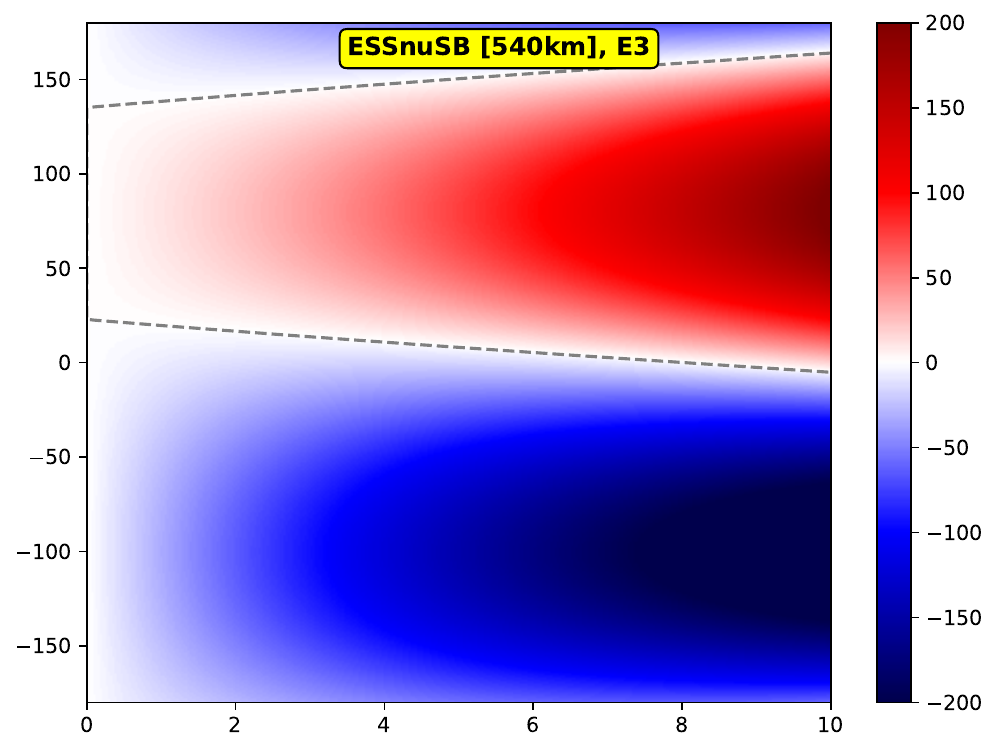}
    \includegraphics[width=0.3259\linewidth]{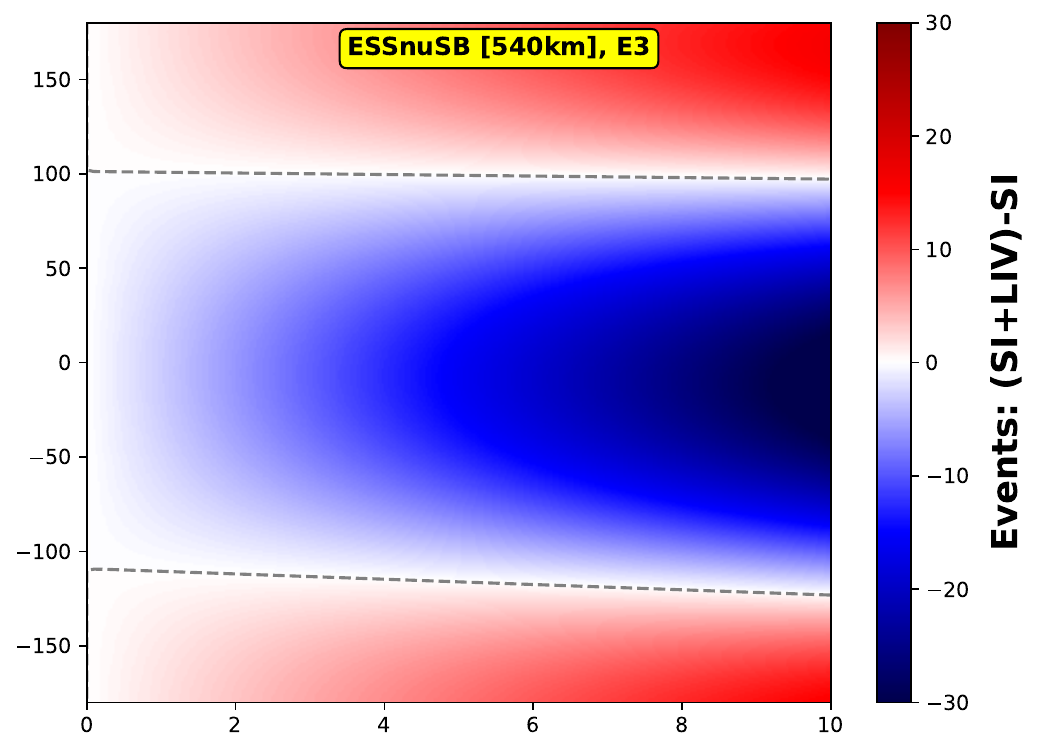}

    \includegraphics[width=0.3298\linewidth]{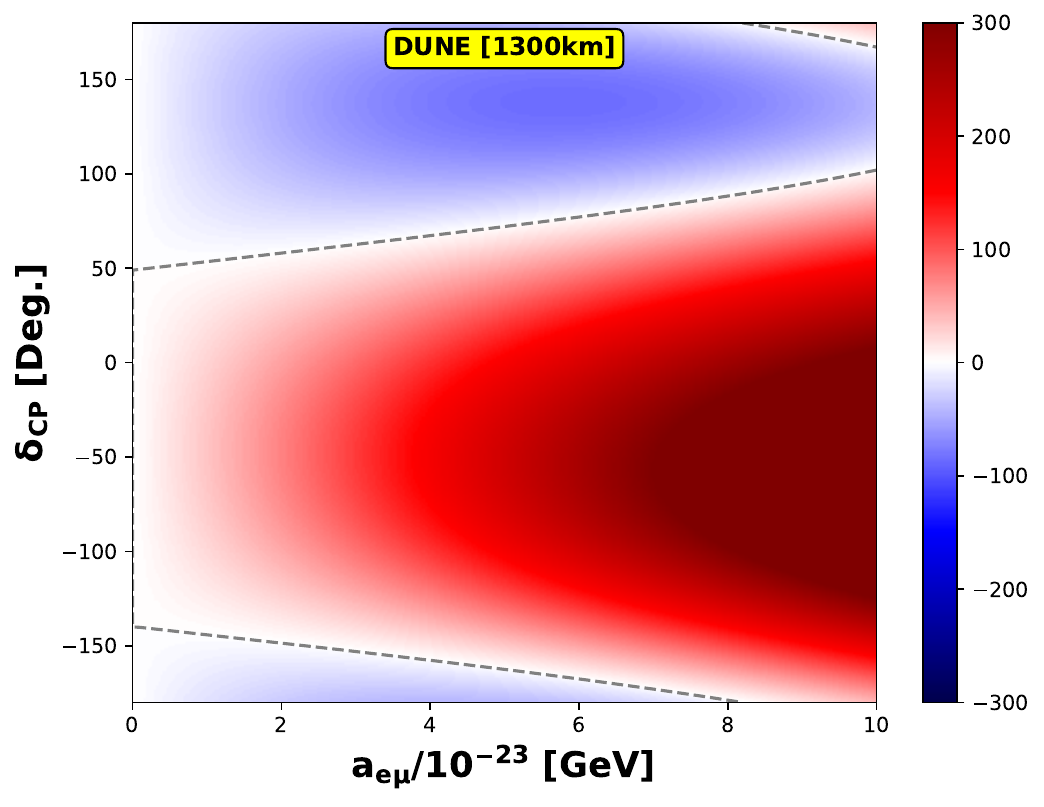}
    \includegraphics[width=0.3141\linewidth]{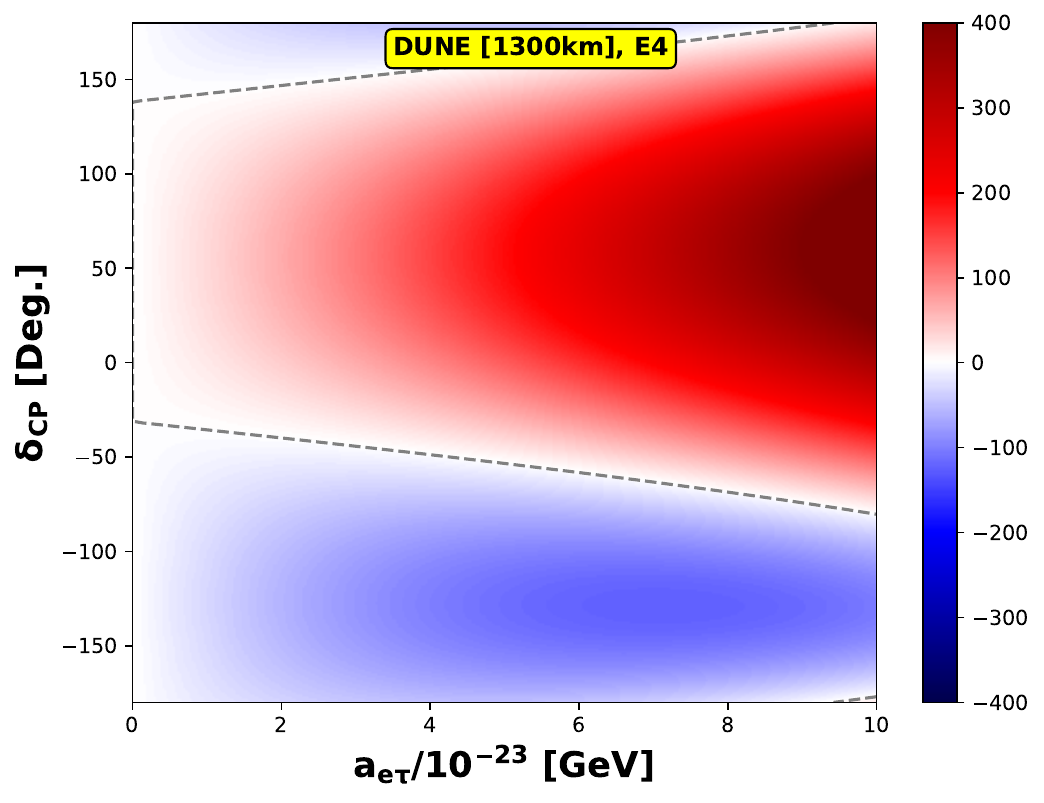}
    \includegraphics[width=0.3259\linewidth]{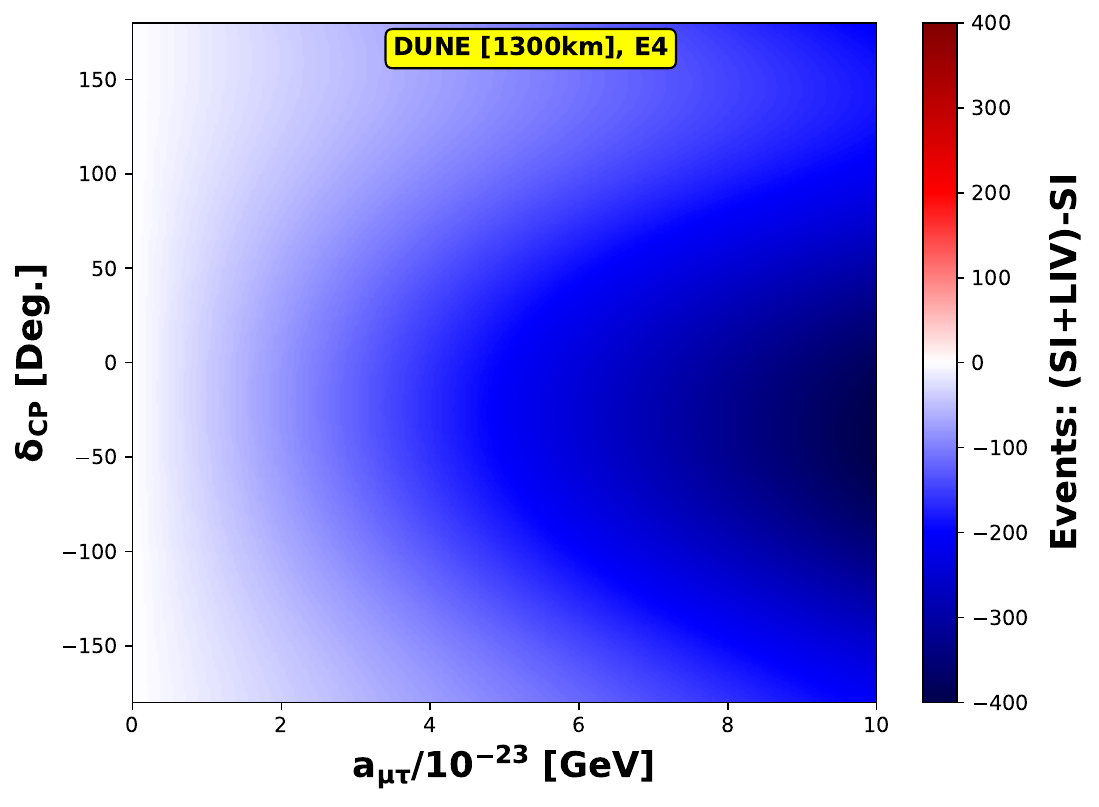}
    
    \caption{The heatmap of change in events ($\Delta N_{events}$ =  $N_{events}(SI+LIV) - N_{events}(SI)$) in $\delta_{CP}$--$a_{\alpha\beta}$ plane for the off diagonal LIV parameters. The top three panels are for ESSnuSB and the bottom panel is for the DUNE experiment. The E1, E2 and E3 correspond to energy values of 0.3 GeV, 0.6 GeV, and 2.5 GeV, respectively, which represent the peak values for the ESSnuSB.}
    \label{fig:dcp_2d_4}
\end{figure}
\end{appendix}

\newpage
\bibliographystyle{JHEP}
\bibliography{main}

\end{document}